%% file: main.tex
\documentclass[10pt,a4paper]{ctexart}

\usepackage[margin=2.2cm]{geometry}
\usepackage{microtype}
\usepackage{amsmath}
\usepackage{amssymb}

\usepackage{graphicx}
\usepackage{booktabs}
\usepackage{tabularx}
\usepackage{array}
\usepackage{caption}
\usepackage{subcaption}
\usepackage{placeins}

\usepackage{enumitem}

\usepackage{listings}
\usepackage{xcolor}

\definecolor{codegray}{RGB}{245,245,245}
\definecolor{codeborder}{RGB}{210,210,210}

\lstdefinestyle{projectstyle}{
  basicstyle=\ttfamily\small,
  backgroundcolor=\color{codegray},
  frame=single,
  rulecolor=\color{codeborder},
  breaklines=true,
  columns=fullflexible,
  keepspaces=true,
  showstringspaces=false
}

\usepackage[hidelinks]{hyperref}

\input{macros}

\title{\textbf{OpenRDMA: A GPU-Native Multi-Path Reliable RDMA Transport Architecture}}
\author{%
\small
\setlength{\tabcolsep}{1.2em}%
\begin{tabular}{@{}ccc@{}}
Ziheng Ji$^{*}$ & Chunyu Liu$^{*}$ & Weihao Peng \\[2pt]
The University of Hong Kong & Xi'an Jiaotong University & Beijing University of Posts and Telecommunications \\[2pt]
\href{mailto:ziheng830@gmail.com}{\texttt{ziheng830@gmail.com}} &
\href{mailto:lcy_web@126.com}{\texttt{lcy\_web@126.com}} & \\[10pt]
Mingheng Mi & Pu Wang & Hai Zhang \\[2pt]
Pazhou Laboratory (Huangpu) & Pazhou Laboratory (Huangpu) & Pazhou Laboratory (Huangpu) \\[2pt]
\href{mailto:mmh_web@163.com}{\texttt{mmh\_web@163.com}} & & \\[10pt]
\multicolumn{3}{c}{\footnotesize $^{*}$These authors contributed equally to this work.}
\end{tabular}%
}
\date{}

\begin{document}

\maketitle

\renewcommand{\abstractname}{Abstract}
\input{sections/00_abstract}

\input{sections/01_introduction}

\input{sections/02_background_motivation}

\input{sections/03_system_overview}

\input{sections/04_design}

\input{sections/05_implementation}

\input{sections/06_evaluation}


\input{sections/08_conclusion}

\renewcommand{\refname}{References}
\markboth{References}{References}
\bibliographystyle{unsrt}
\bibliography{bibliography/references}

\end{document}

%% file: macros.tex
\newcommand{\techterm}[1]{\textbf{#1}}

\newcolumntype{Y}{>{\raggedright\arraybackslash}X}

%% file: sections/00_abstract.tex
\begin{abstract}

As the computing power density of large-scale AI clusters rises sharply, network communication is emerging as the limiting factor in system performance. To meet the demands for overlap between computation and communication in high-frequency collective communication scenarios, GPUs must submit network send requests with extremely low overhead between adjacent computation phases. However, existing communication techniques between GPUs and NICs still incur significant control-path overhead: GPUDirect RDMA (GDR) relies on the CPU for its control path, while InfiniBand GPUDirect Async (IBGDA) requires assembling Work Queue Entries in GPU memory via atomic operations, consuming additional SM resources and memory bandwidth.

To address these issues, we propose \textbf{OpenRDMA}, an open-source multi-path reliable RDMA transport architecture for GPU-friendly communication. OpenRDMA makes three main contributions: \textbf{(1) A GPU-friendly minimalist doorbell mechanism.} To the best of our knowledge, this is the first design that completely offloads the heavy WQE assembly task to dedicated FPGA hardware---the GPU only issues a lightweight doorbell, and the NIC parses it to generate complete WQEs. \textbf{(2) An open-source multi-path out-of-order reliable transport protocol inspired by MRC.} The protocol is designed as an improvement on the Multipath Reliable Connection（MRC） framework and, to address the high degree of reordering caused by packet spraying, achieves in-order delivery through ordered DMA and flag polling. \textbf{(3) A hardware-software co-designed retransmission management mechanism.} Through a division of labor where the GPU initiates requests, the NIC performs fast data transfer, and the CPU software manages retransmissions, complex timeout detection, state tracking, and recovery policies are moved to the software side, preserving the GPU's fast submission path while substantially reducing NIC hardware state overhead, making the architecture deployable on resource-constrained board-level FPGAs.

This paper presents the architectural design and key hardware implementation aspects of OpenRDMA, and the prototype implementation on an FPGA platform is currently underway, offering a new design paradigm for next-generation GPU-friendly RDMA communication.

The current work is published at \url{https://github.com/open-rdma/open-rdma}, which has received 391 GitHub stars as of November 2026.
\end{abstract}

%% file: sections/01_introduction.tex
\section{Introduction}

With GPUDirect RDMA, the NIC can bypass host-memory staging
and move data directly between the network and GPU memory~\cite{gpudirect-rdma}.
GPU-initiated communication, represented by IBGDA,
allows a GPU kernel to directly construct network requests and ring the NIC Doorbell~\cite{ibgda,nccl-device-api,doca-gpunetio}.
This path removes the CPU proxy,
yet still requires the GPU to operate a queue protocol designed for GPU software threads.
A GPU thread must construct work descriptors, advance the work queue,
update the Doorbell record, and learn of and reclaim completed requests
through the completion queue~\cite{ibgda}.
Connection establishment, memory registration, and Queue Pair (QP) initialization are still handled by the CPU/runtime,
but the per-operation fast path still exposes the full Work Queue Entry (WQE)/Completion Queue Entry (CQE) lifecycle.
The implicit premise is that the GPU must drive RDMA queues like a CPU
in order to become an active RDMA endpoint.
This paper challenges this premise.

When a GPU kernel initiates communication, what it actually produces is only an intent---
to carry out a data exchange that has already been established, authorized, and configured.
When it consumes the communication result, what it actually needs is only a fact---
that the corresponding data has arrived completely and can be safely read.
Static information such as addresses, Memory Keys, peers, routes, and permissions
is typically already written into the NIC context by the CPU/runtime
and should not be re-encoded into a complete work descriptor on every operation.
Similarly, the GPU need not parse CQEs, advance the CQ head, and reclaim completion entries
just to determine whether a pre-arranged data exchange is ready.
If the NIC can recover this information from a protected context,
the GPU-facing RDMA interface can be reduced from a queue protocol to
\textbf{a submission doorbell token and a completion flag}.

Simplifying the communication initiation path does not mean discarding transport semantics;
rather, it offloads complex work such as state maintenance and correctness guarantees to the NIC.
This problem is especially difficult in multi-path networks.
To exploit multiple network planes, rails, or fabric paths in parallel,
packets of the same message can be distributed across different paths on a per-packet basis
~\cite{mprdma,mrc-spec,mrc-srv6-paper}.
Because each path has different propagation, queuing, and congestion states,
packet reordering becomes a normal operating condition under packet spraying
~\cite{mprdma,left}.
When the receiver sees a later packet,
it can neither immediately treat intermediate gaps as packet loss,
nor notify the GPU to consume data merely because a packet has arrived.
The system must distinguish packet arrival, payload placement,
and GPU-visible completion.
Premature retransmission would misclassify normal cross-path latency differences as packet loss,
and premature completion may cause the GPU to read payload that is not yet ready.

Based on this observation, we propose OpenRDMA,
a multi-path reliable RDMA NIC architecture for GPU-friendly communication.
The core principle of OpenRDMA is
\textbf{to let the GPU only express communication intent and observe data readiness,
while letting the NIC maintain the complete semantics from packet transmission to message publication in a layered manner}.
The CPU/runtime still handles connection establishment, memory registration, permission configuration,
route preparation, and exception recovery;
once the communication phase begins, the CPU no longer participates in the per-operation fast path.

OpenRDMA realizes this principle across three layers: submission, transmission, and completion.
On the submission side, the CPU/runtime pre-writes addresses, Memory Keys, peers, routes, and permissions
into a protected Slot Table.
The GPU kernel only needs to write an 8-byte Doorbell token via a single store,
after which the NIC recovers and validates the complete request.
On the transmission side, the sender selects a path for each packet,
while the receiver records sparse arrival state in a bitmap
~\cite{mprdma,left,irn,mrc-srv6-paper}
and uses selective feedback and retransmission to distinguish normal reordering from packet loss
~\cite{mrc-spec,uec-spec,irn}.
On the completion side, the NIC updates the GPU-visible completion flag
only after the corresponding payload has been written,
ordering dependencies have been satisfied,
and the platform's write ordering and GPU visibility conditions have been met.
The GPU thus only needs to submit ``what to do'' and poll ``is it ready,''
without managing the full WQE/CQE lifecycle.

The main contributions of this paper are as follows:

\begin{enumerate}[leftmargin=2em, label=\textbf{\arabic*.}]

    \item \textbf{A minimal RDMA work submission mechanism for GPUs.}
    OpenRDMA offloads WQE assembly into the FPGA.
    The GPU only needs to write a lightweight Doorbell token,
    which the NIC parses and expands into a complete WQE,
    thereby reducing submission overhead on the GPU side.

    \item \textbf{A multi-path out-of-order reliable transport protocol based on MRC.}
    OpenRDMA treats packet spraying and out-of-order arrival as normal operating conditions,
    combines selective feedback and retransmission to recover lost data,
    and guarantees ordered data visibility through in-order DMA and GPU-visible Flag polling.

    \item \textbf{A hardware-software co-designed retransmission management mechanism.}
    OpenRDMA adopts a division of labor where the GPU issues requests,
    the NIC executes data transfer, and CPU software manages retransmission.
    This moves complex timeout detection, state tracking, and recovery policies
    to the software side, thereby reducing hardware state overhead in the NIC.
    
\end{enumerate}

Section~\ref{sec:background} introduces the relevant background and problem scope.
Section~\ref{sec:system-overview} describes the end-to-end architecture and semantic layering.
Section~\ref{sec:design} unfolds the core mechanisms.
Section~\ref{sec:implementation} presents the hardware datapath organization and implementation considerations.
Finally, present the conclusion.

%% file: sections/02_background_motivation.tex
\section{Background and Motivation}
\label{sec:background}

This section introduces the background needed to understand OpenRDMA
and analyzes the limitations of existing RDMA protocols in GPU direct
communication and multi-path transport scenarios.
We first review the request submission, data transfer, and completion notification
paths of traditional RDMA, then discuss the interface granularity mismatch
that arises when GPUs initiate communication.
Next, this section explains why per-packet multi-path execution makes reordering
a normal condition and why packet-level reliable delivery does not directly
imply that the GPU can safely consume data.
Finally, this section distills the above issues into the design requirements for OpenRDMA.

\subsection{Traditional RDMA Fast Path}
\label{sec:bg-rdma-fast-path}

RDMA allows the NIC to move data directly between pre-registered and authorized
memory regions, avoiding byte-by-byte data copying by the CPU.
A typical RDMA endpoint uses a Queue Pair (QP) to hold connection and transport
context. A QP typically contains a Send Queue (SQ) and a Receive Queue (RQ),
and is associated with a Completion Queue (CQ).
The SQ is used to submit active operations such as RDMA Write, RDMA Read, and Send;
the RQ prepares receive resources for message operations that require buffers on the
receiver side;
the CQ holds completion records produced by the NIC, allowing software to confirm
whether an operation has completed and to reclaim the corresponding resources
accordingly~\cite{rdma-aware-programming}.

Software describes an RDMA operation through a Work Queue Entry (WQE).
A WQE contains the opcode, local or remote address, data length, Memory Key,
and related control information.
Software writes the WQE into the work queue, advances the producer state,
and notifies the NIC of a new request via a doorbell.
The NIC then fetches and parses the WQE, checks addresses and access permissions,
reads local send data via PCIe DMA Read, or writes the payload to the target
memory via DMA Write on the receiver side.
Once the operation meets the protocol-specified completion conditions,
the NIC can write a Completion Queue Entry (CQE) to the CQ;
software reads the status, advances the CQ consumer state,
and reclaims request resources~\cite{rdma-aware-programming}.

A large RDMA operation is split into multiple network packets.
The reliable transport protocol identifies these packets using Packet Sequence
Numbers (PSNs) and maintains send and receive progress through mechanisms
such as ACKs, NAKs, and timeouts.
In traditional single-path or low-reordering environments,
the cumulative acknowledgment point can effectively indicate that a contiguous
range of PSNs has arrived;
the sender releases the corresponding state accordingly
and retransmits missing data upon detecting anomalies or timeout expiration~\cite{irn}.

The traditional RDMA fast path thus exposes three categories of state
to the request producer:

\begin{enumerate}
    \item WQEs and work queue producer state for describing operations;
    \item PSNs and acknowledgment state for maintaining packet reliability;
    \item CQEs and completion queue state for notifying software and reclaiming resources.
\end{enumerate}

This interface is suitable for general-purpose communication driven by CPU software threads.
The CPU can construct complex descriptors, maintain shared queue pointers,
and perform error handling and resource reclamation based on the status in CQEs.
However, when communication is triggered by GPU kernels at high frequency
and fine granularity, exposing the same queue protocol directly to the GPU
creates a new interface mismatch.

\subsection{GPU-Initiated Communication and Interface Mismatch}
\label{sec:bg-gpu-interface}

GPUDirect RDMA allows third-party NICs to directly access registered GPU memory
via PCIe, enabling payload to move between the network and GPU memory without
host-memory staging~\cite{gpudirect-rdma}.
However, GDR primarily addresses memory reachability on the data path
and does not separately define how the GPU submits network requests
or learns that remote data is ready.

With GDR, the local NIC can directly read payload from the sender GPU memory,
and the remote NIC writes it to the receiver GPU memory;
the entire data transfer process does not traverse host memory~\cite{gpudirect-rdma}.
However, the ability to transfer payload directly does not mean the control path
also bypasses the CPU.
One common approach still uses a CPU proxy: the GPU kernel writes communication
requests into a shared queue or host-visible memory, and the CPU proxy reads
the request, constructs the WQE, submits the doorbell, and processes the CQE
after the operation completes.
This approach can reuse the mature RDMA software stack,
but every operation requires synchronization between the GPU and the CPU,
placing CPU scheduling back on the per-operation critical path~\cite{potluri2017gpu}.

GPU-initiated communication further allows the device-side producer to directly
drive network operations.
The GPUDirect Async Kernel-Initiated Communication（GDAKI） transport, represented by IBGDA, can place both the control path
and the data path of InfiniBand communication onto the GPU,
reducing dependence on a CPU reverse proxy~\cite{ibgda,nccl-device-api,doca-gpunetio,nvshmem-env}.
However, communication is still organized around protocol objects such as QPs,
doorbells, work descriptors, and device-side queue state.
GPU threads or the GPU runtime must participate in request generation,
queue position coordination, memory ordering maintenance, and related state
management~\cite{ibgda,cuda-c-programming-guide}.

\begin{figure}[htbp]
    \centering
    \includegraphics[width=\textwidth]{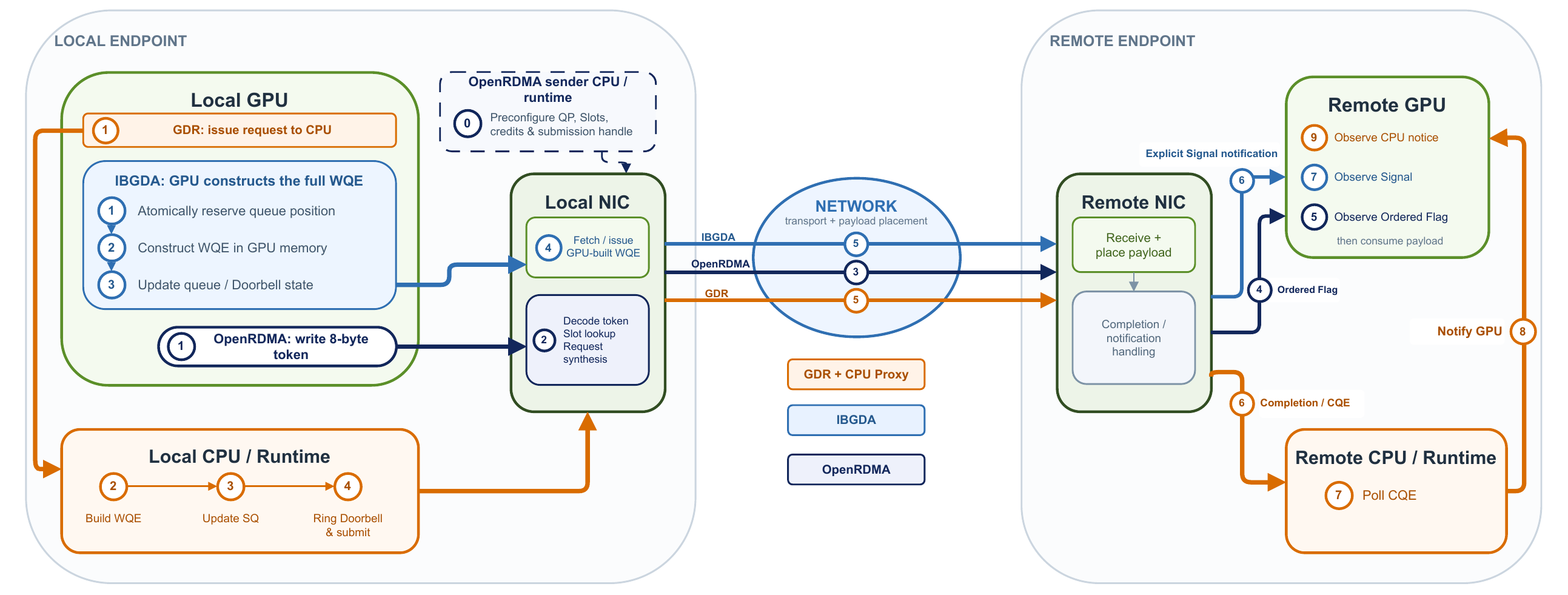}
    \caption{Per-operation control path comparison across traditional CPU-proxy RDMA,
    IBGDA, and OpenRDMA.
    The three paths share payload transport but differ in which entity---
    CPU, GPU, or NIC---bears the primary per-operation control work;
    the dashed step 0 represents pre-configuration performed only by the
    OpenRDMA sender CPU/runtime.}
    \label{fig:end-to-end-path-comparison}
\end{figure}

Figure~\ref{fig:end-to-end-path-comparison} places the CPU proxy, IBGDA, and OpenRDMA
in the same end-to-end topology.
Each of the three per-operation paths is numbered starting from step 1,
with arrows showing the control flow from local GPU submission to remote
completion notification; OpenRDMA additionally uses a dashed step 0
to indicate pre-configuration by the sender CPU/runtime,
which does not enter the per-operation fast path.
The shared network backbone indicates that all three rely on the NIC
to directly move GPU payload.

\FloatBarrier

The CPU-proxy path has the longest per-operation control chain:
the local GPU first issues a request to the CPU, the CPU constructs the WQE,
updates the SQ, and rings the Doorbell; after remote completion,
the CPU must also poll the CQE and notify the GPU.

The IBGDA path removes the CPU reverse proxy, but shifts work such as
atomically reserving a queue position, constructing a full WQE in GPU memory,
and updating queue/Doorbell state onto the GPU.
The GPU-visible signal in the figure denotes an explicit notification of the operation,
rather than treating an ordinary RDMA Write as automatically signaling remote completion.

The OpenRDMA path further shortens the GPU-visible control chain:
the GPU only writes an 8-byte doorbell; the local NIC performs doorbell decode,
Slot lookup, and request synthesis; after data transfer completes and the
specified ordering conditions are met, the remote NIC issues an Ordered Flag,
which the remote GPU then observes.
Thus, the core distinction among the three paths is not whether the NIC can
directly move payload, but which entity constructs the request, maintains queue
state, and issues the remote-visible completion notification for each operation.

This interface does not match the actual needs of many GPU communication workloads.
In collective communication, pipeline parallelism, and repeated data exchanges,
remote addresses, Memory Keys, peers, QPs, access permissions, and path constraints
are typically already determined by the CPU/runtime during the initialization
phase~\cite{rdma-aware-programming,gpudirect-rdma,nccl-docs}.
What changes at runtime is often only the specific operation selected for this round,
the data length, or a small number of dynamic parameters.
What the GPU truly needs to express is: \textbf{to carry out a communication operation
that has already been established, authorized, and configured.}

It does not need to re-encode the complete connection, protection, and routing
context on every submission.
Placing static context repeatedly into each GPU-side request not only increases
descriptor generation and device-memory access, but also forces the GPU producer
to understand a queue protocol originally designed for general-purpose CPU
software stacks.

Interface simplification must also address the resource admission problem.
The fixed capacity of a traditional SQ naturally limits the number of requests
not yet consumed by the NIC; when the GPU-facing interface no longer exposes
a traditional SQ, the system must still prevent the GPU from submitting more
requests than the NIC input buffer, request state, or ordered-completion state
can accommodate.
Therefore, direct GPU submission cannot merely shrink the Doorbell format;
it must also provide an explicit, bounded backpressure mechanism that supports
concurrent producers.

This yields the first design requirement:

\paragraph{R1: Intent-driven submission with bounded admission.}
The GPU fast path should only express pre-authorized communication intent and
a small number of dynamic parameters, without requiring the GPU to construct a
full WQE or maintain the traditional queue lifecycle.
At the same time, the NIC must expose bounded admission state to the GPU
and block new requests when internal resources are insufficient.

\subsection{Multipath Execution and Out-of-Order Reliability}
\label{sec:bg-multipath-reliability}

Modern GPU clusters typically provide multiple network planes, rails, NIC ports,
or fabric paths.
Traditional RDMA usually forces all packets of a connection onto a single path
to avoid reordering at the receiver.
Even when multiple equal-cost paths exist in the network, such a connection can
only use one of them, and its throughput is limited by the available bandwidth
and congestion state of that path.

The impact of path collisions is especially severe in large-model training.
Data parallelism, tensor parallelism, and pipeline parallelism periodically
trigger collective communications such as All-Reduce, All-Gather, Reduce-Scatter,
or All-to-All; each GPU/NIC typically maintains only a small number of fixed
peers/QPs, and each flow continuously carries large blocks of gradients,
parameters, activations, or expert data during a communication phase,
forming a small number of high-rate, link-saturating \emph{elephant flows}
~\cite{pytorch-distributed,pytorch-fsdp,megatron-lm,zero,zero-plus-plus,deepspeed-ulysses,gpipe,megatron-core,deepseek-v3,nccl-docs}.
Large-model training traffic thus simultaneously exhibits low flow entropy,
burstiness, and iteration-to-iteration repetition
~\cite{gangidi2024rdma,demystifying-nccl}.

Networks typically select and pin one path per flow through per-flow ECMP hashing.
Because of the low flow entropy described above, multiple concurrent elephant flows
are more likely to hash onto the same uplink by chance.
Each elephant flow can persist near NIC line rate throughout the entire communication
phase, so collisions do not dissipate quickly as with short flows;
instead, they persistently occupy the bottleneck link and cause queue buildup,
ECN/backpressure, or even packet loss.
Meanwhile, other equal-cost paths may remain lightly loaded.
The completion time of collective communication is bounded by the slowest transfer;
a hotspot created by one collision becomes a straggler, causing participating GPUs
to wait and amplifying local congestion into a delay for the entire training iteration
~\cite{rfc2992,rfc6437,mprdma,gangidi2024rdma,presto}.

Per-packet multi-path transport allows the NIC to select different paths for
different packets of the same RDMA operation, spreading the traffic of one
connection across multiple links.
It cannot guarantee the complete elimination of congestion, but can reduce the
probability that a single path selection creates a hotspot and allow a single
connection to more fully exploit the parallel bandwidth available in the network
~\cite{mprdma,mrc-spec,mrc-srv6-paper,openai-mrc}.

Per-packet path selection also changes the fundamental assumptions of reliable
transport.
Different paths have different propagation delays, switch queuing states,
and instantaneous congestion levels, so a later-sent packet can arrive before
an earlier one.
In this setting, packet reordering is no longer a rare anomaly but a normal
consequence of multi-path execution~\cite{mrc-spec,left}.

For example, the sender has transmitted PSNs 10--20, but the receiver first
receives PSNs 11--20 while PSN 10 is temporarily missing.
PSN 10 may have been genuinely lost, or it may merely be queued on a slower path.
If the receiver only returns a cumulative ACK, the acknowledgment point remains
at PSN 9, and the sender cannot learn that subsequent packets have already arrived.
If the sender retransmits immediately upon first observing a gap, it would
misclassify normal cross-path latency differences as packet loss.

In traditional RDMA reliable protocols, represented by the Reliable Connection (RC)
transport commonly used in RoCE v2, the receiver must process packets in PSN order
and advance receive state according to the next expected PSN.
If the receiver first sees a larger PSN, it typically discards these out-of-order
packets; the sender then retransmits all subsequent data from the first gap onward
via Go-Back-N.
In high-reordering scenarios, this approach wastes already-achieved receive progress,
generates duplicate traffic, and can create head-of-line blocking~\cite{irn,left}.

Multi-path reliable transport therefore cannot rely solely on packet arrival order
to infer transport outcomes.
The receiver needs to record which packets within the current window have arrived,
and the sender needs to record which outstanding packets have received cumulative
or selective acknowledgment.
ACKs can advance the contiguous acknowledgment frontier, SACKs can expose sparse
arrival state beyond that frontier, NAKs can report protocol anomalies that can
be clearly pinpointed, and timeouts or reliability probes provide the ultimate
fallback when the feedback path has made no progress for an extended period
~\cite{mrc-spec,uec-spec,irn}.

A critical distinction is that a zero bit in the receive window only indicates
that the corresponding packet has not yet been recorded; it cannot, by itself,
prove that the packet has been lost.
Whether to trigger retransmission must also consider the acknowledgment progress
beyond that hole, explicit error feedback, or elapsed duration.
Otherwise, the reliability mechanism would repeatedly convert normal multi-path
reordering into unnecessary duplicate traffic.

Per-packet multi-path transport must also have bounded state.
If the sender can advance PSN without limit while the receiver can only store
a fixed-capacity bitmap, a sufficiently large reordering distance will eventually
exceed the hardware tracking range.
The system must therefore constrain in-flight state through a receive window,
Maximum PSN Range (MPR), or an equivalent flow-control boundary,
so that the capacities of the receiver bitmap and sender scoreboard have explicit
upper bounds~\cite{mrc-spec,uec-spec}.

This yields the second and third design requirements:

\paragraph{R2: Multipath as a first-class execution model.}
The NIC shall exploit multiple parallel paths at packet granularity,
without requiring the GPU to select the specific path for each packet
and without relying on packets arriving in send order.

\paragraph{R3: Explicit packet state and selective recovery.}
The receiver and sender shall maintain bounded, explicit per-packet state,
preserve already-achieved out-of-order receive progress, and selectively
recover still-missing packets through ACKs, SACKs, NAKs, and timeouts,
rather than simply rolling back and retransmitting the entire subsequent window.

\subsection{The Gap between Reliable Delivery and Semantic Completion}
\label{sec:bg-semantic-completion}

Packet-level reliability answers whether network packets eventually arrive,
but it does not automatically answer when the remote GPU can safely read
a communication result.
MRC likewise treats the responder's packet placement, ordering, and completion
as protocol semantics distinct from reliable acknowledgment~\cite{mrc-spec,uec-spec}.
On the OpenRDMA receive path, at least three states must be distinguished.

\paragraph{Packet arrival.}
The receiver NIC has received and validated a packet and recorded the
corresponding PSN in its receive state.
An ACK or SACK can feed this fact back to the sender,
but the payload may still reside in a packet buffer or DMA pipeline at this point.

\paragraph{Payload placement.}
The receiver NIC has submitted the payload to the DMA Write path targeting
GPU memory and has reached the placement or ordering point specified by the
architecture.
Because multiple packets and multiple DMA channels can proceed in parallel,
the order in which packets arrive at the NIC does not directly represent the
order in which payload is written to GPU memory.

\paragraph{GPU-visible semantic completion.}
After the GPU observes the completion state under the memory-ordering contract
specified by the runtime, it can safely read the data represented by that operation.
In addition to the current payload having completed placement, this state may
also require that earlier ordered operations on the same QP have satisfied
their publication conditions.

Thus, no direct equivalence holds among the three states:

\begin{equation}
\text{packet arrival}
\;\not\Rightarrow\;
\text{payload placement}
\;\not\Rightarrow\;
\text{GPU-visible completion}.
\end{equation}

If the NIC issues a completion notification immediately upon packet arrival,
the GPU may read data that has not yet been fully written.
If multiple Writes can enter the DMA pipeline out of order,
merely waiting for the last packet of the current message to arrive
also does not guarantee that prior ordering dependencies have been satisfied.

Furthermore, GPUDirect RDMA introduces a data path independent of the GPU
execution pipeline.
The fact that a third-party device has issued PCIe writes does not naturally
imply that a concurrently running GPU kernel has observed all updates in the
expected order; the platform must still provide explicit synchronization and
memory-ordering contracts~\cite{gpudirect-rdma}.
Therefore, the completion condition inside the NIC is responsible for establishing
the device-side payload-before-notification ordering,
while the GPU runtime must still specify the load, acquire, or fence semantics
to be used when polling the completion flag.

A traditional CQE can report operation status to software, but the CQE itself
belongs to a completion protocol oriented toward software queue consumers.
Letting the GPU directly manage the CQ means the GPU side must still parse CQEs,
coordinate the CQ head, and reclaim ring entries,
yet what many GPU kernels truly need is only a monotonic data-ready state.
On the other hand, if the CPU proxy continues to read CQEs and translate them
into GPU-visible state, the CPU re-enters the per-operation completion path.

GPU-friendly communication therefore requires a semantic completion predicate
that is independent of packet acknowledgment.
This predicate should explicitly specify: which packets have arrived,
which payloads have completed placement, and which QP-ordering conditions
have been satisfied; only after this predicate holds may the NIC publish
an observable completion state to the GPU.

This yields the fourth design requirement:

\paragraph{R4: Ordered GPU-visible completion.}
The system must maintain packet reliability, payload placement, and message
completion as separate concerns.
Only after the relevant data has completed placement and satisfied the specified
ordering dependencies may the NIC issue a GPU-visible completion,
so that the GPU can determine a safe consumption point without managing a
traditional CQ.

\subsection{Control--Data Plane Separation}
\label{sec:bg-control-data-separation}

Moving more execution semantics into the NIC does not mean the CPU/runtime
no longer participates in the system.
Memory registration, QP creation, permission configuration, path preparation,
resource reclamation, and fatal error recovery all involve complex policies
and resource lifecycles and are suitable for CPU/runtime control.
In contrast, packet arrival, PSN acknowledgment, DMA placement, and completion
condition evaluation occur at high frequency and directly affect transport
correctness; they are not suitable for waiting on software round-trips in the
common-case fast path.

The placement of reliability state must also respect hardware resource boundaries.
Maintaining unbounded state for every packet, QP, and timer is infeasible;
but handing all acknowledgment and retransmission decisions to the CPU would
introduce PCIe round-trips, software scheduling, and long recovery latency.
The system needs to distinguish two categories of state:

\begin{itemize}
    \item Correctness state that must be maintained in a closed loop within the
          NIC fast path, such as the bounded receive window,
          outstanding confirmation state, DMA placement state, and
          completion-ordering state;
    \item Policy state that can be configured or handled at low frequency by
          the CPU/runtime, such as resource lifecycles,
          routing configuration, protocol parameter updates, and unrecoverable errors.
\end{itemize}

First transmissions and retransmissions also share the same DMA, packet assembly,
and network send resources.
The reliability mechanism determines which packets need to be resent,
congestion control determines whether injection into the network is currently
permitted, multi-path scheduling selects the specific output path,
and resource backpressure protects the NIC's internal bounded queues.
These mechanisms can share the send datapath but must not be conflated into
a single control problem.

This yields the fifth design requirement:

\paragraph{R5: Bounded hardware state with control--data plane separation.}
The NIC shall maintain common-case packet transmission, reliability, and
completion state within explicit window and resource bounds;
the CPU/runtime is responsible for initialization, configuration, and exception
recovery and does not participate in the per-operation submission, acknowledgment,
and completion paths under normal conditions.

\subsection{Design Requirements}
\label{sec:bg-design-requirements}

In summary, OpenRDMA must simultaneously satisfy the following requirements:

\begin{enumerate}
    \item Support direct GPU submission using compact, pre-authorized
          communication intent, with bounded resource backpressure;
    \item Treat per-packet multi-path and out-of-order arrival as the normal
          execution model;
    \item Use explicit packet state to preserve out-of-order progress and
          perform selective recovery;
    \item Decouple packet-level reliable delivery from GPU-visible semantic
          completion;
    \item Establish a clear, bounded state boundary between the NIC fast path
          and the CPU/runtime control path.
\end{enumerate}

The next section describes how OpenRDMA maps these requirements onto an
end-to-end system architecture.
The submission path transforms GPU intent into protected NIC internal requests;
the multi-path transport and reliability path executes and recovers packets;
the completion path issues a safe consumption point to the GPU after payload
placement and intra-QP ordering conditions have been satisfied.

%% file: sections/03_system_overview.tex
\section{System Overview}
\label{sec:system-overview}

OpenRDMA provides GPUs with a doorbell-in, flag-out end-to-end communication path: the CPU handles initialization --- memory registration, QP creation, Slot Table, path, and credit configuration --- while at runtime GPU-submitted data flows through the system along three end-to-end data streams. \textbf{Send path:} after the GPU issues a doorbell, the NIC assembles the WQE, fetches data via DMA, sprays packets across four network channels, and simultaneously writes the WQE's PSN range into the WQE Record to initiate send-side state tracking. \textbf{Receive path:} packets from the four channels merge into the Packet Buffer, are parsed to update the RQ Bitmap, DMA-written into remote GPU memory, and upon ordering completion a Flag is written to notify the GPU that the data is safe to consume. \textbf{Retransmission path:} the receiver-side RQ Bitmap detects holes and generates ACK/SACK/NAK fed into the retransmission logic; the latter queries the Scoreboard and WQE Record to locate missing packets, which are re-injected into the send path under congestion control scheduling, with Timer providing timeout fallback.

\begin{figure}[htbp]
    \centering
    \includegraphics[width=0.7\textwidth]{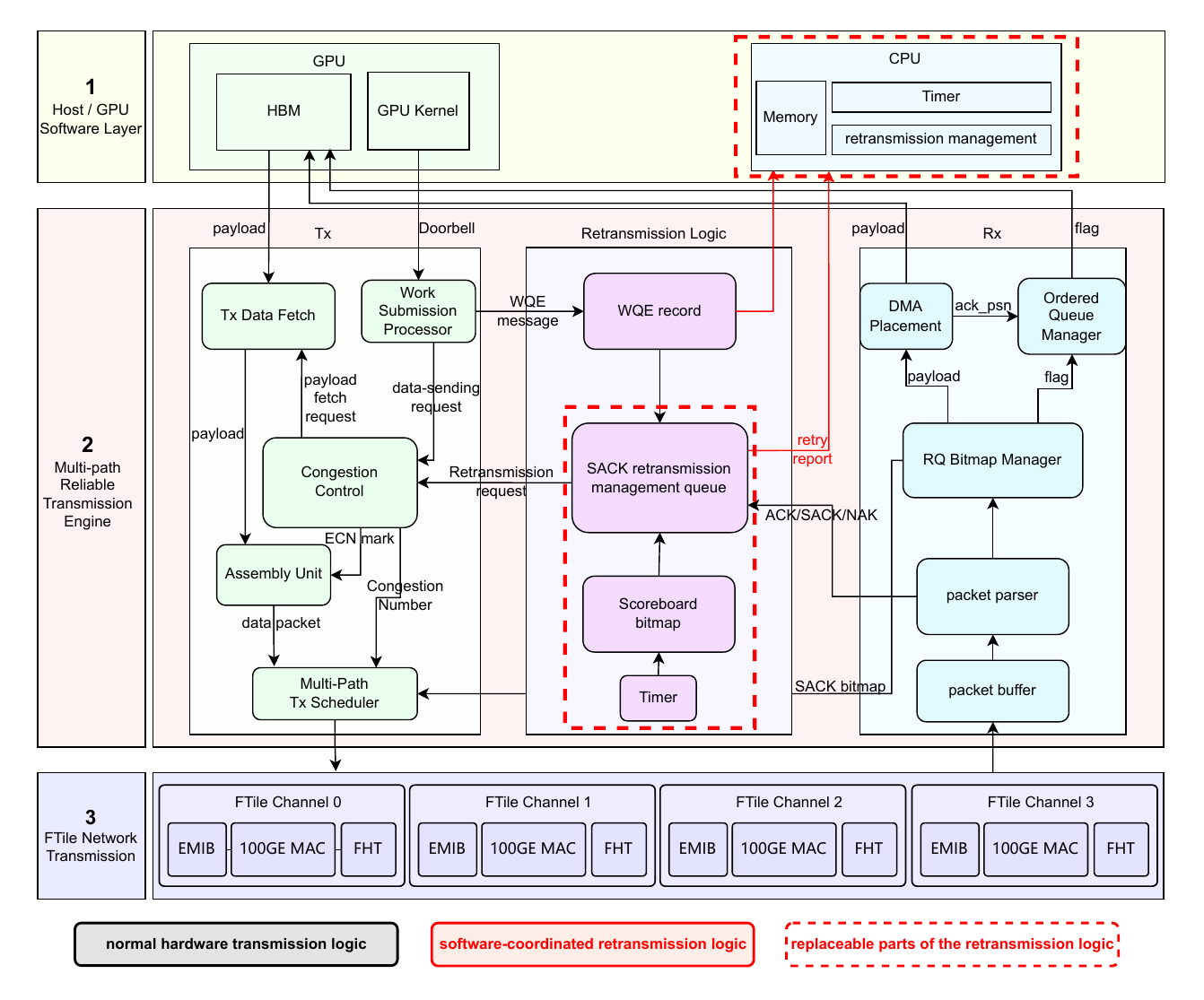}
    \caption{OpenRDMA system architecture overview.}
    \label{fig:arch-overview}
\end{figure}

We designed OpenRDMA as a layered modular system. From top to bottom: the Host/GPU software layer, the multipath reliable transport engine, and the four-channel FTile network transport layer, as shown in Fig.~\ref{fig:arch-overview}. The architecture lets the GPU focus on computation, enabling communication primitives with extremely low GPU-side overhead.

\subsection{Host/GPU Software Layer}

The Host/GPU software layer provides the user interface and resource management, supporting dual-path work submission from both CPU and GPU.
The Host side runs the CPU driver and runtime, responsible for QP lifecycle management, resource allocation and reclamation, Aperture/Slot Table configuration, and software-side retransmission management and Completion processing.
The GPU side runs compute kernels and the communication runtime; the GPU issues doorbells directly to the NIC via MMIO-Write, triggering DMA-Write.
Host and GPU each have independent physical memory (Host Memory and GPU device memory), which the NIC accesses via PCIe DMA Read/Write respectively.

\subsection{Multipath Reliable Transport Engine}

The multipath reliable transport engine is the core functional layer, delivering reliable, efficient multipath data transfer. It divides internally into three functional blocks: the transmit path (Tx), the receive path (Rx), and the retransmission logic.

\subsubsection{Transmit Path (Tx)}

The transmit path organizes Host/GPU data into network packets and schedules them onto the underlying network channels. It contains the following submodules:

\begin{enumerate}
    \item \textbf{Work Submission Processor:}
    Monitors doorbell signals, performs Aperture decoding and parallel Slot Table lookup, then assembles the minimal metadata into a complete WQE. It subsequently parses the WQE's opcode, address, key, and other fields, allocates PSN sequence numbers, issues a data fetch request to Tx Data Fetch, and sends a WQE message to the WQE Record module in the retransmission logic to register the pending send task.

    \item \textbf{Tx Data Fetch:}
    Reads payload data from GPU device memory over the PCIe interface. Upon receiving a data send request from the Work Submission Processor, Tx Data Fetch initiates a DMA read to move data from GPU memory into the NIC.

    \item \textbf{Congestion Control:}
    Schedules first transmission requests and retransmission requests in a unified manner based on network congestion, issues data send requests to Tx Data Fetch, and feeds congestion-related information back to the multipath transmit scheduler.

    \item \textbf{Assembly Unit:}
    Prepends protocol headers to the payload data from Tx Data Fetch according to the network protocol format and generates complete packets. It also embeds ECN marks from Congestion Control into the corresponding packet fields to signal path congestion to the receiver.

    \item \textbf{Multi-Path Tx Scheduler:}
    Employs an EV (Entropy Value)-based per-packet path-spraying mechanism. The control plane pre-configures SRv6 path templates and the EV mapping table; at data-plane transmission time, selecting one EV per packet determines both the outer SRv6 address and the local output channel, allowing packets from the same QP to enter different fabric paths in parallel.
\end{enumerate}

\subsubsection{Receive Path (Rx)}

The receive path accepts packets from the underlying network interfaces, parses protocol information, and delivers payloads to GPU device memory. It contains the following submodules:

\begin{enumerate}
    \item \textbf{Packet Buffer:}
    Temporarily stores packets received from the FTile network channels, providing buffering for subsequent protocol parsing.

    \item \textbf{Packet Parser:}
    Parses protocol headers in the Packet Buffer, extracts headers and payload, and generates bitmap information for the RQ Bitmap Manager. ECN marks extracted from received packets are echoed back to the sender via SACK, where they are delivered to the congestion control module to trigger path switching.

    \item \textbf{RQ Bitmap Manager:}
    Merges multi-channel bitmaps, maintains the receive queue bitmap state, records received and expected PSNs, and detects packet loss and out-of-order arrival, providing receiver-side state for the reliable transport mechanism.

    \item \textbf{Ordered Queue Manager:}
    Implements the ordering guarantee for ordered writes: it ensures that the current Flag is written to remote GPU memory only after all preceding Writes within the QP have completed. The written Flag serves as a ready signal, notifying the remote GPU that the data preceding this Flag is safe to consume.

    \item \textbf{DMA Placement:}
    Writes parsed payload data to designated locations in GPU device memory via PCIe DMA, completing the final delivery of data from the network to GPU memory.
\end{enumerate}

\subsubsection{Retransmission Logic}

The retransmission logic adopts a \techterm{hardware-software co-design} as its primary strategy: hardware maintains fast-path state, and the CPU executes low-frequency control decisions. In Fig.~\ref{fig:arch-overview}, the WQE Record and Scoreboard Bitmap are fixed hardware modules that permanently reside in the FPGA as state storage; the components marked in red are \techterm{optional retransmission logic} --- implemented by default in CPU software to save FPGA resources, but offloadable to the NIC when hardware resources are sufficient to pursue the lowest recovery latency.

\begin{enumerate}
    \item \textbf{WQE Record:}
    Receives WQE messages from the Work Submission Processor and records \texttt{\{start\_psn, psn\_count, DMA metadata\}} tuples at WQE granularity, providing the SACK Retransmission Management Queue with PSN-to-DMA-metadata reverse lookup.

    \item \textbf{Scoreboard Bitmap:}
    Maintains the acknowledgment status of every outstanding PSN on the send side. ACK/SACK are consumed by hardware internally to update the Scoreboard; the SACK Retransmission Management Queue reads it only on demand after receiving a retransmission signal, obtaining the current ACK bitmap to determine packet loss.

    \item \textbf{SACK Retransmission Management Queue:}
    Residing in NIC hardware, this module parses ACK/SACK/NAK to obtain the ACK boundary and selective ACK ranges, then extracts the SQ Scoreboard bitmap and WQE metadata, assembles them into a retransmission status report, and pushes it to the CPU via a single PCIe write. The CPU determines packet loss from the report and assembles retry requests to issue downstream. The hardware send path never waits for retransmission and can continuously advance new data; software independently tracks packets awaiting retransmission in the background. When logic resources are sufficient, loss detection and request assembly can be moved into this queue, allowing hardware to close the retransmission loop and eliminate PCIe latency.

    \item \textbf{Timer:}
    Implemented by default in CPU software. The runtime maintains a per-QP software deadline for each QP with outstanding packets: upon expiry, it re-scans the Scoreboard and generates retries. When FPGA resources are sufficient, this can be replaced by a hardware Grouped Timer, with timeout scanning driven internally by the NIC.
\end{enumerate}

The detailed design of software-coordinated retransmission and the full hardware scheme is presented in Sections~\ref{sec:packet-reliability} and~\ref{sec:software-recovery}. Both designs share the same WQE Record and Scoreboard hardware interfaces, so relocating the SACK Retransmission Management Queue and Timer does not require changes to any other module.

\subsection{FTile Network Transport Layer}

The FTile network transport layer is the physical interface between the system and the external network. Four independent FTile channels each contain a complete Ethernet physical layer: EMIB (Embedded Multi-die Interconnect Bridge), 100GE MAC/PCS/FEC (Ethernet MAC + Physical Coding Sublayer + Forward Error Correction), and FHT (FTile High-Speed Serial Transceiver). The four FTile channels operate in parallel, scheduled and allocated in a unified manner by the Multi-Path Tx Scheduler above, achieving link aggregation and multipath transmission to deliver an aggregate 400 Gb/s of physical network bandwidth.

\subsection{Chapter Summary}

This chapter used the three end-to-end data streams as an organizing framework, unfolding OpenRDMA's layered architecture and module responsibilities level by level.
The Host/GPU software layer supports dual-path work submission from both CPU and GPU, with the GPU directly triggering hardware WQE assembly through a minimal doorbell.
The multipath reliable transport engine, at the core, achieves EV-table-based multipath packet spraying, receiver-side Bitmap reliable transport, and selective retransmission through the coordination of the transmit path, receive path, and retransmission logic.
The FTile network transport layer provides four channels of 100GE physical network access, delivering an aggregate 400 Gb/s of bandwidth.
The next chapter analyzes the key designs within the multipath reliable transport engine --- the minimal doorbell mechanism, the ordering scheme, and the retransmission strategy --- in depth.

%% file: sections/04_design.tex
\section{OpenRDMA Design}
\label{sec:design}

This chapter presents the core design of OpenRDMA, systematically addressing the key challenges in realizing a fast path for GPU-initiated communication. First, once packets are reliably delivered, how do we ensure the remote GPU can safely consume data in the correct order? We address this in Section~\ref{sec:gpu-visible-ordering} with a GPU-visible ordering guarantee. Second, Section~\ref{sec:gpu-work-submission} describes how the GPU escapes the burden of per-operation CPU proxy involvement while achieving bounded submission under constrained NIC resources. Building on this, how can the NIC fully exploit multiple parallel paths without imposing path management overhead on the GPU? Section~\ref{sec:multipath-execution} presents our multipath execution scheme. Furthermore, in the face of the natural reordering introduced by multipath, how do we distinguish it from genuine packet loss while preserving the reception progress already made? Section~\ref{sec:packet-reliability} discusses the packet reliability mechanism. At the same time, NIC state overhead and software recovery cost are in tension; Section~\ref{sec:software-recovery} analyzes this tradeoff and presents our chosen strategy. Finally, how do we apply unified constraints to first transmissions and retransmissions while clearly separating three distinct concerns --- congestion control, path selection, and local backpressure? Section~\ref{sec:multipath-scheduling} provides our congestion control scheme and multipath scheduling policy. The following sections elaborate on each of these designs in turn.

\input{sections/04_design/04_1_gpu_visible_ordering_completion}
\input{sections/04_design/04_2_gpu_work_submission}
\input{"sections/04_design/04_3_SRv6-Based_Multipath_Packet_Spraying"}
\input{sections/04_design/04_4_packet_level_reliability}
\input{sections/04_design/04_5_software_coordinated_recovery}
\input{sections/04_design/04_6_congestion_control}

\subsection{Chapter Summary}
\label{sec:design-summary}

The six design choices in OpenRDMA do not exist in isolation; they follow a clear chain of reasoning in which each link depends on the last. GPU-visible ordering (\S\ref{sec:gpu-visible-ordering}) first defines the boundary of receiver-side correctness --- data must not only be received by the NIC but also be made safe for the GPU to consume --- and this serves as the foundation for all subsequent designs. Building on this, GPU direct submission (\S\ref{sec:gpu-work-submission}) frees the GPU from per-operation CPU proxy involvement and achieves bounded submission under limited NIC resources through a Credit doorbell mechanism. Once the submission path is unblocked, the bottleneck in end-to-end performance shifts to network utilization itself --- multipath packet spraying (\S\ref{sec:multipath-execution}) uniformly spreads a single QP's traffic across all available planes by encoding EVs in SRv6, solving the problem of throughput collapse under flow collisions that plagues traditional single-path RDMA.

The out-of-order arrival introduced by multipath imposes new requirements on the reliability mechanism. Per-packet reliability feedback (\S\ref{sec:packet-reliability}) uses scoreboard-based selective ACK/SACK and static timeout retransmission to complete loss detection and repair in a purely hardware loop. However, hardware state overhead scales linearly with window depth and QP count, posing a challenge for FPGA resources. Software-coordinated recovery (\S\ref{sec:software-recovery}) hands retransmission off to the CPU, with hardware maintaining reliability state only within the normal window, thereby saving on-chip storage. Finally, congestion control and scheduling (\S\ref{sec:multipath-scheduling}) abandons traditional per-path congestion windows and instead leverages multipath redundancy to convert congestion into path switching --- switching paths rather than reducing rate.

In summary, OpenRDMA's design follows three core principles: (1) hardware handles the steady-state fast path --- ordering, submission, multipath forwarding, and routine loss recovery all close within the ASIC/FPGA; (2) software handles rare retransmission events --- crash recovery and QP lifecycle management are taken over by the CPU to reduce hardware complexity; (3) congestion control and fault avoidance share the same EV replacement mechanism --- regardless of the trigger, the system always chooses to ``switch paths rather than reduce rate.''

%% file: sections/04_design/04_1_gpu_visible_ordering_completion.tex
\subsection{GPU-Visible Ordered Message Completion}
\label{sec:gpu-visible-ordering}

GPU communication typically overlaps with computation. The challenge is not only to place data in
remote GPU memory, but also to determine when a GPU kernel may safely consume it. The receiving
kernel can continue computing while the NIC moves a payload. Before consuming the result, however,
the kernel needs an indication that the data is ready. Conventional RDMA completion interfaces
primarily target CPU-managed completion queues (CQs). They therefore do not directly provide a
lightweight data-ready notification to the GPU~\cite{rdma-aware-programming,potluri2017gpu}.

A conventional CPU-proxy path coordinates these steps through a host CQ. The GPU kernel first
writes a communication request to a queue shared by the CPU and GPU. The CPU proxy then retrieves
the request and submits it to the NIC. When the request completes, the proxy polls a completion
queue entry (CQE), which is the completion record generated by the NIC. It then writes the
completion state back to GPU-visible memory. The CPU thus participates in both submission and
completion notification for every request, as Figure~\ref{fig:end-to-end-path-comparison} in
Section~2 shows.

Even when the NIC moves the payload directly, this control path traverses
\(\text{GPU}\rightarrow\text{CPU}\rightarrow\text{NIC}\rightarrow\text{CPU}
\rightarrow\text{GPU}\). IBGDA also removes the CPU reverse proxy by moving per-operation network
control to the GPU. However, IBGDA requires the GPU to atomically reserve a queue position,
construct a complete WQE in GPU memory, and update the queue/Doorbell state and lifecycle for every
operation~\cite{ibgda}. OpenRDMA instead issues a compact doorbell. In the IBGDA path
shown in the figure, the sending GPU explicitly issues an NVSHMEM put-with-signal operation. This
operation writes the payload and then atomically updates a signal object in the remote PE's
symmetric memory. The receiving GPU waits on that object before consuming the payload
~\cite{nvshmem-signaling}.

OpenRDMA moves complete-request construction from the GPU into the NIC and publishes completion
notifications directly to the remote GPU. On the per-message fast path, the GPU writes only an
8-byte Doorbell. The local NIC uses the doorbell to look up a preconfigured Slot and synthesize
the complete request. The remote NIC places the payload in remote GPU memory and publishes the
Ordered Flag only after the ordering condition is met. The CPU/runtime handles initialization and
failure recovery but does not enter this path during normal operation. The remainder of this
section defines how the remote NIC publishes the Ordered Flag and the ordering semantics that this
notification must satisfy.

This completion path must address four issues over an out-of-order data path. Direct data placement
allows out-of-order payloads to reach their final locations without reconstructing submission
order. Relaxed Writes and Ordered Writes separate data movement from completion notification at
the operation level. A publication condition prevents multipath execution from exposing a
GPU-visible Flag before its payload. Finally, bounded per-QP state limits the resources occupied by
Flags that are not yet publishable. The following four parts describe these mechanisms and their
boundaries.

\techterm{Out-of-Order Execution and Direct Data Placement.}
Completion under multipath, out-of-order execution requires a distinction between placing
individual payloads and declaring a group of payloads ready. OpenRDMA permits different Writes to
be processed out of order without requiring the receiver to reconstruct their submission order.
The receiver uses direct data placement (DDP): each Write carries a target address or offset and
places its data directly at the final location. DDP allows independent payloads to progress
concurrently, but completing one payload write does not indicate that all payloads are ready for
GPU consumption~\cite{rdma-aware-programming,gpudirect-rdma,mrc-spec}.

\techterm{Relaxed Writes and Ordered Writes.}
OpenRDMA uses two Write types to add a data-ready notification without sacrificing DDP parallelism.
A Relaxed Write, represented by \texttt{WRITE\_RELAXED}, places only its payload at the target
location and can progress independently of other Relaxed Writes. An Ordered Write, represented by
\texttt{WRITE\_ORDERED\_FLAG}, writes a 4-byte Flag to a preconfigured address to notify the remote
GPU that data is ready. A queue pair (QP) represents one RDMA communication connection. Within the
same QP, the NIC writes an Ordered Write's Flag only after writing all preceding Writes. This
ordering guarantee is local to a QP and does not extend across QPs.

\techterm{GPU-Visible Flags.}
An Ordered Write does not move another payload; instead, it publishes a GPU-visible Flag. The Flag
is a separate ordered-completion notification issued by the sender rather than metadata attached
to each payload packet. The Flag carries a 4-byte value, but its arrival at the remote NIC alone
does not indicate that all preceding Writes on the same QP have been written. The remote NIC writes
the value to remote GPU memory only after confirming that it has written all preceding Writes on
the same QP. At this point, the Flag becomes a GPU-observable data-ready notification. The Flag and
its covered payloads use the same QP but remain distinct network requests.

Send order alone cannot guarantee publication order. Although a Flag follows its payloads in send
order, latency differences across paths may cause it to arrive before some payload packets. The
receiving NIC therefore does not immediately write an early Flag to GPU memory. Instead, it
temporarily retains the Flag address and value without buffering the complete Flag packet.
Meanwhile, payload packets that have arrived continue to their final locations through DDP. The
NIC publishes the retained Flag only after confirming that it has written all preceding Writes on
the QP. This wait covers the interval in which the Flag has arrived but earlier payloads remain in
flight.

Once the preceding payloads satisfy this condition, the NIC publishes the pending Flag to the
remote GPU. The complete Flag path is
\(\text{GPU}\rightarrow\text{local NIC}\rightarrow\text{remote NIC}
\rightarrow\text{remote GPU}\). The remote GPU polls the Flag without waiting for a CPU proxy to
poll a CQE or copy completion state. After observing the Flag, the GPU can conclude that the
Relaxed Writes covered by that Flag and all earlier Ordered Writes on the same QP have been
written. Packet arrival, payload DMA placement, Flag publication, and GPU observation are four
distinct completion boundaries.

The \emph{ordered-write completion point} is the boundary at which the NIC can confirm that all
preceding Writes on the same QP have been written. The NIC may then begin writing the current Flag.
After the Flag is written, the remote GPU must still observe the new value before consuming the
data.

This causal relationship is semantically analogous to release--acquire ordering. The NIC writes
all preceding Writes on a QP before writing the current Flag. This behavior is analogous to a
release that completes earlier writes before publishing a notification. The GPU observes the Flag
before reading the payload. This behavior is analogous to an acquire that begins data consumption
after observing the notification~\cite{cuda-c-programming-guide,gpudirect-rdma}.

\begin{figure}[htbp]
    \captionsetup{name=Figure}
    \centering
    \includegraphics[width=0.6\textwidth]{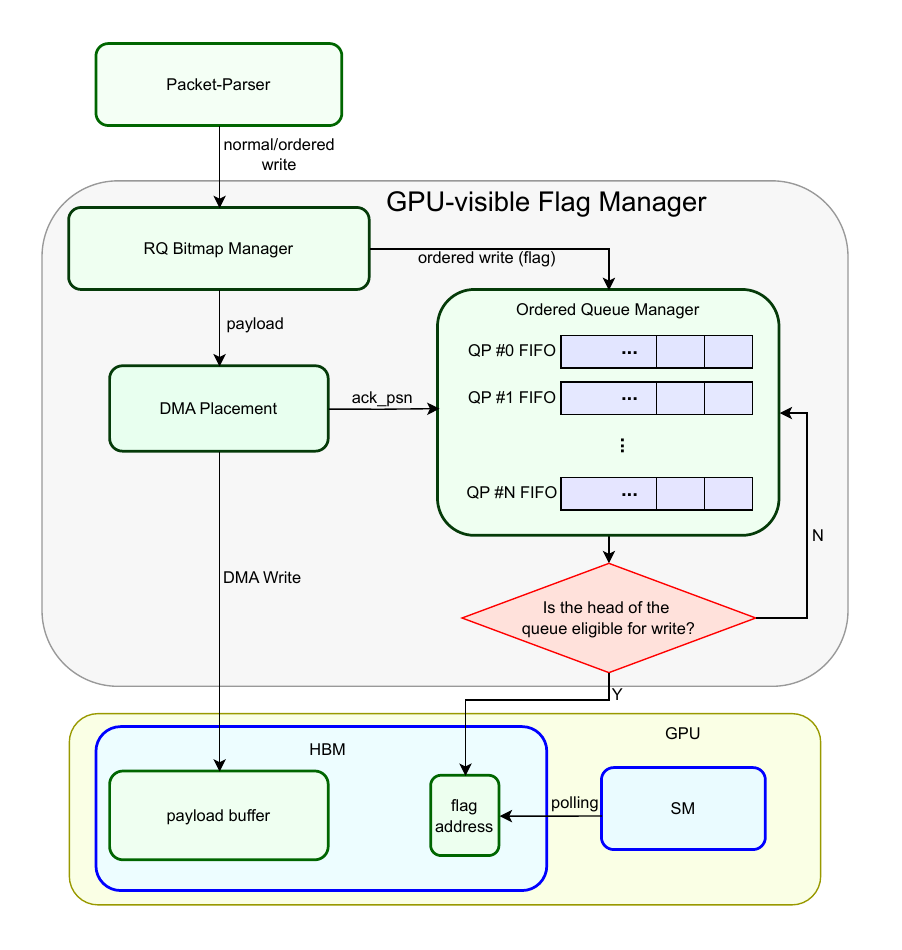}
    \caption{Flag Manager architecture.}
    \label{fig:flag-manager}
\end{figure}

These semantics require the receiver to retain Flags that have arrived but cannot yet be
published. The Flag Manager in Figure~\ref{fig:flag-manager} implements this function.

\techterm{Bounded Per-QP Ordered State.}
On the receive queue (RQ) side, the receiving NIC maintains a finite-capacity Ordered Write FIFO
for each QP. A FIFO removes entries from its head in arrival order; here, each entry represents a
Flag that has not yet been written. The NIC publishes the head Ordered Write only after all of its
preceding Writes have been written. Consequently, each QP has a fixed upper bound on the number of
Ordered Writes that may wait concurrently. This bound determines the resource backpressure
required during submission.

The local NIC sends an independent Ordered Flag request. The remote NIC uses the Ordered Write FIFO
to ensure that it writes preceding Writes on the same QP before the Flag. The remote GPU polls the
Flag to determine when it may consume the data. OpenRDMA preserves out-of-order processing and DDP
while removing the per-message CPU proxy from the completion-notification path. These semantics
introduce an explicit resource constraint. Every unpublished Flag occupies a finite FIFO entry, so
the submitter must obtain the corresponding resource credit before issuing an Ordered Write.

\FloatBarrier

%% file: sections/04_design/04_2_gpu_work_submission.tex
\subsection{GPU-Initiated Work Submission and Resource Backpressure}
\label{sec:gpu-work-submission}

With OpenRDMA, the GPU submits each request by writing a single 8-byte Doorbell instead of a
complete work queue entry (WQE). The GPU does not construct a complete WQE or maintain a send queue
(SQ) on the fast path.

Relaxed Writes and Ordered Writes define communication semantics that a GPU can consume. The
submission path must convert those semantics into operations that the GPU can initiate, and the NIC
must admit these operations within bounded resources. Conventional CPU-driven RDMA uses an SQ both
to describe requests and to limit outstanding work. An SQ is a fixed-size circular table. Its head
points to the position already consumed by the NIC, and its tail identifies the next position to
be filled by the driver. The driver writes a complete WQE into an available entry. The WQE
specifies the operation type, address, length, and memory-access key. The driver then advances the
tail and writes the NIC's Doorbell register, prompting the NIC to fetch the new WQE. Because the SQ
has fixed capacity, the driver can derive the available entries from its head and tail. It can
thereby bound the number of outstanding requests~\cite{rdma-aware-programming}.

Moving this interface to the GPU presents two alternatives. The GPU can notify a CPU proxy that
continues to use the SQ. Alternatively, it can bypass the proxy, maintain the SQ, and construct
complete WQEs. GPU-initiated communication systems such as IBGDA move the network-control path onto
the GPU and remove the CPU reverse proxy from per-operation submission. The GPU must still
coordinate queue positions, construct WQEs in GPU memory, and maintain the queue and Doorbell state
~\cite{ibgda,nccl-device-api,doca-gpunetio,nvshmem-env}. OpenRDMA instead limits the GPU fast path
to the Doorbell write.

\begin{figure}[htbp]
    \captionsetup{name=Figure}
    \centering
    \includegraphics[width=\textwidth]{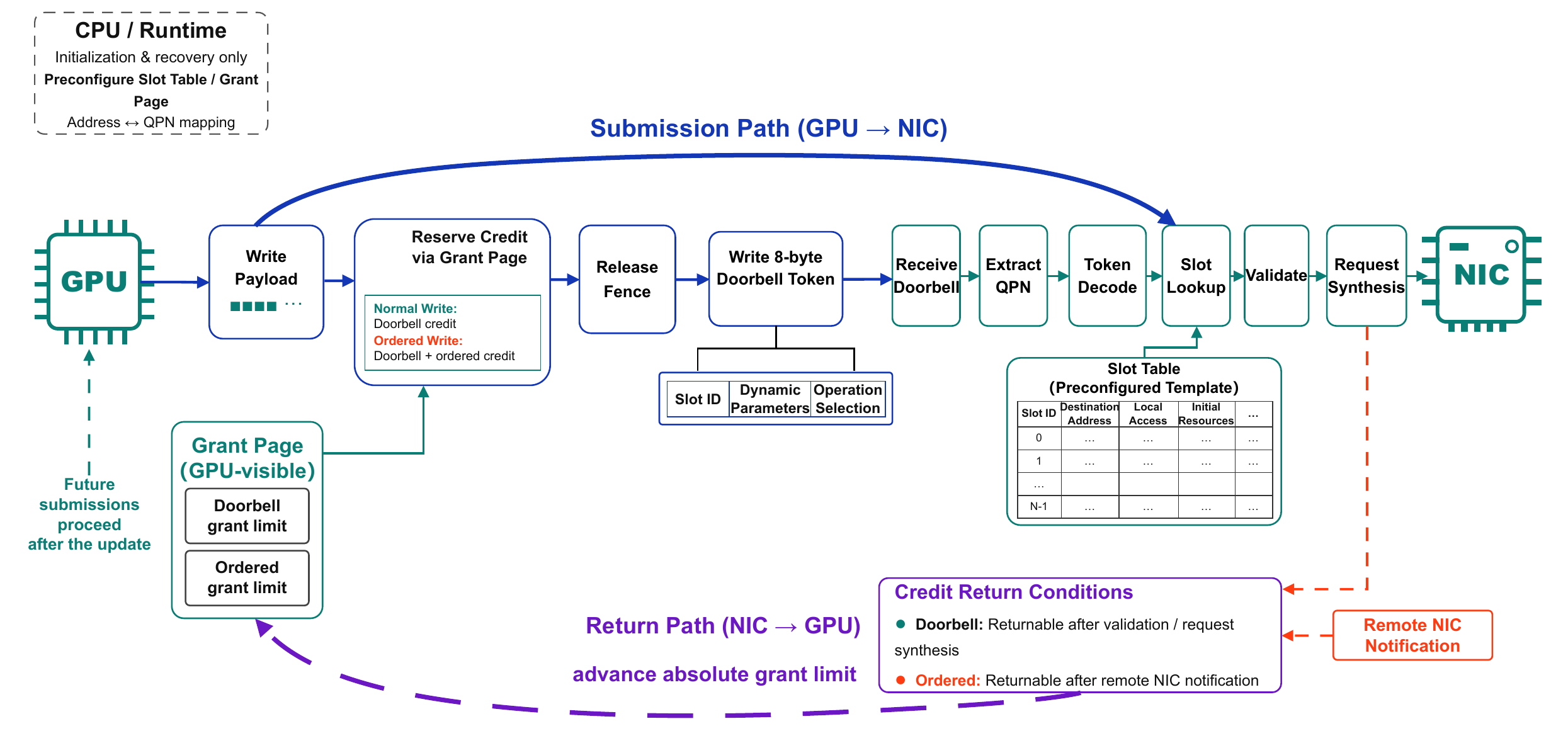}
    \caption{OpenRDMA's GPU submission and credit/grant workflow. The CPU/runtime preconfigures QPs,
    Slots, and initial credits. The GPU reserves the required credits before writing a Doorbell. The NIC receives and completes the request, then advances the GPU-readable absolute
    grant limit when the corresponding queue entry can be reused.}
    \label{fig:gpu-submission-credit-loop}
\end{figure}

A complete-WQE interface does not naturally match GPU execution. A GPU kernel organizes work into
warps, which are groups of GPU threads that execute synchronously. It processes data blocks or
gradient buckets in parallel. Constructing a complete WQE for each communication operation would
require collecting the address, key, length, and operation type from these parallel threads. The
GPU would then arrange these fields in the prescribed descriptor layout. OpenRDMA does not require
the GPU to assemble this complete WQE on the fast path
~\cite{cuda-c-programming-guide,ibgda}.
Figure~\ref{fig:gpu-submission-credit-loop} summarizes initialization, GPU submission, and NIC
credit return.

GPU-initiated work submission presents two coupled problems. First, the GPU must submit a
communication operation without constructing a complete WQE. Second, without a conventional SQ,
the GPU must determine whether the bounded NIC queues can admit another request. The following two
parts describe the Slot-triggered GPU Doorbell and credit/grant backpressure.

\techterm{Slot-Triggered GPU Doorbells.}
The CPU/runtime first registers the communication memory and creates the QPs. It then represents
each permitted operation as a Slot. A Slot is a preconfigured record that stores the address, key,
maximum length, and permitted operation types. Integer IDs index all Slots in a single global Slot
Table. The runtime combines a Doorbell address and a Slot ID into the submission handle
\texttt{\{gpu\_doorbell\_addr, slot\_id\}}. The GPU can select this handle directly.

To initiate a request, the GPU obtains the Doorbell address and Slot ID from the submission handle.
It inserts the length or Flag value for the current operation and writes an 8-byte
\emph{Doorbell}, a compact submission descriptor consumed by the NIC. The opcode selects the
operation type, the 18-bit \texttt{slot\_id} selects a Slot, and a 32-bit field carries the length
or Flag value. The remaining bits support debugging or are reserved for future extensions. The
Doorbell address directly identifies the QP number (QPN) used by the submission. The doorbell
therefore need not repeat the address, key, QPN, or complete WQE.

Section~\ref{sec:gpu-visible-ordering} defines the ordering semantics of Relaxed and Ordered
Writes. Here, the relevant distinction is the credit requirement of each operation. A
Relaxed Write does not occupy the receiver's Ordered Write FIFO, whereas an Ordered Write occupies
one entry for the destination QP while its 4-byte Flag waits to be written. The Doorbell address
provides the QPN directly. All 18 bits of \texttt{slot\_id} therefore index the global Slot Table
rather than encoding a QP.

The Doorbell remains compact because the NIC recovers the remaining request information from
the Slot. After receiving a doorbell, the NIC verifies that the Slot exists and is valid. It also
checks that the selected operation is permitted and that the length does not exceed the configured
limit. The NIC then combines the address-derived QPN, doorbell parameters, and Slot address and key to
construct an internal request. The GPU therefore selects only a configured operation and does not
maintain the complete RDMA request format.

The CPU/runtime creates and updates QPs and Slots. During reconfiguration, it stops new
submissions, waits for the NIC to finish already accepted requests, and then rebuilds the relevant
QP and Slot metadata. The NIC validates each submission and rejects requests with an invalid address
or key. The CPU therefore handles only infrequent configuration and recovery, while the GPU and NIC
perform per-message submission.

\techterm{Credit/Grant Backpressure.}
The Slot-triggered Doorbell solves direct submission but does not expose a conventional SQ to the
GPU, so the GPU can no longer bound submissions by counting free SQ entries. OpenRDMA instead uses
a credit as permission to occupy one entry in a finite queue. The GPU may write a Doorbell only
after reserving the required credits. The NIC returns those credits when the corresponding entries
can be reused. This mechanism protects both the NIC's Doorbell receive buffer and the per-QP
Ordered Write FIFO defined in Section~\ref{sec:gpu-visible-ordering}
~\cite{rdma-aware-programming,mrc-spec}.

The NIC publishes an absolute grant limit through a GPU-readable grant page. This limit is the
cumulative number of submissions permitted so far. The GPU maintains a local head that records its
number of reservations. The difference between the grant limit and local head is the currently
available capacity. Multiple GPU producers update the local head through atomic reservation. This
indivisible check-and-increment step prevents two producers from consuming the same credit. Because
the grant limit increases monotonically, repeated or out-of-order reads cannot create duplicate
credits.

OpenRDMA maintains two credit classes. All QPs share module-level Doorbell credits, which bound the
number of doorbells whose ingress processing remains incomplete. The NIC returns one Doorbell
credit when the corresponding Doorbell receive-buffer entry becomes reusable. Each QP also
maintains separate ordered credits. Their total does not exceed the capacity of that QP's Ordered
Write FIFO at the receiver. Each Ordered Write consumes one ordered credit for its destination QP,
whereas a Relaxed Write consumes none.

The receiver returns an ordered credit only after four conditions are met. The Ordered Write must
reach the head of its queue, all preceding Writes must be written, the write path must accept the
Flag-write request, and the corresponding FIFO entry must be reusable. This feedback advances the
absolute grant limit for the associated QP. It indicates only that the FIFO entry has been
released. It does not imply that the Flag write has completed or that the remote GPU has observed
the Flag. If the receiver encounters a full FIFO despite correct credit accounting, the credit
counters and queue occupancy have diverged. The NIC must reject the request and report an error rather
than treat the full queue as ordinary backpressure.

The two credit classes bound only unprocessed doorbells and occupied Ordered Write FIFO entries.
They do not limit payload bytes, link rate, or network congestion. The CPU/runtime preconfigures
operations and initial credits. The GPU reserves credits before submitting directly through a
doorbell. The NIC completes each request and returns capacity when internal queue entries are
released. OpenRDMA avoids complete-WQE construction on the GPU and replaces the implicit admission
control of a conventional SQ with explicit credits.

\FloatBarrier

%% file: sections/04_design/04_3_SRv6-Based_Multipath_Packet_Spraying.tex
\subsection{Out-of-Order Multipath Packet Execution}
\label{sec:multipath-execution}

In a cross-node All-Reduce, the GPU needs to continuously send large blocks of gradient data to a remote endpoint, while the fabric typically provides multiple network planes, rails, and spine paths.
If each flow is always bound to a single path, the bandwidth of other idle links cannot be utilized by the same transfer.
However, if the GPU kernel is required to manage multipath on its own---maintaining multiple connections and selecting the destination for each packet---path decisions intrude into the GPU's most frequent submission path, introducing scheduling overhead and programming complexity at the communication library level.
OpenRDMA therefore addresses not ``whether multiple paths exist in the network,'' but how to let the NIC automatically exploit these paths without exposing per-packet path management to the GPU.

Our design adopts SRv6-based packet spraying.
The control plane prepares an SRv6 path template and EV profile for each destination based on the fabric topology.
In the data plane, when sending each packet, one active EV is selected, and the NIC derives the corresponding SRv6 outer destination address from that EV.
Topology-aware path construction occurs on the low-frequency control path, while the high-frequency send path only handles lightweight EV selection and channel mapping.

\begin{figure}[t]
    \centering
    \includegraphics[width=0.48\linewidth]{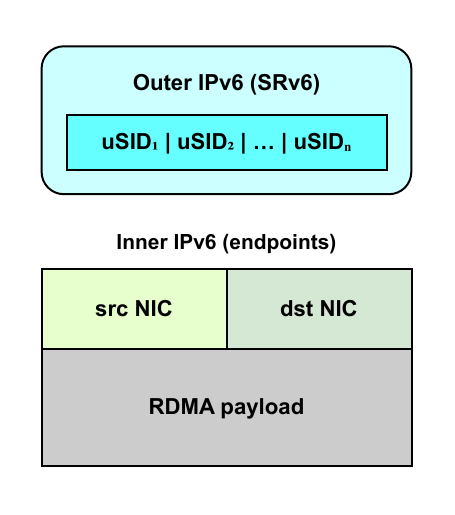}
    \caption{Encapsulation relationship among the SRv6 outer path, inner RDMA endpoints, and payload.}
    \label{fig:srv6-encap}
\end{figure}

\paragraph{Using SRv6 to express deterministic paths.}
OpenRDMA chooses SRv6 over per-packet ECMP or MPLS-style label stacks because SRv6 reuses the existing IPv6 forwarding plane without introducing additional label space, and uSID addressing allows multi-hop paths to be compressed into a single destination address field, naturally matching the NIC fast path's single-EV table lookup model.
Each end-to-end path is described by an SRv6 segment list: the switching nodes along the path are encoded into the outer IPv6 destination address, and switches forward hop-by-hop according to preconfigured SRv6 forwarding tables---the source controls the path, and intermediate nodes only execute preconfigured uSID behaviors without participating in per-packet path computation.
Figure~\ref{fig:srv6-encap} shows the SRv6-encapsulated RDMA packet format: the outer IPv6 destination address carries the uSID-encoded segment list, the inner layer preserves the source/destination NIC endpoint addresses and RDMA payload, and the switches along the path only process the outer source route.

This encapsulation separates ``how to traverse the fabric'' from ``whom to finally deliver to''---after the destination NIC decapsulates at the path endpoint, it still identifies the communication endpoint according to the inner address.
Therefore, the same destination can correspond to multiple different outer paths without changing the remote NIC targeted by the upper-layer RDMA operation.

\begin{figure}[t]
    \centering
    \includegraphics[width=\linewidth]{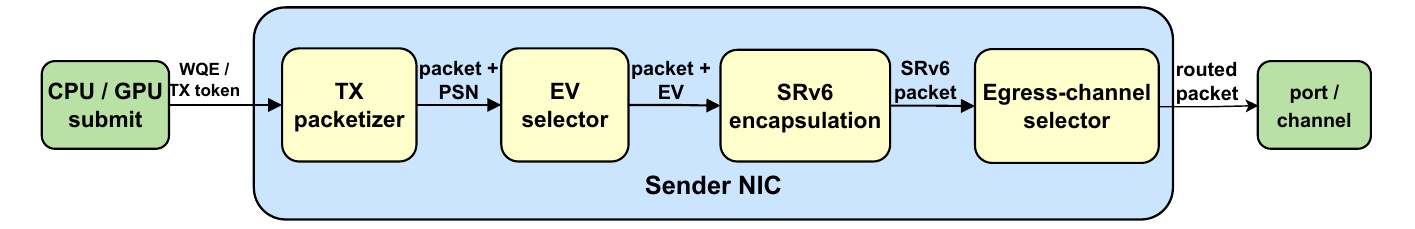}
    \caption{Sender-side per-packet EV selection, SRv6 encapsulation, and output channel selection pipeline.}
    \label{fig:ev-pipeline}
\end{figure}

\begin{figure}[t]
    \centering
    \includegraphics[width=0.58\linewidth]{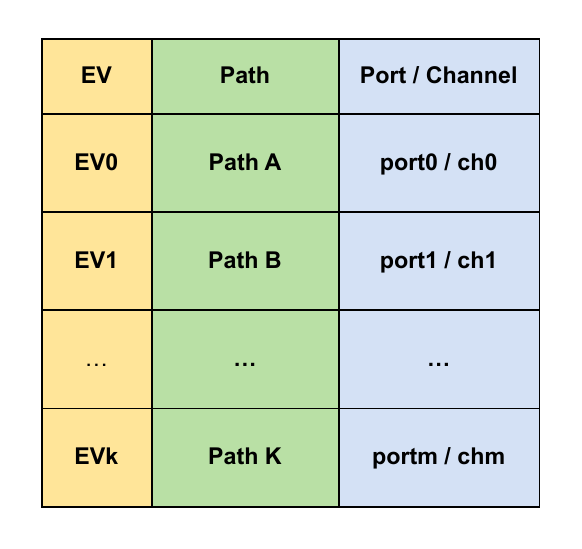}
    \caption{Mapping from the active EV set to SRv6 paths and local ports/channels.}
    \label{fig:ev-table}
\end{figure}

\paragraph{Using EV to compress per-packet path selection and spraying.}
Figure~\ref{fig:ev-pipeline} shows the sender-side per-packet EV selection hardware pipeline, and Figure~\ref{fig:ev-table} shows the mapping from the active EV set to SRv6 paths and local ports.

Even if paths are already described by SRv6, if each packet queries the complete segment list in the fast path, the send path would still be too heavy.
OpenRDMA uses Entropy Value (EV) as a lightweight path identifier.
At QP setup, the NIC maintains a set of active EVs for the target node, covering different planes, rails, and intermediate switching paths as evenly as possible.
Each EV corresponds to a deterministic SRv6 source route, a local NIC port, and a TX channel.

At send time, the NIC selects one EV from the current active EV set, which determines the packet's outer SRv6 address and local output channel.
Adjacent PSNs can select different EVs, so packets of the same QP can simultaneously enter different planes, rails, and spine paths without waiting for the previous packet to leave the network.
The EV can be encoded in the UDP source port, the IPv6 Flow Label, or an explicit field defined by the protocol.
The low-frequency control plane is responsible for preparing the path set and EV profile.
The high-frequency data plane only needs a single EV-to-(route, port) mapping.
The GPU fast path is completely unaware of per-packet routing.
The key here is not for the GPU to understand EV, but for the NIC to use a single compact identifier to simultaneously complete path selection, port mapping, and path identification in subsequent feedback.

\paragraph{Out-of-order arrival is a normal execution result.}
A direct consequence of packet-level spraying is that packets of the same RDMA operation may traverse different paths and arrive out of order.
OpenRDMA does not require the receiver to first reorder packets before writing them into memory.
Instead, each data packet carries the address and permission information needed to complete direct placement.
After validating the packet, the receiver can write the payload to its final target location.
The packet order in the network and the placement order in memory are decoupled: packets can arrive out of order, but the payload is still written to the correct address.
Whether reordering causes missing packets, duplicate packets, or completion semantic issues is handled by the subsequent packet-level reliability and ordered completion mechanisms.

\paragraph{EV as path health status.}
EV is not only a path selection identifier, but also an index for path health status.
The receiver carries or echoes the relevant EV in feedback, allowing the sender to associate congestion and packet loss with specific paths.
If a path exhibits ECN marking, the sender can temporarily reduce the weight of the corresponding EV or skip that EV.
If a packet is lost on a path, the sender removes its EV from the active set and selects a spare EV as a replacement.
Recovery detection and EV re-enabling for failed paths are further handled by the congestion control mechanism.

Through the combination of SRv6 source routes and EV-based packet spraying, OpenRDMA converts parallel links in multi-plane, multi-rail fabrics into end-to-end bandwidth, while keeping paths identifiable, feedback-capable, and bypassable.
Per-packet path selection, health maintenance, and feedback processing are all completed inside the NIC.
The GPU only submits communication operations.

%% file: sections/04_design/04_4_packet_level_reliability.tex
\subsection{Packet-Level Reliability Feedback}
\label{sec:packet-reliability}

Consider a sequence of packets sprayed across multiple paths.
The sender has transmitted PSN 10 through PSN 20,
yet the receiver first sees PSNs 11--20 while PSN 10 remains missing.
PSN 10 may merely be queued on a slower path, or it may have been genuinely lost.
If the receiver only returns a cumulative ACK, the acknowledgment point stays at
PSN 9, and the sender cannot learn that the subsequent ten packets have already
arrived; if the sender retransmits immediately upon seeing a hole, it would
misclassify normal cross-path latency differences as packet loss.
Worse still, ACKs, SACKs, or NAKs themselves can also be lost, causing both sides
to temporarily lose synchronization even when data transfer has made progress.

OpenRDMA bases reliability on explicit packet-level state rather than inferring
transport outcomes from packet arrival order.
The receiver uses a receive bitmap to store sparse arrival state within the receive
window, and the sender uses an SQ Scoreboard to store the acknowledgment status
of each outstanding PSN.
ACKs and SACKs convey cumulative and selective acknowledgment information between
the two sets of state, NAKs report anomalies that can be clearly pinpointed,
and grouped timers provide the ultimate fallback when the feedback path fails for
an extended period.
Recovery requests generated by SACK holes, NAKs, and grouped timeouts all converge
into a single retransmission scheduler, which rechecks the Scoreboard before actual
retransmission, thereby forming a unified packet-level reliability closed loop
within the NIC.

\begin{figure}[!t]
    \centering
    \includegraphics[
        width=0.82\textwidth
    ]{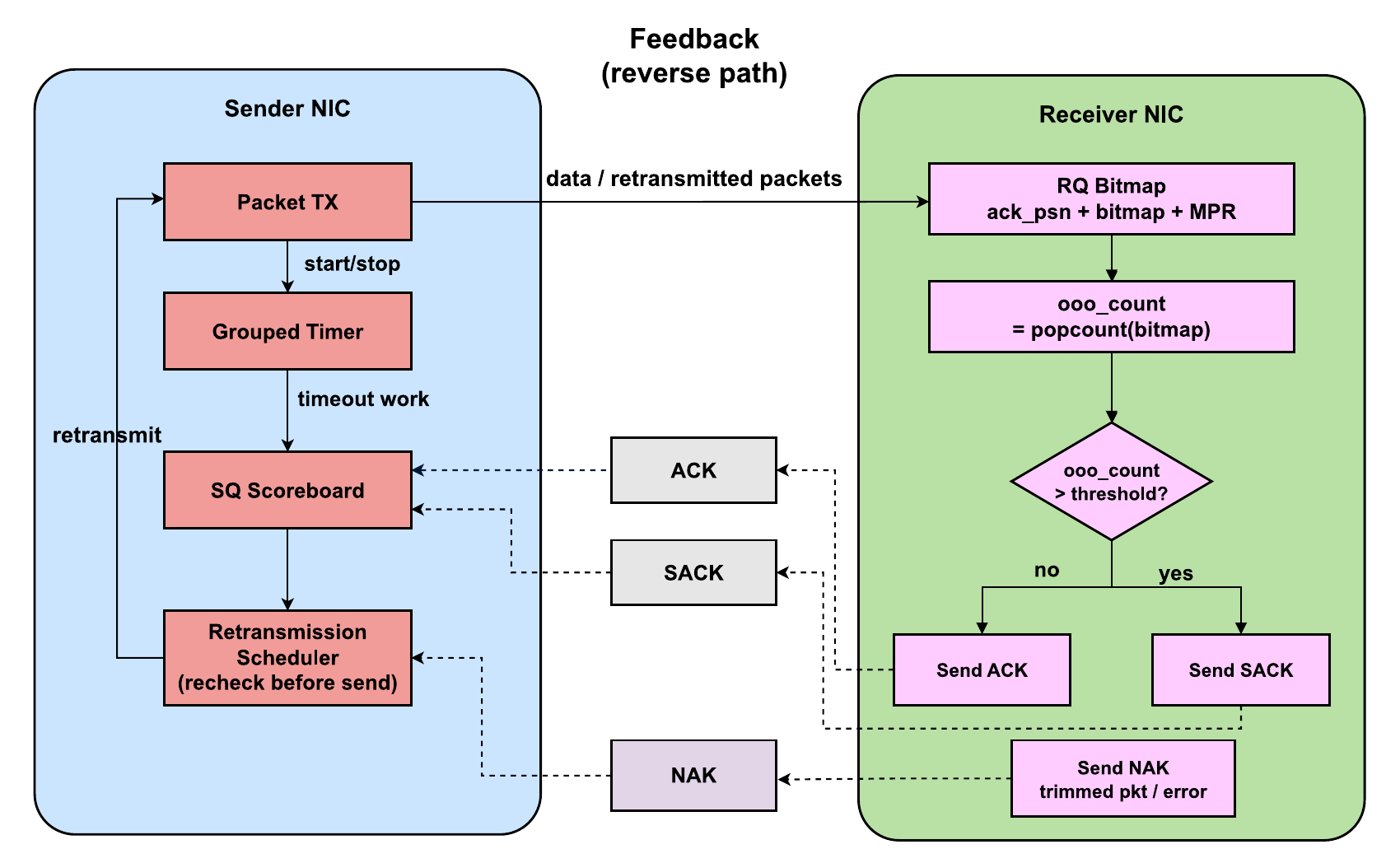}
    \caption{The packet-level reliability feedback closed loop in OpenRDMA.
    The receiver NIC uses the RQ Bitmap to record \texttt{ack\_psn}, sparse
    arrival state, and MPR, and generates ACKs or SACKs based on the degree of
    reordering; clearly pinpointable anomalies are returned via NAKs.
    The sender NIC merges feedback into the SQ Scoreboard, from which the
    retransmission scheduler selectively retransmits still-unacknowledged packets;
    the grouped timer provides a fallback when feedback is absent for an extended period.}
    \label{fig:packet-level-reliability-feedback}
\end{figure}

Figure~\ref{fig:packet-level-reliability-feedback} summarizes this closed loop.
The receiver first maintains \texttt{ack\_psn}, sparse arrival state, and MPR
through the RQ Bitmap, and selects either a lightweight ACK or a bitmap-carrying
SACK based on the current \texttt{ooo\_count}; for trimmed packets or clearly
pinpointable errors, the receiver directly generates a NAK.
The sender uses this feedback to monotonically update the SQ Scoreboard,
from which the retransmission scheduler generates necessary retransmission requests;
if feedback is absent for an extended period, the grouped timer triggers a rescan
of the current Scoreboard.

\paragraph{Bounded receive window.}
For each QP, the receiver maintains a cumulative acknowledgment point
\texttt{ack\_psn} and a receive bitmap covering the contiguous PSN range beyond it.
When a legitimate packet first arrives, the corresponding bitmap bit is set;
duplicate packets do not modify the state again.
Only when contiguous 1's form starting from the lowest position of the window
does the receiver consume those bits, advance \texttt{ack\_psn}, and slide the
receive window.
Early-arriving high PSNs can thus be preserved as valid progress rather than being
discarded because a lower PSN has not yet arrived.

The receiver simultaneously declares the highest PSN boundary it can currently
track through the MPR (Maximum PSN Range).
The sender must not send new PSNs exceeding the current MPR upper bound,
and retransmissions can only target PSNs within the current outstanding window
that have not yet been acknowledged.
MPR constrains the reordering state that the network can produce to a bounded
window, so that fixed-capacity receive bitmaps and sender Scoreboards suffice
to cover all valid protocol state.
Packets falling below the lower window boundary are treated as duplicates;
packets exceeding the MPR upper bound cannot enter the normal receive state
update flow.

\paragraph{From ACK to lazy SACK.}
Under in-order arrival or low-reordering conditions, there is no need to carry
selective acknowledgment information in every feedback message; a lightweight
cumulative ACK suffices to advance the sender's cumulative acknowledgment point.
As reordering increases, the receiver maintains \texttt{ooo\_count}, representing
the number of out-of-order packets already stored above \texttt{ack\_psn} after
the window has been slid forward:

\begin{equation}
\texttt{ooo\_count} = \operatorname{popcount}(\texttt{receive\_bitmap})
\end{equation}

\texttt{ooo\_count} increases only when a bitmap bit that was originally 0
is first set to 1; duplicate packets are not counted again.
When the contiguous 1's at the low end of the window are cumulatively acknowledged
and removed from the bitmap, \texttt{ooo\_count} decreases correspondingly.
Thus, this count always reflects the number of packets within the current receive
window that have arrived but have not yet been covered by a cumulative ACK.

When \texttt{ooo\_count} exceeds a configured reordering threshold, the receiver
enters lazy SACK mode.
A SACK carries the cumulative acknowledgment point \texttt{ack\_psn}, a bitmap
base or offset, and a fixed-width receive bitmask relative to that position;
if the receive window boundary changes, the feedback can also carry a new MPR.
The fixed-width bitmask describes only the selective receive state within the
range covered by this SACK, without requiring the entire receive bitmap to be
placed into a single control packet.

A SACK need not be sent immediately after every out-of-order packet arrival.
The receiver can coalesce state over a period of time and return a SACK when:
\texttt{ooo\_count} exceeds the threshold, the feedback aggregation timer expires,
a retransmitted packet arrives, or a periodic refresh is triggered.
This exposes persistent holes to the sender in a timely manner without generating
a control packet for every out-of-order packet.

\texttt{ooo\_count} deliberately uses the count of stored packets rather than
\texttt{max\_rcv\_psn - ack\_psn}.
Under multi-channel parallel reception, a single high PSN may occasionally arrive
first; a distance-based metric would amplify a few distant arrivals into severe
reordering, whereas a count-based metric reflects only how many packets have
actually been stored beyond the cumulative acknowledgment point.
Thus, \texttt{ooo\_count} describes the reordering pressure across the entire
receive window and is used to decide whether to send a SACK, but it does not
directly judge whether a particular hole represents a loss.

\paragraph{Recovering sender state from SACK.}
The sender maintains an outstanding Scoreboard for each QP.
Upon receiving an ACK or SACK, the hardware first processes \texttt{ack\_psn},
releasing all packet state that has been cumulatively acknowledged; it then maps
the bitmask to the corresponding PSNs according to the bitmap base in the SACK
and monotonically merges the positions where the bit is 1 as selectively received.
Already-acknowledged Scoreboard state is never rolled back to unacknowledged due
to stale SACKs, duplicate SACKs, or 0 bits in subsequent bitmasks.

A PSN becomes a hole candidate only when all of the following conditions are met:
the PSN has been sent, it remains within the outstanding range, it falls within
the effective coverage range of the current SACK bitmask, and the corresponding
bit is 0.
PSNs outside the bitmask coverage range have unknown status and cannot be judged
as holes based solely on this SACK.
The upper bound of the current scan range is:

\begin{equation}
\texttt{scan\_end} = \min(\texttt{last\_sent\_psn},\; \texttt{sack\_base\_psn} + \texttt{sack\_width} - 1)
\end{equation}

A hole still cannot be immediately treated as a loss.
OpenRDMA counts the number of selectively acknowledged packets beyond that hole
and within the current SACK scan range:

\begin{equation}
\begin{aligned}
&\texttt{post\_hole\_sacked\_count}(\texttt{hole\_psn},\; \texttt{scan\_end}) \\
&\quad = \operatorname{count}\{\text{sacked packets with } \texttt{hole\_psn} < \texttt{psn} \leq \texttt{scan\_end}\}
\end{aligned}
\end{equation}

Only when this local post-hole acknowledgment count exceeds a preset threshold
does the sender mark the corresponding PSN as a retransmission candidate.
Intuitively, the more subsequent packets that have already traversed the network
and been acknowledged by the receiver, the less credible the explanation that
the earlier hole is merely queued on a slow path.

This local criterion is more precise than using the global \texttt{ooo\_count}:
the former answers ``how much acknowledgment progress has appeared beyond this
hole,'' while the latter answers ``how out-of-order the entire receive window
currently is.'' The two serve respectively to control SACK feedback frequency
and to judge the loss likelihood of a specific packet; they are not interchangeable.

\paragraph{Pinpoint anomalies and retransmission deduplication.}
For anomalies that the receiver or lower layers can clearly pinpoint---such as
the arrival of a trimmed packet header, a packet checksum failure, or temporary
insufficient receiver resources---OpenRDMA uses a NAK with a reason code pointing
to a specific PSN, allowing the sender to initiate targeted recovery without
waiting for subsequent SACKs to accumulate evidence.

Trimmed packets and silent drops must be distinguished.
The former still deliver a recognizable packet header to the receiver, so the
receiver can extract the PSN and return a NAK; if a packet is completely dropped
in the network with no header reaching the receiver, the receiver cannot
immediately generate a targeted NAK---the loss can only be discovered through
subsequent SACK holes, periodic feedback, or sender timeout.

PSNs produced by SACK holes and NAKs are only marked as retransmission candidates,
not immediately transmitted.
The hardware guarantees that at most one pending retransmission request exists
for the same PSN and re-reads the Scoreboard before actual transmission.
If a new ACK or SACK has already acknowledged that PSN, the old retransmission
request is canceled.
Thus, feedback processing, Scoreboard updates, and retransmission scheduling can
proceed in parallel without readily amplifying duplicate traffic.

\paragraph{Grouped timer fallback.}
SACKs and NAKs alone cannot close the protocol, because feedback packets
themselves can also be lost.
The most direct approach is to maintain an independent timer for each outstanding
packet, but timer entries, cancellation logic, and expiry queues would all grow
linearly with window size and QP count.
OpenRDMA implements timeout semantics using grouped timers: a fixed-size segment
of contiguous PSNs shares one timer group, thereby reducing the timer state that
must be maintained.

A timer group is closed and its timer started when any of the following conditions
is met: the group is full; the last packet of the current WQE or message has been
reached; or the group has not filled up within a maximum aggregation time.
The last condition protects low-rate send and sender-idle scenarios, preventing
an unfilled tail group from never receiving timeout protection.

Timer expiry does not directly retransmit all packets in the group; it only
generates one group-level retransmission work item.
When this work item is scheduled, it rescans the current Scoreboard, skips PSNs
that have already been cumulatively confirmed or selectively received, and
generates retransmission candidates only for PSNs that remain outstanding within
the group.
As shown in Figure~\ref{fig:packet-level-reliability-feedback}, grouped timers
share the same retransmission scheduler with ACK/SACK/NAK feedback, rather than
forming an independent full-retransmission path.

A timer event carries the identifier and generation of the corresponding group.
If the group has already left the outstanding range by the time the event arrives,
or if a new generation of state for the same group has already begun, the event
is lazily deemed invalid.
Duplicate timeout work items for the same group are also deduplicated, preventing
stale timer events from acting on reused state.

If a group times out consecutively, its subsequent timeouts are lengthened
according to a configured backoff policy, avoiding a sustained retransmission
storm when the feedback path is persistently unavailable.
When the retry count exceeds the QP-configured retry limit, the sender transitions
that QP to an error state and reports the failure upward, rather than retransmitting
indefinitely.
Grouped timers thus trade fixed-width Scoreboard scans for fewer timer states,
while preserving selective retransmission and eventual failure semantics.

\paragraph{State convergence after feedback loss.}
When feedback packets are lost, the protocol still has multiple natural convergence
paths.
Subsequent cumulative ACKs or SACKs can cover previously lost acknowledgment
progress; the receiver's SACK refresh timer also periodically re-exposes
\texttt{ack\_psn}, the receive bitmask, and necessary MPR updates.
If none of this feedback ever arrives, the sender's grouped timer eventually
triggers a Scoreboard rescan and selective retransmission.

The receiver identifies duplicate packets via the receive bitmap; it does not
repeat state updates with side effects, but can return an ACK or SACK again
to help the sender regain synchronization.
SACK holes, NAKs, and grouped timeouts thus handle three categories of evidence
respectively: the loss inference formed when subsequent packets have arrived,
clearly identified pinpoint anomalies at the receiver, and send stagnation caused
by a prolonged absence of reliable feedback.

The three recovery paths share the same sender Scoreboard and recheck acknowledgment
state before actual packet retransmission, enabling OpenRDMA to complete
packet-level out-of-order recording, loss judgment, selective retransmission,
and feedback-loss recovery entirely within the NIC, while avoiding unnecessary
retransmissions caused by duplicate or stale feedback.

This closed loop is responsible only for packet delivery and retransmission
recovery.
The receiver-side receive bitmap provides the completion layer with the fact that
packets have arrived, while the sender-side Scoreboard is used solely for
retransmission decisions and send-state reclamation.
A receive bit being set does not indicate that DMA placement of the corresponding
payload has completed, nor that the entire RDMA operation has become visible to
the GPU.
Message-level completion must still combine packet arrival, DMA completion,
and ordered-completion conditions, and is separately released by the GPU-visible
ordered completion mechanism defined in Section~\ref{sec:gpu-visible-ordering}.

%% file: sections/04_design/04_5_software_coordinated_recovery.tex
\subsection{Software-Coordinated Recovery}
\label{sec:software-recovery}

The complete hardware reliability path described in Section~\ref{sec:packet-reliability} closes three loops within the NIC: packet loss detection, selective retransmission, and timeout recovery. To implement this closed loop, the sender-side NIC must maintain four types of critical hardware structures. The WQE Record holds request metadata, the SQ Scoreboard tracks PSN acknowledgment status, the SACK retransmission queue performs hole scanning and scheduling, and the Grouped Timer provides a timeout safety net. Their storage and logic complexity grows linearly with the number of QPs and window depth, placing a heavy burden on resource-constrained FPGAs.

We observed that GPU-driven communication is inherently asymmetric. GPU kernels generate communication operations at extremely high frequency — each All-Reduce chunk or doorbell granularity may trigger a network submission — and WQE synthesis, address translation, permission checking, and packet generation sit on the per-packet fast path, where NIC hardware delivers the highest efficiency. Packet loss, however, is a low-probability event in production fabrics; retransmission decisions occur orders of magnitude less frequently than normal sends. This points to a different retransmission strategy: \textbf{keep high-frequency WQE synthesis in NIC hardware to guarantee fast-path latency; move low-frequency retransmission management — SACK processing, hole detection, timeout management, and retry counting — to CPU software to save FPGA resources}. More importantly, this division fully decouples retransmission progress from normal sends. The hardware send path never stalls waiting for retransmission to complete and can continuously consume new WQEs to advance the data stream; software independently tracks all outstanding packets in the background and completes recovery without blocking first transmissions. The result is a hybrid reliability architecture in which hardware handles line-rate sending and software handles asynchronous retransmission.

\begin{figure}[htbp]
    \centering
    \includegraphics[width=1.0\textwidth]{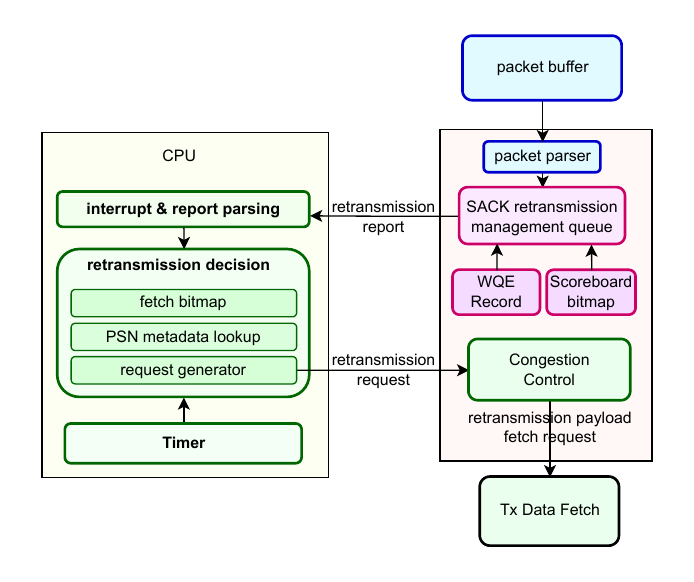}
    \caption{Schematic diagram of the software collaborative retransmission module}
    \label{fig:software-co-retry}
\end{figure}

\techterm{We designed a scheme where the NIC reports facts and the CPU makes recovery decisions,} as shown in Fig.~\ref{fig:software-co-retry}. The software-coordinated approach does not change the wire protocol — PSN, ACK/SACK/NAK formats, and receiver-side deduplication remain identical to the pure hardware scheme — it only relocates the sender-side reliability state.

The NIC retains the receiver-side fast-path state — RQ Bitmap, ACK/SACK/NAK generation — and the ability to execute CPU-issued retries. The key change is on the send side: the NIC no longer scans the Scoreboard to detect loss, and no longer maintains the SACK retransmit management queue or the Grouped Timer. Upon receiving ACK/SACK/NAK feedback, the NIC parses the cumulative ACK boundary, selective ACK ranges, and error causes. Simultaneously, it reads the complete ACK bitmap of current outstanding PSNs from the local hardware SQ Scoreboard and extracts request metadata summaries (source address, destination address, length, key, etc.) for hole-related PSNs from the WQE Record. It packages all this information into a retransmission status report and pushes it to CPU memory via a single PCIe write. After receiving the report, the CPU performs post-hole SACK count scanning locally — a PSN is declared lost only when a sufficient number of subsequent packets beyond the hole have been acknowledged, an explicit NAK targets that PSN, or a software timer triggers a scan. For confirmed lost PSNs, the CPU directly assembles retry requests using the WQE metadata already contained in the report and issues them to the NIC. The send-side NIC's new responsibility simplifies to receiving CPU-generated retry requests, rechecking QP state, window, credit, and congestion control constraints before transmission, confirming the PSN has not been acknowledged by a late ACK, and sending it through the TX datapath.

This design compresses the software-involved PCIe overhead into a single status report transfer. With this in place, the hardware send path and retransmission are decoupled: the NIC can always consume doorbells to generate new PSNs and send them without waiting for any retransmission to complete; software independently tracks PSNs awaiting retransmission using the complete reports it receives and advances retries asynchronously. When upper-layer semantics do not require in-order delivery, this property minimizes the throughput penalty of retransmission, preserving link utilization to the greatest extent possible.

We also replace the Grouped Timer with per-QP software clocks. The runtime maintains a software deadline for each QP with outstanding packets: upon timeout, the CPU follows the same flow described above, obtaining the bitmap and metadata from the most recently received report (or triggering a NIC state push if the report is stale), and performs batch scanning and retry generation within a budget. This scheme moves all timer management, reload, and expiry processing to CPU software, eliminating the timer array, expiry queue, and deduplication logic from the FPGA.

\techterm{Engineering tradeoffs.}
On the NIC send side, the software-coordinated scheme retains the WQE Record, SQ Scoreboard, and a streamlined SACK retransmit management queue (used only for feedback parsing and information extraction), while removing the retransmit request assembly logic, the Grouped Timer, and its associated state machines, scheduling queues, and deduplication logic. The added software path includes: one PCIe transfer for the NIC to assemble and push the retransmission status report, CPU interrupt handling, local loss detection and retry request generation, per-QP software timer management, and the PCIe bandwidth for issuing retries to the NIC. The key latency increment is one PCIe report transfer plus one CPU processing cycle.

This division stems from two core requirements. First, GPU doorbell submission at high frequency constitutes the normal-case fast path — WQE assembly, Scoreboard maintenance, and feedback parsing must remain in NIC hardware to avoid PCIe latency amplification. Second, loss recovery is the low-frequency exception path — moving retransmission decisions and request assembly into the CPU not only saves substantial FPGA BRAM and state machine resources, but more fundamentally removes the hardware send path's dependence on retransmission progress. Software independently tracks all packets awaiting retransmission in the background, allowing hardware to continuously advance normal transmission without stalling, thereby sustaining near line-rate send behavior in scenarios without ordering requirements. If hardware resources are abundant, keeping the full SACK retransmit management queue and Grouped Timer in the NIC — with hardware closing the loop on retransmission detection and request assembly — eliminates the PCIe and software processing overhead described here and yields the lowest recovery latency. This is precisely the full hardware scheme described in Section~\ref{sec:packet-reliability}.

%% file: sections/04_design/04_6_congestion_control.tex
\subsection{Congestion Control and Transmission Scheduling}
\label{sec:multipath-scheduling}

In the multipath transport architecture, EV-driven path selection determines each packet's network path, ECN congestion signals drive rapid avoidance of failed paths, and FTile channel backpressure schedules first transmissions and retransmissions --- together these three mechanisms determine end-to-end throughput and latency. This section focuses on the design of congestion control.

\begin{figure}[htbp]
    \centering
    \includegraphics[width=1.0\textwidth]{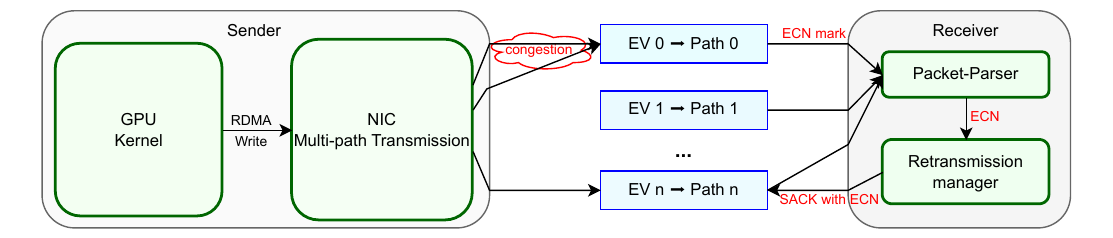}
    \caption{Congestion condition.}
    \label{fig:congestion}
\end{figure}

\begin{figure}[htbp]
    \centering
    \includegraphics[width=1.0\textwidth]{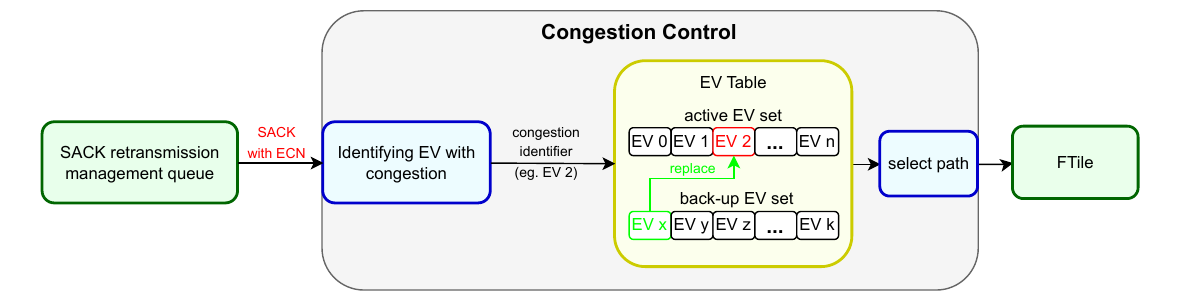}
    \caption{Congestion control and transmission scheduling module.}
    \label{fig:congestion-control}
\end{figure}

As shown in Fig.~\ref{fig:congestion}  and Fig.~\ref{fig:congestion-control}, OpenRDMA does not adopt traditional per-path congestion windows (cwnd) and rate adjustment mechanisms such as AIMD. Instead, it exploits the natural redundancy of multipath: upon receiving a congestion signal, it directly \textbf{switches paths} rather than reducing rate. This design choice rests on three observations: (1) a typical MRC deployment provides on the order of 100 EV paths, and switching paths preserves throughput while reducing rate loses it; (2) FTile channel FIFOs already provide a hardware-level natural rate ceiling --- a full FIFO automatically backpressures, eliminating the need for protocol-layer inflight accounting; (3) GPU Credits independently throttle at the submission layer --- if the NIC cannot keep up, it withholds Credits, causing the GPU to naturally stop sending, decoupled from network-layer congestion.

\textbf{ECN congestion signal processing.}
The receiver-side Packet-Parser extracts the ECN marking bits during header parsing and echoes them back to the sender via the corresponding EV field in the SACK control packet.
Upon receiving an ECN-marked SACK, the sender-side SACK retransmit management module executes the following steps:

\begin{enumerate}[leftmargin=2em, label=(\arabic*)]
    \item \textbf{Identify the problematic EV:} Extract the EV value marked as congested by ECN, or declared faulty by MRC due to packet loss, from the SACK echoed EV field.
    \item \textbf{Remove the EV:} Remove this EV from the QP's active EV set. The EV is subsequently managed for recovery by the path probe mechanism.
    \item \textbf{Replace with a spare EV:} Select an unused EV from the spare EV set and add it to the active EV set. The new EV maps to a different physical path, bypassing the congestion point or failed link.
    \item \textbf{Retransmit affected packets:} Packets sent over the problematic EV and not yet acknowledged by SACK are re-sent over the replacement EV by the retransmit management module.
\end{enumerate}

This EV removal procedure uses a unified processing path for both ECN congestion and packet loss failures --- regardless of the trigger, the active EV set replacement logic and the subsequent path probe recovery flow are identical.

\textbf{Path probe and failed-path recovery.}
An EV marked as congested or faulty is not permanently removed.
When a data packet is lost without being trimmed, MRC declares that transport path as failed and immediately deactivates the corresponding EV; the same applies when an ECN mark is received.
However, not all packet loss results from path failure --- packets may also be lost due to bit errors, transient queue overflows, and other causes --- and permanently deactivating an EV after a single loss would gradually exhaust the available EV resources.

We therefore designed a \textbf{Path Probe} mechanism in OpenRDMA: MRC periodically sends probe requests over EV paths that have been marked as unavailable.
Probe packets are lightweight MRC control packets that travel along exactly the same paths as data packets, carrying EV information for the peer to echo back.
If a probe packet is successfully acknowledged by the peer within the timeout window, a probe success counter is incremented; when the consecutive success count exceeds a configurable threshold (default 3), the path is declared recovered and the EV is re-admitted to the active EV set.
If probe packets continue to fail, the EV remains unavailable and the Clustermapper system~\cite{mrc-srv6-paper} performs further diagnosis of the underlying link failure.

Path probes share the same EV-based forwarding mechanism as normal data transmission, so probe results accurately reflect the true reachability of the data-plane path --- rather than relying on control-plane link state advertisements.
The probe send rate is governed by a hardware rate limiter (at most one probe per marked path per millisecond), designed to keep probe overhead within a small fraction of link bandwidth.

ECN path switching solves the problem of \textit{where} to send. The remaining challenge is \textit{when} to send --- how to schedule the two classes of demand, first transmissions and retransmissions, onto the four FTile physical channels in a unified manner.

\textbf{The Send Engine} is the core scheduler of the TX datapath, managing two sources of packet transmission requests:

\begin{itemize}[leftmargin=2em]
    \item \textbf{First transmission requests:} originating from the WQE parsing module, these enter the transmit pipeline after EV selection and FTile channel arbitration.
    \item \textbf{Retransmission requests:} originating from the SACK retransmit management module (SACK-driven selective retransmission, including re-sending of affected packets after ECN path switching) and the Timer retransmit management module (timeout-driven fallback retransmission).
\end{itemize}

\textbf{FTile FIFO backpressure-driven scheduling.}
The Send Engine executes the following decisions every cycle:

\begin{enumerate}[leftmargin=2em, label=(\arabic*)]
    \item \textbf{Poll FTile channels:} Traverse the four FTile channels in priority order, skipping channels whose FIFOs are full.
    \item \textbf{Arbitrate requests:} For channels with non-full FIFOs, select from the request sources by priority: retransmission requests $\;>\;$ first transmission requests. Within the same priority, round-robin arbitration is used.
    \item \textbf{EV selection and address generation:} For the selected request, pick an EV from the active EV set and generate the corresponding destination address through EV mapping.
    \item \textbf{Initiate DMA transfer:} Dispatch the request to the DMA control logic, which starts a PCIe DMA Read to fetch the payload data.
\end{enumerate}

\textbf{Why retransmission takes priority.}
Retransmission requests are prioritized over first transmission requests for three reasons: (1) a lost packet blocks the receiver's PSN progress, and delaying retransmission causes receiver-side buffer accumulation; (2) retransmissions typically involve few packets and consume a limited share of bandwidth; (3) timely retransmission prevents the receiver from triggering Go-Back-N fallback due to timeout, which would cause far more unnecessary retransmissions.

\textbf{The backpressure propagation chain.}
When all four FTile channel FIFOs are full, the Send Engine stops pulling new requests from the WQE parsing module and the retransmit management modules, forming a staged backpressure chain from \textbf{FTile FIFO $\;\rightarrow\;$ Send Engine $\;\rightarrow\;$ WQE queue}.
WQE queue buildup causes the NIC to stop returning GPU Credits, which in turn prevents the GPU from submitting new doorbells --- forming a complete end-to-end backpressure chain from the network egress to the GPU submission ingress.

This ECN path switching and unified scheduling design brings three advantages to OpenRDMA:

\begin{enumerate}[leftmargin=2em, label=(\arabic*)]
    \item \textbf{Path-level congestion isolation:} ECN-driven path switching isolates congestion on one path from the sending capability of other paths --- the congested path is replaced by a spare EV while the remaining paths continue transmitting, so that single-path congestion is not expected to reduce aggregate throughput. Deterministic EV mapping makes the behavior of the replacement path predictable.
    \item \textbf{Simplified hardware implementation:} Unlike traditional AIMD congestion control, which must maintain per-path cwnd, inflight, credit, and other runtime state, ECN path switching requires only an active EV set bitmap and a configurable cooldown timer, keeping BRAM and logic resource overhead low.
    \item \textbf{End-to-end backpressure without software intervention:} The staged backpressure chain --- FTile FIFO $\;\rightarrow\;$ Send Engine $\;\rightarrow\;$ WQE queue $\;\rightarrow\;$ GPU Credit --- is designed to coordinate deceleration across all pipeline stages under congestion, with no congestion-window computation step requiring software intervention.
\end{enumerate}

In summary, OpenRDMA's congestion control does not rely on per-path cwnd maintenance or rate adjustment. Instead, it leverages multipath redundancy to convert congestion into path switching, supplemented by path probes for automatic failed-path recovery. On the send side, first transmissions and retransmissions are scheduled in a unified manner driven by FTile FIFO backpressure, with retransmission receiving priority to release receiver-side progress as quickly as possible. The entire control loop backpressures from the network egress to the GPU submission ingress, operating as a hardware-closed loop with no CPU involvement in congestion decisions.

%% file: sections/05_implementation.tex
\section{Protocol and Hardware Design}
\label{sec:implementation}

Section~\ref{sec:design} explained why OpenRDMA adopts GPU Doorbells, multi-path
out-of-order execution, packet-level reliability, and Ordered Flags; this section
further describes how these design choices are translated into executable protocol
objects, NIC state, and FPGA datapaths.
To avoid writing each module as an independent interface specification, this
section unfolds the implementation along the lifecycle of a GPU operation from
local submission to remote visible completion.

First, Section~\ref{sec:impl-work-submission} converts a GPU Doorbell into a
validated internal request;
Section~\ref{sec:impl-tx-datapath} then completes PSN allocation, DMA,
segmentation, and multi-path transmission.
After packets arrive at the remote side,
Section~\ref{sec:impl-rx-datapath} merges four-way out-of-order input and performs
direct placement;
Section~\ref{sec:impl-reliability} then closes the feedback and retransmission
loop based on receive evidence.
Reliable data arrival does not yet mean the GPU can safely consume the data,
so Section~\ref{sec:impl-ordered-completion} further converts DMA progress into
Ordered Flag publication.
Finally, Section~\ref{sec:impl-resource-timing} discusses, from a full-chain
perspective, the state scale, resource lifecycle, and timing constraints that
collectively govern the modules above.

This lifecycle contains multiple completion boundaries that must not be conflated.
Writes, opcodes, and requests describe operation-level work; PSNs and
ACK/SACK/NAK describe packet-level reliability;
\texttt{ack\_psn} only records receive evidence, while \texttt{dma\_ack\_psn}
records the ordered retire watermark after DMA.
Credit return only indicates that the corresponding queue resource can be reused;
Flag DMA completion and GPU observation must still follow before semantic
completion that the GPU can consume is formed.
ACK/SACK/NAK wire layouts and reliability thresholds are reserved as protocol
parameters; queue depths, resource scales, and critical paths are determined by
synthesis and stress testing.

\input{sections/05_implementation/05_1_work_submission_pipeline}
\input{sections/05_implementation/05_2_tx_datapath_implementation}
\input{sections/05_implementation/05_3_rx_datapath_implementation}
\input{sections/05_implementation/05_4_reliability_retransmission_implementation}
\input{sections/05_implementation/05_5_ordered_completion_implementation}
\input{sections/05_implementation/05_6_resource_and_timing_considerations}

%% file: sections/05_implementation/05_1_work_submission_pipeline.tex
\subsection{Submission Interface and Admission Pipeline}
\label{sec:impl-work-submission}

Section~\ref{sec:gpu-work-submission} defines direct GPU submission and credit-based
backpressure. The implementation must safely transform an 8-byte Doorbell write into an executable
internal request. It must preserve a compact GPU fast path while admitting only requests with
reserved credits, valid capabilities, and fully specified transport semantics.

The OpenRDMA GPU fast path does not construct complete WQEs in HBM. A host-assisted compatibility
path may express equivalent high-level operations through a platform-specific input format, but it
does not change the GPU doorbell ABI described below. Every input path must produce the same internal
request representation before the common WQE Parse boundary. The requests then share the PSN, DMA,
reliability, and completion paths.

This section follows a Doorbell operation from configuration through admission in the lifecycle order summarized in Table~\ref{tab:impl-submission-stages}.
The four parts describe the end-to-end roles and the GPU-side submission and credit reservation
process. They then describe NIC-side doorbell decode and request construction, followed by the admission,
error, and WQE Parse boundaries.

\begin{table}[htbp]
  \captionsetup{name=Table}
  \centering
  \caption{Stages of the submission and admission pipeline.}
  \label{tab:impl-submission-stages}
  \small
  \setlength{\tabcolsep}{4pt}
  \renewcommand{\arraystretch}{1.15}
  \begin{tabularx}{\linewidth}{
      >{\centering\arraybackslash}p{0.08\linewidth}
      >{\raggedright\arraybackslash}p{0.20\linewidth}
      >{\raggedright\arraybackslash}p{0.20\linewidth}
      Y
  }
    \toprule
    Order & Stage & Input & Main operation \\
    \midrule

    1 & Configuration and handle setup
    & Application buffers and runtime configuration
    & Registers memory, creates QPs and Slots, maps the Doorbell BAR, initializes credits, and
      publishes the submission handle. \\

    2 & GPU payload and credit reservation
    & Payload, submission handle, and grant page
    & Completes all payload writes, makes the payload visible to the NIC, reserves the required credits,
      and writes one 8~B Doorbell. \\

    3 & Doorbell decode and request construction
    & Doorbell address and 64-bit doorbell
    & Extracts the QPN, decodes the doorbell, looks up the Slot or Slot pair, and constructs one or
      two internal requests. \\

    4 & Admission and WQE Parse
    & Constructed internal request
    & Completes format, capability, and capacity checks, then passes accepted work to WQE Parse for
      PSN allocation. \\

    \bottomrule
  \end{tabularx}
\end{table}

\subsubsection{End-to-End Roles and Flow}

The first stage separates control-plane setup from per-request execution.
Table~\ref{tab:impl-submission-roles} shows how the CPU/runtime, GPU producer, and NIC front-end
divide the end-to-end path. During configuration, the CPU/runtime establishes capabilities and
credit limits. Once the system enters RUN, the per-request fast path traverses only the GPU and
NIC.

\begin{table}[htbp]
  \captionsetup{name=Table}
  \centering
  \caption{Responsibilities in the GPU submission path.}
  \label{tab:impl-submission-roles}
  \small
  \begin{tabularx}{\linewidth}{@{}l X X@{}}
    \toprule
    Participant & Configuration and recovery & Responsibilities during RUN \\
    \midrule
    CPU/runtime &
    Registers memory, creates QPs and Slots, maps the Doorbell BAR, initializes credits, and
    performs stop/drain/reset. &
    Does not participate in individual GPU fast-path requests; monitors errors and handles
    recovery. \\
    GPU producer &
    Receives the indivisible submission handle
    \texttt{\{gpu\_doorbell\_addr, slot\_id\}}. &
    Completes all payload writes, reserves credits, and encodes and writes the 8~B doorbell; polls
    the subsequent Flag when semantic completion is required. \\
    NIC front-end &
    Maintains QP bounds, the Slot Table, grants, and the error epoch. &
    Extracts the QPN from the address, validates the doorbell and Slot, constructs an internal
    request, and passes it to WQE Parse. \\
    \bottomrule
  \end{tabularx}
\end{table}
\FloatBarrier

Once configuration completes and the system enters RUN, the GPU producer begins the per-request
path.

\subsubsection{GPU-Side Submission and Credit Reservation}

Before writing the Doorbell, the GPU producer completes the payload writes, makes the payload
visible to the NIC, and reserves the required credits:

\begin{enumerate}[leftmargin=2.2em,itemsep=0.2em]
  \item Complete all writes to the payload referenced by the request and make the payload visible
        to the NIC.
  \item For an Ordered Flag or combined operation, atomically reserve one ordered credit for the
        destination QPN; a relaxed operation skips this step.
  \item Atomically reserve one Doorbell credit.
  \item Perform one aligned 8~B store to the Doorbell address of the corresponding QPN.
\end{enumerate}

Each credit represents permission to occupy one entry in a bounded NIC queue. Reserving the
required credits before writing the Doorbell prevents the request from exceeding the corresponding
queue capacities. Table~\ref{tab:impl-credit-types} shows the queue capacity represented by each
credit class. These credits do not represent payload-byte budgets, link rates, or congestion
windows.

\begin{table}[htbp]
  \captionsetup{name=Table}
  \centering
  \caption{Indexing and return conditions for GPU-visible credits.}
  \label{tab:impl-credit-types}
  \small
  \begin{tabularx}{\linewidth}{@{}l
      >{\raggedright\arraybackslash}X
      >{\raggedright\arraybackslash}X
      >{\raggedright\arraybackslash}X@{}}
    \toprule
    Type & Capacity represented & Consumption rule & Grant-advance condition \\
    \midrule
    Doorbell credit &
    Sender admission-buffer entries; the baseline configuration shares one head/grant across all
    QPNs. &
    Each 8~B Doorbell consumes one credit, independent of opcode and payload length. &
    Advances after the aperture front-end consumes the corresponding admission entry. \\
    Ordered credit &
    Remote per-QPN Ordered Queue entries, selected by the QPN in the Doorbell address. &
    An Ordered Flag or combined operation consumes one credit; a relaxed operation consumes none. &
    Advances through feedback after the corresponding Ordered Queue entry satisfies its barrier,
    is accepted by the Flag DMA path, and is dequeued. \\
    \bottomrule
  \end{tabularx}
\end{table}
\FloatBarrier

Each head/grant pair is a 64-bit cumulative watermark that must not wrap within one RUN lifecycle.
Multiple producers use compare-and-swap to atomically advance the local head from $h$ to $h+1$. A
reservation succeeds only when $h < \textit{grant limit}$. The ordered-credit reservation precedes
the Doorbell-credit reservation. A producer may therefore hold an acquired ordered credit while
waiting for a Doorbell credit. The normal path does not roll back cumulative counters. After a
fatal error, the runtime stops all producers. It resets the counters only after draining or
flushing the admission, Ordered Queue, and feedback paths.

When the Ordered Queue releases an entry, it returns the absolute watermark
\texttt{\{qpn, ordered\_grant\_limit\}}. The sender accepts only larger values, so duplicate or
out-of-order feedback cannot return a credit twice. This grant proves only that the queue slot can
be reused; it does not establish Flag DMA completion or GPU observation. After completing these
reservations, the GPU performs only one 8-byte store.

Once the NIC observes this store, the front-end decodes its address and value, then constructs the
internal request using the corresponding Slot metadata.

\subsubsection{NIC-Side Doorbell Decode and Request Construction}

The Doorbell write address identifies the request's QP. The doorbell carries only the operation
selection and per-invocation parameters. The runtime maps the Doorbell for $qpn$ as
\[
  \texttt{gpu\_doorbell\_addr}
  = \texttt{gpu\_doorbell\_base} + (\texttt{qpn} \ll 3).
\]
The NIC extracts the QPN from the BAR-relative transaction address and verifies that the address
is aligned to 8~B and that $0 \leq \texttt{qpn} < \texttt{MAX\_QP}$. The QPN therefore occupies
no doorbell bits. The NIC interprets the doorbell as a little-endian 64-bit value with the following
layout:

\begin{center}
  \small
  \begin{tabular}{@{}c c c c c@{}}
    \toprule
    \texttt{seq[63:60]} & \texttt{op[59:58]} & \texttt{slot\_id[57:40]} &
    \texttt{len\_or\_imm32[39:8]} & \texttt{flags[7:0]} \\
    \midrule
    4 bit & 2 bit & 18 bit & 32 bit & 8 bit \\
    \bottomrule
  \end{tabular}
\end{center}

This division between the address and doorbell eliminates a QPN field from the doorbell. It also prevents
the GPU from resubmitting metadata already stored in the Slot.
Table~\ref{tab:impl-doorbell-fields} defines each field and its acceptance conditions. The NIC
rejects every invalid reserved value before producing an internal request.

\begin{table}[htbp]
  \captionsetup{name=Table}
  \centering
  \caption{Fields in the 64-bit Doorbell.}
  \label{tab:impl-doorbell-fields}
  \small
  \begin{tabularx}{\linewidth}{@{}l c X X@{}}
    \toprule
    Field & Width & Interpretation & Validity constraint \\
    \midrule
    \texttt{seq} & 4 &
    Debug value supplied by the producer. &
    Any 4-bit value; does not affect admission, authorization, duplicate suppression, or reset
    correctness. \\
    \texttt{op} & 2 &
    Selects an operation in Table~\ref{tab:impl-doorbell-ops}. &
    \texttt{00}--\texttt{10} are valid; \texttt{11} is reserved. \\
    \texttt{slot\_id} & 18 &
    Capability index into the global physical Slot Table. &
    Must be less than \texttt{MAX\_SLOT}, and the entry must be valid. \\
    \texttt{len\_or\_imm32} & 32 &
    Encodes the length for a payload operation and the complete 32-bit Flag value for a standalone
    Ordered Flag. &
    For a payload, 0 encodes 16~MiB; other valid values range from 1 to 16~MiB and must not exceed
    \texttt{slot.max\_len}. \\
    \texttt{flags} & 8 &
    Reserved for future extensions. &
    Must currently be 0. \\
    \bottomrule
  \end{tabularx}
\end{table}

The opcode determines whether the NIC looks up one Slot or two adjacent Slots. It also determines
whether the operation consumes an ordered credit. Table~\ref{tab:impl-doorbell-ops} maps the
three valid operations to their output requests and ordered-credit consumption.

\begin{table}[htbp]
  \captionsetup{name=Table}
  \centering
  \caption{GPU Doorbell operations and their credit requirements.}
  \label{tab:impl-doorbell-ops}
  \small
  \begin{tabularx}{\linewidth}{@{}l l X c X@{}}
    \toprule
    Opcode & Name & Slot lookup and output request & Ordered credit & Semantics \\
    \midrule
    \texttt{00} & \texttt{WRITE\_RELAXED} &
    Looks up one Payload Slot and produces one payload request. &
    0 &
    Permits segmentation, out-of-order transmission, and out-of-order completion. \\
    \texttt{01} & \texttt{WRITE\_ORDERED\_FLAG} &
    Looks up one Flag Slot and produces one 4~B Ordered Flag request. &
    1 &
    Publishes the Flag only after preceding Writes on the same QP cross the ordered-write
    completion point. \\
    \texttt{10} & \texttt{WRITE\_RELAXED\_THEN\_FLAG} &
    Looks up an even-numbered Payload Slot and its adjacent Flag Slot, producing a payload request
    and an Ordered Flag request. &
    1 &
    One Doorbell triggers a related payload and its subsequent completion Flag. \\
    \texttt{11} & reserved &
    Does not look up or submit a Slot. &
    -- &
    Rejects the operation and reports an unsupported opcode. \\
    \bottomrule
  \end{tabularx}
\end{table}

\texttt{MAX\_SLOT} is an even physical capacity satisfying
$2 \leq \texttt{MAX\_SLOT} \leq 2^{18}$. Every \texttt{slot\_id} bit directly indexes one global,
flat Slot Table; no bits encode a QP group. The CPU/runtime preconfigures each Slot with the
permitted operation kind, local and remote addresses, key, maximum length, and peer metadata
needed to construct a request. In the baseline configuration, a trusted runtime pairs each
Doorbell address with the correct Slot handle. An optional \texttt{owner\_qpn} check can strengthen
this constraint without changing the doorbell ABI.

After decoding the doorbell, the NIC uses \texttt{slot\_id} to recover request information that the GPU
does not repeat on the fast path. A regular operation reads the single entry selected by
\texttt{slot\_id}. A combined operation requires an even \texttt{slot\_id}. It reads the selected
Payload Slot and the Flag Slot at \texttt{slot\_id + 1}. Before submitting either request
downstream, the NIC validates both entries. These checks cover the range, valid bit, operation
kind, capability, key, and length. Only after every check succeeds does the front-end atomically
produce one relaxed payload request and one Ordered Flag request on the same QP. A single doorbell can
therefore produce one or two requests. Only a fully validated result enters the common processing
pipeline.

Request construction produces candidate work rather than admitted work. The next stage applies the
atomic validation boundary before the request reaches WQE Parse.

\subsubsection{NIC Admission Rules and Errors}

The NIC front-end must finish the address, format, Slot, and capability checks before exposing a
constructed request to downstream logic. It performs the following steps in order:

\begin{enumerate}[leftmargin=2.2em,itemsep=0.2em]
  \item Check the BAR range and 8~B alignment of the Doorbell transaction, then extract the QPN
        from the address.
  \item Decode the doorbell as a little-endian 64-bit value and reject invalid \texttt{flags} or a
        reserved opcode.
  \item Check the \texttt{slot\_id} range and valid bit, then perform one lookup or two adjacent
        lookups as required by the operation.
  \item Validate the Slot kind, capability, key, length, and optional \texttt{owner\_qpn}.
  \item After every check succeeds, construct an internal request from the address-derived QPN,
        doorbell parameters, and Slot metadata.
  \item Pass the QPN, addresses, key, length, operation attributes, and association tag to the
        common WQE Parse stage.
\end{enumerate}

Only after these checks does WQE Parse allocate a PSN range and pass the tagged work to the TX
Engine in Section~\ref{sec:impl-tx-datapath}. At this point, one compact GPU Doorbell write has
become bounded, validated internal work. Admission proves only that the request has been accepted;
it does not imply packet transmission, payload DMA completion, Ordered Flag publication, or remote
GPU observation.
\FloatBarrier

%% file: sections/05_implementation/05_2_tx_datapath_implementation.tex
\subsection{TX Datapath and Multipath Packetization}
\label{sec:impl-tx-datapath}

The TX datapath converts the validated internal requests produced by
Section~\ref{sec:impl-work-submission} into packets that can be transmitted
in parallel across four network paths.
First-transmission requests and retransmission requests generated by
Section~\ref{sec:impl-reliability} share the same back-end pipeline,
differing only in entry source and scheduling priority.

To scale data processing bandwidth, datapaths such as payload DMA, packet assembly,
and FTile output are replicated per channel; protocol state that must remain
consistent---QP context, WQE Record, reliability windows, and congestion-control
state---is maintained through centralized pipelines, banking, and arbitration.
Table~\ref{tab:tx-pipeline-stages} summarizes the five main stages of the TX datapath.


\par\medskip
\noindent
\begin{minipage}{\linewidth}
    \centering

    \refstepcounter{table}
    \label{tab:tx-pipeline-stages}
    \addcontentsline{lot}{table}{
        \protect\numberline{\thetable}
        TX pipeline 的主要处理阶段
    }

    \small
    \textbf{Table~\thetable\quad Main processing stages of the TX pipeline.}
    \par\medskip

    \begingroup
    \setlength{\tabcolsep}{4pt}
    \renewcommand{\arraystretch}{1.15}

    \begin{tabularx}{\linewidth}{
        >{\centering\arraybackslash}p{0.08\linewidth}
        >{\raggedright\arraybackslash}p{0.19\linewidth}
        >{\raggedright\arraybackslash}p{0.19\linewidth}
        >{\raggedright\arraybackslash}X
    }
        \toprule
        Order & Stage & Input & Main Operations \\
        \midrule

        1
        & PSN Allocation \& WQE Record
        & Validated internal request
        & Allocates a contiguous PSN range and records request metadata
          needed for retransmission and resource reclamation.
        \\

        2
        & Page Table（PGT） Translation \& DMA Read
        & GPU address and memory capability
        & Completes address translation and permission checking, and
          asynchronously reads payload.
        \\

        3
        & Segmentation \& PSN Tagging
        & Payload and PSN range
        & Segments by MTU, fills in PSN and message boundary markers.
        \\

        4
        & EV Selection \& Channel Mapping
        & Packet metadata
        & Selects an active EV, queries the Path Table, and determines
          the SRv6 outer header and FTile channel.
        \\

        5
        & Packet Assembly \& Send
        & Header, payload, and output channel
        & The Payload/Ordered TX Engine completes packet assembly and writes
          to the corresponding FTile FIFO.
        \\

        \bottomrule
    \end{tabularx}

    \endgroup
\end{minipage}
\par\medskip

When a request enters the TX pipeline, the first step is to establish its identity
in protocol space: allocate a PSN range and record request metadata in the WQE
Record for subsequent use by the retransmission and reliability engines.

\techterm{PSN range allocation and WQE Record.}

WQE Parse first allocates a contiguous PSN range based on the request length and
MTU, and binds that range and the original operation to a WQE Record.
Each Record stores four categories of information:
\texttt{start\_psn} records the start of the PSN range, used to map a missing PSN
back to its original message;
\texttt{psn\_count} records the number of packets covered by this request;
DMA metadata stores addresses, keys, and lengths, used to reconstruct DMA Reads
for both first transmissions and retransmissions;
and a QP context reference links to the reliability window, Scoreboard, and
send state.

The WQE Record avoids replicating the full request descriptor for every outstanding
PSN.
When a SACK hole, NAK, or timeout work item arrives, the reliability engine can
locate the owning message and its payload offset from the PSN.

Record release is driven solely by packet-level acknowledgment progress:
the corresponding PSN range must have been cumulatively acknowledged, or every
PSN in the range must have been selectively acknowledged.
Local PCIe DMA Read completion only indicates that the payload has been fetched
from GPU memory and cannot serve as grounds for early WQE Record release
~\cite{rdma-aware-programming,mrc-spec}.

PSNs and WQE Records are only protocol-level metadata.
The actual payload still resides in GPU device memory; the NIC must actively read
the data into on-chip buffers via PCIe DMA before it can enter the subsequent
segmentation and packet assembly flow.

\techterm{PGT address translation and DMA read.}

A GPU request carries a pre-registered memory capability, not a raw physical
address that the NIC can directly use.
The DMA control path processes it in the following order
~\cite{rdma-aware-programming,gpudirect-rdma}:

\begin{enumerate}[leftmargin=2em]
    \item Query the PGT or equivalent address translation state to translate the
          GPU virtual/IO address into a PCIe address accessible by the NIC;

    \item Check whether memory registration, keys, and related lifecycle state
          remain valid;

    \item Issue a PCIe DMA Read and use an internal tag to associate the
          asynchronously returned payload with the original request and PSN range;

    \item Check the DMA return status and actual length. If the returned length
          differs from what the request declared, halt the current request and
          enter the unified error path.
\end{enumerate}

Host payload uses host address mapping and does not traverse the GPU PGT.
After address translation and permission checking complete, DMA Read and subsequent
packetization advance in a decoupled manner through bounded buffers, thereby
avoiding PCIe access latency from serially blocking the entire TX pipeline.

The payload returned by DMA can be up to several MB, far exceeding the MTU limit
of a single packet.
The NIC must segment it into a series of packets that can be independently
transmitted and retransmitted in the network, and mark each packet with its
position within the original message.

\techterm{Segmentation and packet assembly.}

The Segmentation module divides the payload into multiple packets according to
the MTU and assigns PSNs sequentially from the already-allocated PSN range.
This module does not maintain a per-QP global PSN counter: PSN ranges are
centrally allocated by WQE Parse, and Segmentation only consumes the local range
of the current request.
This separation decouples PSN state management from the payload data pipeline,
allowing DMA Reads and packet assembly for multiple requests to proceed in parallel.

Each packet uses two fields to mark message boundaries: the first packet sets
\texttt{isfirst}, the last sets \texttt{islast}, a single-packet message sets
both, and intermediate packets set neither.
In addition to boundary markers, the packet header also carries protocol fields
needed for receiver-side direct placement and reliability processing, as shown
in Table~\ref{tab:packet-header-fields}.


\par\medskip
\begin{center}
\refstepcounter{table}
\label{tab:packet-header-fields}
\addcontentsline{lot}{table}{
    \protect\numberline{\thetable}
    TX packet header 的关键字段
}

\small
\textbf{Table~\thetable\quad Key fields of the TX packet header.}
\par\smallskip

\begingroup
\setlength{\tabcolsep}{4pt}
\renewcommand{\arraystretch}{1.15}
\begin{tabularx}{\linewidth}{
    >{\raggedright\arraybackslash}p{0.19\linewidth}
    >{\raggedright\arraybackslash}p{0.23\linewidth}
    >{\raggedright\arraybackslash}X
}
    \toprule
Field & Source & Purpose \\
    \midrule

    QPN
        & Doorbell address or CPU WQE
        & Identifies the communication endpoint and protocol state to which
          the packet belongs.
        \\

        PSN
        & PSN range allocated by WQE Parse
        & Identifies packet order, tracked by the receiver bitmap and sender
          Scoreboard.
        \\

        Opcode
        & Doorbell \texttt{op} or CPU WQE
        & Distinguishes relaxed payload, Ordered Flag, and combined operation.
        \\

        \texttt{isfirst}/\texttt{islast}
        & Segmentation
        & Marks the start, end, or single-packet boundary of a message.
        \\

        Remote placement info
        & Slot Table or CPU WQE
        & Carries the target address, key, and offset, supporting receiver-side
          direct data placement.
        \\

        ECN
        & Congestion-control interface
        & Carries congestion-related semantics and is returned by the receiver
          through a separate feedback path.
    \\

    \bottomrule
\end{tabularx}
\endgroup
\end{center}
\par\medskip

After segmentation completes, each packet has an independent PSN, boundary
markers, and protocol header fields, but is not yet a complete packet ready
for network transmission.
Final header population, payload concatenation, and packet assembly are performed
by the four parallel Payload TX Engines.

\techterm{Four-way Payload TX Engine.}

The four Payload TX Engines each face one of the four FTile physical channels
and execute header population, payload concatenation, and packet output in parallel.
This replication scales data processing bandwidth rather than replicating full
protocol state.

State such as QP context, reliability window, PSN allocation,
WQE Record, and SQ Scoreboard is uniformly checked by the shared TX front-end
before packets are distributed to specific channels, with consistent access
provided through banking or arbitration.
The Payload TX Engine itself performs only data operations that are as stateless
as possible and does not independently advance QP or reliability state.

When backpressure occurs on one FTile FIFO, Route Channel Selection can redirect
subsequent packets that have not yet entered a specific Payload TX Engine to
other available channels.
Packets that have already completed path binding are not remapped due to
subsequent Path Table changes; switching channels also must not modify the QPN,
skip congestion-control admission, or bypass the packet-level reliability window.

After packet assembly, each packet has a complete header and payload, but still
lacks one critical piece of information: which of the four 100GE channels to
traverse.
EV/Path Table and Route Channel Selection perform this mapping, translating the
design principles of Section~\ref{sec:multipath-execution} into the hardware
pipeline.

\techterm{EV/Path Table and route channel selection.}

Section~\ref{sec:multipath-execution} defines the design principles of EV-based
packet spraying.
In the hardware implementation, the control plane pre-configures an active EV set
and a Path Table for each destination~\cite{rfc8754,rfc8986,mrc-srv6-paper}.
Each Path Table entry contains at least

\begin{center}
\texttt{\{SRv6 outer destination, local FTile port, TX channel\}}.
\end{center}

Route Channel Selection completes path selection in the following order before a
packet enters the Payload TX Engine:

\begin{enumerate}[leftmargin=2em]
    \item Select an EV based on the current send type and the active EV set;

    \item Query the Path Table using the destination and EV;

    \item Obtain the outer SRv6 address, local FTile port, and TX channel;

    \item Attach the path mapping result to the packet metadata and feed it into
          the corresponding TX Engine.
\end{enumerate}

Adjacent PSNs can select different EVs, so packets of the same QP can be sprayed
across different fabric paths.
Path selection supports two basic profiles: the basic profile rotates evenly among
active EVs, suitable for scenarios where paths are approximately symmetric and
congestion prior information is unavailable; the enhanced profile adjusts the
selection weight of each EV based on ECN frequency, loss history, path
availability, and skips paths that are temporarily
unavailable or overloaded.

Packets that have already entered the Payload TX Engine use the path mapping
determined at entry.
Subsequent updates to the Path Table or active EV set affect only new packets
that have not yet completed path selection, thereby preventing the same packet
from being repeatedly redirected within the pipeline.

Packets from different sources ultimately share Route Channel Selection and FTile
output.
Table~\ref{tab:route-channel-priority} gives the input priorities in the current
design.


\par\medskip
\begin{center}
\refstepcounter{table}
\label{tab:route-channel-priority}
\addcontentsline{lot}{table}{
    \protect\numberline{\thetable}
    Route Channel Selection 的输入优先级
}

\small
\textbf{Table~\thetable\quad Input priorities for Route Channel Selection.}
\par\smallskip

\begingroup
\setlength{\tabcolsep}{4pt}
\renewcommand{\arraystretch}{1.15}
\begin{tabularx}{\linewidth}{
    >{\centering\arraybackslash}p{0.12\linewidth}
    >{\raggedright\arraybackslash}p{0.30\linewidth}
    >{\raggedright\arraybackslash}X
}
    \toprule
Priority & Input Source & Description \\
        \midrule

        0 (highest)
        & ACK/SACK/NAK feedback
        & High-priority reliability feedback; does not wait for payload DMA
          completion.
        \\

        1
        & Ordered TX Engine
        & Ordered Flag or other ordered-control packet.
        \\

        2
        & Retransmission packet
        & Retry generated by SACK hole, NAK, or grouped timeout.
        \\

        3 (lowest)
        & First-transmission packet
        & New payload request generated by WQE Parse.
    \\

    \bottomrule
\end{tabularx}
\endgroup
\end{center}
\par\medskip

The path selection mechanism described above applies equally to ordinary payload
and Ordered Flag: both share the EV lookup, Route Channel Selection, and FTile
output infrastructure.
The distinction is that Ordered operations require independent semantic management
on the send side: they do not require physical ordering guarantees from the sender;
the ordering responsibility is entirely delegated to the receiver's Ordered Queue.

\techterm{Ordered TX Engine.}

The Ordered Flag reuses the same send infrastructure---EV lookup, Route Channel
Selection, and FTile output---while a dedicated Ordered TX Engine maintains its
request metadata.
For a combined operation \texttt{WRITE\_RELAXED\_THEN\_FLAG}, one Doorbell
atomically produces two logical requests: the payload request enters the Payload
TX Engine, and the Flag request enters the Ordered TX Engine.

The two types of requests use the same QP and occupy consecutive protocol ordering
positions, but each may select a different EV and can arrive out of order in the
network due to differing path and queue states.
The sender does not force the Flag packet to be bound to the payload packets in
physical send completion order; the receiver re-establishes the semantic dependency
between the two based on the barrier information stored in the Ordered Queue,
thereby guaranteeing that the Flag cannot cross the ordered completion point ahead
of the preceding payload it covers.

Table~\ref{tab:relaxed-vs-ordered-tx} summarizes the core differences between
relaxed payload and Ordered operations in the TX path.


\par\medskip
\begin{center}
\refstepcounter{table}
\label{tab:relaxed-vs-ordered-tx}
\addcontentsline{lot}{table}{
    \protect\numberline{\thetable}
    普通 Write 与 Ordered operation 在 TX path 中的差异
}

\small
\textbf{Table~\thetable\quad
Differences between ordinary Write and Ordered operation in the TX path.}
\par\smallskip

\begingroup
\setlength{\tabcolsep}{4pt}
\renewcommand{\arraystretch}{1.15}
\begin{tabularx}{\linewidth}{
    >{\raggedright\arraybackslash}p{0.22\linewidth}
    >{\raggedright\arraybackslash}p{0.35\linewidth}
    >{\raggedright\arraybackslash}X
}
    \toprule
Dimension
        & Ordinary Write (\texttt{WRITE\_RELAXED})
        & Ordered Flag / Combined \\
        \midrule

        Send module
        & Payload TX Engine
        & Ordered TX Engine
        \\

        Path selection
        & Supports per-packet EV spraying
        & Also supports independent EV selection
        \\

        Sender-side ordering
        & Does not restore packet or message order
        & Does not depend on sender-side physical ordering
        \\

        Ordering responsibility
        & No additional Flag publication barrier
        & Guaranteed by receiver-side Ordered Queue and DMA watermark
        \\

        Ordered credit
        & Not consumed
        & Each Ordered Flag or combined operation consumes 1
        \\

        Receiver processing
        & Payload written to target location via DDP; consumption determined by
          upper-layer synchronization or a subsequent Flag
        & Receiver publishes GPU-visible Flag only after preceding Writes have
          crossed the barrier
    \\

    \bottomrule
\end{tabularx}
\endgroup
\end{center}
\par\medskip

Regardless of whether a packet originates from the Payload TX Engine, the Ordered
TX Engine, or the retransmission path, it must ultimately pass through the same
send gate before entering the network.
First transmissions and retransmissions do not maintain two independent back-end
mechanisms.

\techterm{Unified entry for first transmissions and retransmissions.}
Requests are uniformly converted into packet work and undergo the same send gating
before entering the DMA, packetization, and path selection pipelines:

\begin{enumerate}[leftmargin=2em]
    \item The QP must be in the active state;

    \item The target PSN must lie within the current MPR or reliability window;

    \item The congestion-control interface must grant network injection permission
          for the current packet;

    \item The target FTile FIFO or an alternative channel must have available space;

    \item The current destination must have at least one available EV.
\end{enumerate}

Retransmission packets can receive bounded priority to fill receive-window holes
and advance the cumulative acknowledgment point as quickly as possible.
The arbiter must simultaneously reserve a minimum service fraction for first
transmissions, preventing sustained retry traffic from starving new requests
over the long term.

Three categories of control boundaries must be distinguished here.
FTile FIFO backpressure only protects the local output FIFO from overflowing;
Doorbell credit only protects NIC admission state and bounded request buffers;
congestion control is responsible for regulating network injection rate to avoid fabric congestion~\cite{intel-ftile-ethernet,dctcp,dcqcn,timely,hpcc,swift}.

The first two mechanisms belong to local bounded-resource protection; they do not
infer end-to-end congestion and do not determine network fairness.
Both first transmissions and retransmissions must pass through the independent
congestion-control interface; retry packets having higher priority, FTile FIFO
having available space, or the GPU having obtained Doorbell credit cannot justify
bypassing network injection limits.

At this point, a request has completed the full TX flow from PSN allocation,
DMA read, segmentation and packet assembly, and path selection through to final
transmission.
This pipeline serves both first-transmission and retransmission requests, with
unified gating ensuring that every packet satisfies the three constraints of
congestion control, reliability, and local resources before entering the network.

%% file: sections/05_implementation/05_3_rx_datapath_implementation.tex
\subsection{RX Datapath and Direct Placement}
\label{sec:impl-rx-datapath}

The RX datapath simultaneously faces four FTile inputs, out-of-order arrival
caused by packet spraying, and shared QP state that must be maintained across
the four parallel channels.
Its core responsibilities are threefold: (1)~leave reliability evidence for every
arriving packet as quickly as possible, driving ACK/SACK feedback; (2)~write
payload directly to its final location in remote GPU memory (DDP); (3)~preserve
the global ordering information necessary for subsequent ordered completion.
Table~\ref{tab:rx-pipeline-stages} gives an overview of the five RX pipeline stages.

\begin{table}[htb]
    \centering
    \caption{RX pipeline stages.}
    \label{tab:rx-pipeline-stages}
    \small
    \setlength{\tabcolsep}{4pt}
    \renewcommand{\arraystretch}{1.15}
    \begin{tabularx}{\linewidth}{
        >{\centering\arraybackslash}p{0.08\linewidth}
        >{\raggedright\arraybackslash}p{0.18\linewidth}
        >{\raggedright\arraybackslash}p{0.15\linewidth}
        Y
    }
        \toprule
        Order & Stage & Input & Operation \\
        \midrule

        1 & Four-way Receive \& Buffer
        & Network packet stream ($\times$4)
        & Receives packets; buffers absorb rate differences.
        \\

        2 & Packet Parsing
        & Raw packet $\times$4
        & Extracts QPN, PSN, opcode, boundary markers, placement info, EV, ECN.
        \\

        3 & Bitmap Update \& Feedback Generation
        & Parsed data packet
        & Updates RQ Bitmap; advances \texttt{ack\_psn}; generates ACK/SACK.
        \\

        4 & Direct Data Placement (DDP)
        & Payload + placement info
        & Four-way parallel PCIe DMA Write; writes to final destination in
          target GPU memory.
        \\

        5 & Ordered Retire
        & Four-way DMA completion requests
        & Merges by \texttt{global\_seq}; advances \texttt{dma\_ack\_psn}.
        \\

        \bottomrule
    \end{tabularx}
\end{table}

After packets arrive at the NIC from the network, they first enter the four
receive channels.
Each channel buffers and parses independently.
This stands in sharp contrast to the TX pipeline:
TX centrally allocates PSNs and then distributes to four channels; RX receives
in parallel across four channels and then converges at a merge point.

\techterm{Four-way receive and packet parsing.}
Each FTile channel corresponds to one RQ receive entry and input buffer.
The buffer absorbs short-term rate differences among the Parser, bitmap update,
and DMA backpressure, but cannot compensate for a sustained situation where the
service rate falls below the input rate: a processing budget of 24/32 cycles is
the red line for not blocking Ethernet input.
Four Packet-Parsers disassemble headers in parallel, separate data packets from
ACK/SACK/NAK control packets, and extract QPN, PSN, opcode,
\texttt{isfirst/islast}, remote placement information, EV, and ECN markers.

Control packets do not enter the payload DMA path.
The Parser sends ACK/SACK/NAK directly to the feedback processing queue of
Section~\ref{sec:impl-reliability}; data packets simultaneously enter both the
bitmap update and DMA placement paths.

After the Parser has extracted the header information, the first priority for a
data packet is not to write the payload, but to immediately leave reliability
evidence: inform the sender before payload DMA completes.
This task is handled by the RQ Bitmap.

\techterm{RQ Bitmap management and feedback generation.}
After a data packet passes basic validity checks, it immediately updates the
receive bitmap of the corresponding QP---this action occurs before payload DMA
completes, with the goal of leaving reliability evidence as quickly as possible.
The specific steps:

\begin{enumerate}[leftmargin=2em]
    \item Set the bitmap bit corresponding to the arriving PSN; identify
          duplicates (already set) and out-of-window packets.
    \item Starting from the lowest position of the window, contiguous 1's are
          consumed, advancing \texttt{ack\_psn} and sliding the window.
    \item Select the feedback type based on the degree of reordering: return a
          lightweight cumulative ACK when reordering is low; return a
          bitmap-carrying lazy SACK when the reordering degree exceeds the
          threshold; return clearly pinpointable anomalies via NAK.
\end{enumerate}

ACK/SACK is generated and sent back immediately after bitmap update, without
waiting for payload DMA Write to complete.
This establishes the first critical layering boundary: \texttt{ack\_psn} only
indicates packet-level arrival progress; it does not indicate that the payload
has been written to GPU memory, and still less can it directly issue a
GPU-visible completion Flag~\cite{mrc-spec,gpudirect-rdma}.

The bitmap and ACK answer the question of ``whether the packet has arrived.''
The next question is ``to which address should the payload be written.''
Placing data at the correct location in GPU memory is the responsibility of DDP.
Because packet spraying causes out-of-order arrival, DDP does not require
reordering before writing; instead, each packet independently carries its target
address and writes directly to its final location.

\techterm{Four-way Direct Data Placement (DDP).}
Validated data packets enter the four DMA Placement modules according to the
remote address, offset, and key they carry.
Packets are not reordered by network arrival order.
Different lanes can issue PCIe DMA Writes in parallel, with each packet
independently writing to its final target address.
This design ensures that the out-of-order nature of packet spraying does not incur
additional payload copies or reorder buffers
~\cite{rdma-aware-programming,gpudirect-rdma,mprdma,left}.

Each DMA lane produces completion requests in input order and writes them to its
lane's output FIFO.
Table~\ref{tab:ddp-lane-output} lists the information carried by each lane's
output interface.

\begin{table}[htb]
    \centering
    \caption{Information carried by DMA lane output FIFO entries.}
    \label{tab:ddp-lane-output}
    \small
    \setlength{\tabcolsep}{4pt}
    \renewcommand{\arraystretch}{1.15}
    \begin{tabularx}{\linewidth}{
        >{\raggedright\arraybackslash}p{0.25\linewidth}
        Y
    }
        \toprule
        Field & Description \\
        \midrule

        \texttt{global\_seq}
        & Globally monotonic sequence number; assigned after Bitmap merge and
          before DMA dispatch, used by Ordered Retire for merging.
        \\

        \texttt{ack\_valid} / \texttt{ack\_psn\_after\_merge}
        & Whether the current request advanced \texttt{ack\_psn} and the value
          after advancement; used to advance \texttt{dma\_ack\_psn[qpn]} only
          after the request has been dequeued from Ordered Retire.
        \\

        QPN, PSN range, DMA completion status
        & Identifies the owning QP and the corresponding packet range; DMA
          success/failure flag.
        \\

        \bottomrule
    \end{tabularx}
\end{table}

After the four DMA lanes independently complete their writes, completion requests
are scattered across the output FIFOs of four lanes.
What the Ordered Queue needs, however, is not the local progress of one lane,
but the global order across all four lanes.
Completion requests from the four lanes must therefore be merged.
This work is done by the Single Ordered Retire module.

\techterm{Single Ordered Retire and global order merging.}
The output FIFOs of the four lanes each preserve the completion order within their
lane.
A single Ordered Retire module selects from the four FIFO heads the entry whose
\texttt{global\_seq} equals the current \texttt{retire\_head} for dequeuing;
if that entry has not yet arrived, all subsequent requests wait in their
respective FIFOs.

When a request is dequeued, if its \texttt{ack\_valid} is 1,
\texttt{dma\_ack\_psn} of that QP is advanced.
\texttt{dma\_ack\_psn} is the watermark after packets have undergone DMA placement
and been confirmed in global order, used by the Ordered Queue of
Section~\ref{sec:impl-ordered-completion} to evaluate the barrier condition for
Flags.
After Retire dequeuing, requests continue into Completion Management, which
generates CPU/GPU-visible completion notifications.

All of the above RX stages share one engineering constraint: four Parsers may
access the bitmap of the same QP in the same cycle, and four DMA lanes may
simultaneously complete requests for different QPs.
The tension between parallelism and consistency must be resolved through banking
and arbitration.

\techterm{Shared state access.}
Four Parsers may access the bitmap of the same QP in adjacent cycles, and four
DMA lanes may simultaneously complete requests for different QPs.
Entry work queues, banking, and pipelined arbitration are responsible for
serializing conflicts on the same state item, rather than compressing all RX
processing into a single serial path.
The implementation must separately validate two access patterns: random multi-QP
access, which exposes bank distribution problems, and sustained high-pressure
single-QP access, which exposes hotspot arbitration bottlenecks.

%% file: sections/05_implementation/05_4_reliability_retransmission_implementation.tex
\subsection{Reliability and Recovery Engines}
\label{sec:impl-reliability}

The reliability engine converts evidence from the receive side into sender-side
state and retransmission work.
ACKs, SACKs, and NAKs respectively carry cumulative progress, selective
acknowledgment, and pinpoint anomalies; the specific header bit layouts and
trigger thresholds are configured as protocol parameters and do not alter the
state transition semantics.
OpenRDMA supports two recovery profiles: full hardware recovery and
software-coordinated recovery.
The two share identical transport semantics, differing only in state location,
recovery latency, and verification cost (Table~\ref{tab:recovery-profiles}).

\begin{table}[htb]
    \centering
    \caption{Division of responsibilities between the two recovery profiles.}
    \label{tab:recovery-profiles}
    \small
    \setlength{\tabcolsep}{4pt}
    \renewcommand{\arraystretch}{1.15}
    \begin{tabularx}{\linewidth}{
        >{\raggedright\arraybackslash}p{0.30\linewidth}
        >{\raggedright\arraybackslash}p{0.32\linewidth}
        >{\raggedright\arraybackslash}p{0.30\linewidth}
    }
        \toprule
        Responsibility & Hardware profile & Software-coordinated profile \\
        \midrule

        RX bitmap, ACK/SACK/NAK generation
        & NIC hardware
        & NIC hardware (same)
        \\

        SQ Scoreboard maintenance
        & NIC hardware bitmap
        & NIC hardware bitmap(same)
        \\

        SACK hole inference and loss judgment
        & NIC hardware (local post-hole ack threshold)
        & CPU
        \\

        Per-QP timer / retry counter
        & NIC grouped timer
        & CPU per-QP timer
        \\

        Retransmission request generation
        & NIC retransmission scheduler
        & CPU generates retry send descriptor
        \\

        Retransmission execution (MPR/credit/CC check)
        & Unified TX Engine
        & Unified TX Engine (same)
        \\

        \bottomrule
    \end{tabularx}
\end{table}

In both profiles, the FPGA maintains the persistent sender-side scoreboard (WQE Record and Bitmap). In the software-coordinated profile, the NIC merges the bitmap with WQE metadata into a complete state report for the CPU, which then decides retransmissions.

ACK, SACK, NAK: the three feedback types carry information at different
granularities.
ACK advances the cumulative acknowledgment point and releases acknowledged PSNs;
SACK additionally exposes a selective acknowledgment bitmap under reordering
conditions;
NAK points to clearly identified recoverable anomalies.
The specific header bit layouts and trigger thresholds are protocol parameters
and do not affect the state transitions below~\cite{mrc-spec,uec-spec}.

\techterm{Logical content of ACK / SACK / NAK.}
The three feedback types each carry different information granularity:

\begin{enumerate}[leftmargin=2em]
    \item \textbf{ACK}: carries the QPN and cumulative \texttt{ack\_psn},
          releasing all prior contiguous PSNs.
          The default feedback under low-reordering conditions; can aggregate
          acknowledgments for multiple packets.
    \item \textbf{SACK}: in addition to ACK, carries a receive bitmap starting
          from the acknowledgment boundary, allowing the sender to mark
          selectively received PSNs.
          Sent only when the reordering degree exceeds a threshold (lazy SACK),
          and can compensate for feedback packet loss through periodic refresh.
    \item \textbf{NAK}: points to a clearly identified recoverable anomaly and
          its target PSN, such as the arrival of a trimmed packet header or a
          checksum failure.
          Unrecoverable anomalies such as permission errors and DMA failures do
          not enter NAK but go directly to a fatal error report.
\end{enumerate}

Path information such as EV/ECN carried in the feedback is delivered to the
congestion control policy through a separate CC interface and does not alter
reliability state transitions.

Feedback is generated by the receiver-side RQ Bitmap Management,
whose factual basis is the per-QP receive bitmap, \texttt{ack\_psn},
and \texttt{ooo\_count}.
These states only record packet arrival and do not wait for DMA.

\techterm{Receiver side: RQ Bitmap factual state.}
The per-QP RQ Bitmap Management maintains \texttt{ack\_psn}, the out-of-order
bitmap, \texttt{ooo\_count}, and duplicate/window judgments.
\texttt{ooo\_count} counts the number of packets that have arrived above
\texttt{ack\_psn} and is used to decide whether to switch from ACK to SACK mode.
The receiver side generates feedback based on events such as cumulative
advancement, accumulated reordering, or SACK refresh timer expiry.
Duplicate packets can again prompt the receiver to return ACK/SACK but must not
repeat state updates with side effects beyond the bitmap.

\texttt{ack\_psn} only indicates packet-level arrival progress.
Even after ACK/SACK has been returned, subsequent placement permission checks or
payload DMA may still fail, so this watermark cannot directly issue a GPU-visible
Flag.

After evidence from the receive side reaches the sender via ACK/SACK/NAK feedback,
the sender digests this information using the SQ Scoreboard.
It maintains the acknowledgment status of each outstanding PSN and drives
retransmission decisions.

\techterm{Sender side: SQ Scoreboard and WQE Record reverse lookup.}
The hardware profile maintains an SQ Scoreboard bitmap for the outstanding PSNs
of each QP.
Upon feedback arrival:

\begin{enumerate}[leftmargin=2em]
    \item Merge the cumulative ACK, advance the cumulative acknowledgment point,
          and release acknowledged PSN ranges.
    \item Set the selectively acknowledged bits in the SACK bitmap.
    \item PSNs that have been sent but remain 0 in the bitmap are hole candidates;
          they are marked as retransmission candidates after satisfying the local
          post-hole acknowledgment threshold.
    \item PSNs pointed to by NAKs directly enter the retransmission candidate
          queue.
\end{enumerate}

The Scoreboard only indicates the acknowledgment status of PSNs; retransmission
must also reconstruct the original packet.
The WQE Record stores, per message, the PSN range (\texttt{start\_psn},
\texttt{psn\_count}), DMA metadata, and request identity;
the retransmission module uses QPN + PSN to reverse-lookup the owning range and
payload offset, generating retransmission work that retains the original PSN and
sets the \texttt{is\_retry} flag.
The Record's lifetime follows the acknowledgment status of all its PSNs and must
not be released early upon first DMA Read completion~\cite{mrc-spec,irn}.

The \texttt{retransmit\_pending} flag and the pre-send secondary Scoreboard check
prevent the same PSN from being concurrently re-enqueued by multiple feedback
sources; if a newly arrived ACK/SACK has already acknowledged that PSN, the old
retransmission candidate is canceled.

The Scoreboard and WQE Record provide the state needed for retransmission,
but the generation, scheduling, and deduplication of retransmission requests
still require dedicated management mechanisms.
The SACK work queue handles feedback-driven retransmission, while Grouped Timeout
serves as the fallback when feedback is lost.

\techterm{SACK work queue and Grouped Timeout.}
SACK/NAK processing is decoupled from BRAM lookups through a work queue: when
multiple QPs simultaneously produce holes, they can be queued in a bounded manner
and backpressured without blocking normal reception by the RX Parser.
Grouped timeout stores expiry work for a group of contiguous PSNs (a fixed-size
timer group); upon scheduling, it rescans the latest Scoreboard and retransmits
only PSNs that remain outstanding, skipping those already acknowledged by
subsequent feedback.

Retransmission requests generated by SACK holes, NAKs, and grouped timeouts
ultimately converge into a unified retransmission path.
The three recovery paths share a single entry point:
the TX Engine / Congestion Control Module of Section~\ref{sec:impl-tx-datapath}.

The above describes the full hardware recovery path.
All reliability state and decisions are closed within the NIC.
If FPGA resources are constrained, however, retransmission decisions can be moved
to the CPU, with the NIC retaining only observation and execution capabilities.
The software-coordinated recovery profile is designed for precisely this purpose.

\techterm{Software-coordinated recovery profile.}
The software profile places the sender-side Scoreboard, hole inference, per-QP
timer, retry counter, and error transitions on the CPU/runtime; the NIC retains
only reliability observation and execution capabilities.

After the NIC accepts a GPU doorbell or host WQE, it provides the CPU with an
accepted-request report sufficient to establish a replay shadow, containing the
QPN, PSN range, request identity, packetization/DMA metadata, and generation.
The normal GPU fast path does not wait for the CPU to consume the report; if the
shadow backlog exceeds the recoverable range, backpressure or fail-stop must be
applied.

Remote feedback is parsed by the NIC into a meta report containing the
acknowledgment boundary, bitmap range, and event cause.
A CPU worker updates the shadow scoreboard, maintains per-QP timers, and generates
send descriptors with the \texttt{is\_retry} flag when policy conditions are met.
The descriptor retains the original QP and PSN; before execution the NIC still
checks QP state, MPR window, credit, CC permission, and resources.
After a Slot is rebuilt, retries from an old generation must be rejected.

Neither profile requires perfectly canceling all in-flight retries: before
transmission, the latest Scoreboard is used to cancel as many as possible;
duplicates already sent are suppressed by the receiver PSN window.
The worst case is extra link traffic, not duplicate issuance of DMA completions
or Ordered Flags.

%% file: sections/05_implementation/05_5_ordered_completion_implementation.tex
\subsection{Ordered Completion Protocol and Hardware}
\label{sec:impl-ordered-completion}

Packet-level reliability does not by itself establish GPU-safe completion. Multipath execution
may deliver an Ordered Flag while preceding payloads are still progressing through other paths or
DMA lanes. The receiver therefore cannot publish the Flag from packet-arrival evidence alone. It
must derive a release condition from the actual write progress of all four DMA lanes.

The implementation handles this condition through three ordering and release stages. Per-lane
output FIFOs preserve each DMA lane's completion order and carry a global sequence number. A
single Ordered Retire stage merges the four streams and advances the per-QP
\texttt{dma\_ack\_psn}. The per-QP Ordered Queue then evaluates each Flag's barrier before
submitting it to the Flag DMA path. Completion Management separately handles CPU CQE generation
and Flag DMA completion or error state, while the runtime handles final GPU observation.
Table~\ref{tab:ordered-completion-stages} summarizes the three stages and the subsequent completion
reporting boundary. The following parts describe their ordering state, release conditions, and
completion boundaries.

\begin{table}[htb]
    \captionsetup{name=Table}
    \centering
    \caption{Ordering stages and completion-reporting boundary.}
    \label{tab:ordered-completion-stages}
    \small
    \setlength{\tabcolsep}{4pt}
    \renewcommand{\arraystretch}{1.15}
    \begin{tabularx}{\linewidth}{
        >{\centering\arraybackslash}p{0.09\linewidth}
        >{\raggedright\arraybackslash}p{0.18\linewidth}
        >{\raggedright\arraybackslash}p{0.15\linewidth}
        Y
    }
        \toprule
        Role & Component & Input & Main operation \\
        \midrule

        Stage 1 & Per-lane completion buffering
        & DMA-complete requests from four lanes
        & Buffers completion metadata in each lane's input order and carries \texttt{global\_seq}
          and
          \texttt{ack\_valid/ack\_psn\_after\_merge}. \\

        Stage 2 & Single Ordered Retire
        & Heads of the four FIFOs
        & Dequeues the FIFO head whose \texttt{global\_seq} equals \texttt{retire\_head} and
          updates \texttt{dma\_ack\_psn[qpn]} when \texttt{ack\_valid} is set. \\

        Stage 3 & Per-QP Ordered Queue release
        & Head Flag metadata, \texttt{dma\_ack\_psn}, and Flag DMA readiness
        & Checks the head Flag's barrier, submits the Flag to the DMA path, and returns the ordered
          credit after the request is accepted and the entry is dequeued. \\

        Reporting & Completion Management
        & Retired request state and DMA completion status
        & Produces CPU CQEs and handles Flag DMA completion or errors. Final GPU-visible completion
          requires GPU observation of the Flag. \\

        \bottomrule
    \end{tabularx}
\end{table}

The first three stages establish ordered Flag release. The per-lane FIFOs and Single Ordered Retire
produce the progress watermark consumed by the Ordered Queue. Completion reporting follows this
release path but does not add another ordering condition.

A Flag may be published only after all preceding Writes on the same QP have crossed the
ordered-write completion point. Because the four DMA lanes complete independently, their
lane-local orders cannot establish this global condition. OpenRDMA therefore assigns a globally
monotonic \texttt{global\_seq} to every accepted data request. It assigns this value after RQ
Bitmap merge and before DMA fan-out.

\techterm{Global Sequence and Per-Lane Output FIFOs.}
After RQ Bitmap merge and before dispatch to the four DMA lanes, each accepted data request
receives a \texttt{global\_seq}. If the request advances the cumulative packet watermark, it also
carries \texttt{ack\_valid=1} and \texttt{ack\_psn\_after\_merge}. After completing its PCIe DMA
Write, each lane places the request's completion metadata into an ordinary FIFO output buffer in
input order. A future lane implementation that permits out-of-order completion must reconstruct
lane-local order before inserting entries into this FIFO.

The global sequence establishes a common order across the four FIFOs, but does not itself retire
requests. The per-lane FIFOs preserve the local evidence required for a global merge. The
downstream Single Ordered Retire stage performs this merge by examining only the four FIFO heads.
A request becomes eligible to retire when its \texttt{global\_seq} equals the current
\texttt{retire\_head}. This retirement boundary also separates packet progress from DMA progress.
\texttt{ack\_psn} is available after Bitmap merge. In contrast, \texttt{dma\_ack\_psn} cannot
advance until the corresponding request has completed DMA and retired in global order.

\techterm{Single Ordered Retire.}
The retire unit maintains \texttt{retire\_head}. When one FIFO head has
\texttt{global\_seq == retire\_head}, the unit dequeues that request and advances the retire head.
If the request carries \texttt{ack\_valid=1}, the unit also updates
\(\texttt{dma\_ack\_psn[qpn] = ack\_psn\_after\_merge}\). Thus,
\texttt{ack\_psn} can drive ACK/SACK feedback immediately after bitmap merge, while
\texttt{dma\_ack\_psn} advances only after the corresponding DMA operation completes and retires
in global order.

This single-retire design avoids a globally searchable reorder buffer, but introduces head-of-line
blocking across QPs. If an early request has not completed, later requests in other lanes cannot
retire even when they belong to different QPs. The first implementation accepts this tradeoff
between state size and cross-QP parallelism. Per-QP retirement or virtual output queues remain
possible extensions. Table~\ref{tab:ack-vs-dma-ack} summarizes the distinct roles of the two
progress points.

\begin{table}[htb]
    \captionsetup{name=Table}
    \centering
    \caption{Roles of \texttt{ack\_psn} and \texttt{dma\_ack\_psn}.}
    \label{tab:ack-vs-dma-ack}
    \small
    \setlength{\tabcolsep}{4pt}
    \renewcommand{\arraystretch}{1.15}
    \begin{tabularx}{\linewidth}{
        >{\raggedright\arraybackslash}p{0.18\linewidth}
        >{\raggedright\arraybackslash}p{0.35\linewidth}
        >{\raggedright\arraybackslash}p{0.39\linewidth}
    }
        \toprule
        Progress point & Advance condition & Purpose \\
        \midrule

        \texttt{ack\_psn}
        & Advances immediately after bitmap merge without waiting for DMA.
        & Provides the cumulative receive point for ACK/SACK feedback and allows the sender to
          release acknowledged PSNs. \\

        \texttt{dma\_ack\_psn[qpn]}
        & Advances after the corresponding request completes DMA and retires by
          \texttt{global\_seq}.
        & Provides the Ordered Queue barrier watermark and confirms that the covered payloads have
          been written to GPU memory. \\

        \bottomrule
    \end{tabularx}
\end{table}

Retirement supplies the progress value consumed by the third stage. It does not itself publish a
Flag.

Once \texttt{dma\_ack\_psn} advances, the receiver knows the DMA progress of each QP, but the
watermark alone does not trigger publication. The per-QP Ordered Queue performs the release. Its
head Flag can enter the DMA path only when
\(\texttt{barrier\_psn <= dma\_ack\_psn[qpn]}\)
and the Flag DMA path is ready.

\techterm{Per-QP Ordered Queue.}
The receiver maintains a finite-capacity Ordered Write FIFO for each active QP. Multipath
forwarding may deliver an Ordered Flag packet before some preceding payload packets. After the
Packet Parser completes validity and duplicate checks, it inserts only the Flag's
\texttt{barrier\_psn}, address, key, value, and request identity into the corresponding QP's FIFO.
The original packet buffer can then be released. Payload packets continue through DDP to their
final locations and never enter this FIFO. The queue therefore stores deferred-publication
metadata rather than complete packets or preceding payloads.

The head entry is submitted only when
\(\texttt{barrier\_psn <= dma\_ack\_psn[qpn]}\)
and the Flag DMA path is ready. After the write path accepts the Flag request, the Ordered Queue
dequeues the entry and returns an ordered credit carrying the absolute
\texttt{\{qpn, ordered\_grant\_limit\}} watermark. Acceptance means only that downstream logic has
taken responsibility for the request and that the queue slot can be reused. It does not mean that
the Flag DMA has completed or that the remote GPU has observed the Flag.

The queue slot can now be reused, but the notification is not yet complete. The Flag must still
cross the DMA-completion and consumer-observation boundaries.

A DMA error must stop the corresponding retirement and Flag release until the host performs stop,
drain, and reset. Per-QP ordered credits protect the Ordered Queue capacity during normal
operation. If the hardware nevertheless detects \texttt{ORDERED\_QUEUE\_FULL}, the credit and
queue lifecycle have diverged. The NIC reports an error and enters fail-stop rather than treating
the full queue as ordinary backpressure.

The complete lifecycle exposes five distinct boundaries. Advancing \texttt{ack\_psn} establishes
only contiguous packet-level receive evidence, not payload DMA completion. Advancing
\texttt{dma\_ack\_psn} establishes that the cumulative watermark has completed DMA and retired in
global order. It does not publish a Flag whose barrier remains unsatisfied. Acceptance by the Flag
DMA path releases the Ordered Queue slot and permits the credit return. It does not establish Flag
DMA completion. Flag DMA completion establishes that the NIC/PCIe write path has completed the Flag
write. It does not establish that the GPU cache path has observed the value. Only GPU observation
through the platform-defined polling sequence allows the GPU to consume the payloads covered by
that Flag. It provides no completion guarantee for other QPs or for memory operations outside the
Flag's scope.

OpenRDMA exposes two consumer paths at different completion boundaries. A conventional CPU
consumes CQEs from a host CQ ring. A GPU kernel instead polls a preconfigured Flag address
directly.

\techterm{CPU CQEs and GPU Flags.}
OpenRDMA supports both notification paths:

\begin{enumerate}[leftmargin=2em]
    \item \textbf{CPU CQE path}: Completion Management writes the request identity, status, and
          transport result into a host CQ ring. The driver polls the CQE, advances the consumer
          head, and reclaims the entry.
    \item \textbf{GPU Flag path}: The Ordered Queue submits a Flag write directly to a
          preconfigured GPU memory address. The GPU kernel polls that address without parsing or
          reclaiming a CQE.
\end{enumerate}

The two paths can share DMA-write infrastructure, but differ in queue consumers and reclamation
responsibilities. Completion Management handles the NIC-visible boundary, while the runtime
handles the final GPU-observation boundary. An Ordered Flag establishes NIC/PCIe write-order
completion. It does not automatically establish GPU cache-coherent visibility. The runtime must
poll the Flag with the cache-bypass or volatile load required by the target platform. It must also
use the corresponding acquire/fence sequence. If the platform cannot provide and validate this
contract, the driver must select the CPU/CQE fallback. It must not interpret a Flag value observed
through an ordinary cached load as successful completion
~\cite{gpudirect-rdma,cuda-c-programming-guide}.

\FloatBarrier

%% file: sections/05_implementation/05_6_resource_and_timing_considerations.tex
\subsection{Resource, Lifecycle, and Timing}
\label{sec:impl-resource-timing}

OpenRDMA cannot simply replicate every module four times.
Four-way 100GE demands that payload movement, Parsers, and DMA lanes remain
continuously parallel; if protocol states such as QP context, Slot Table,
RQ Bitmap, Scoreboard, WQE Record, Timer, and Ordered Queue were replicated
without constraint, they would introduce merge conflicts and BRAM bloat.
The implementation principle is \textbf{to replicate bandwidth paths, to
centralize or hierarchically maintain state that must be consistent, and to
isolate shared access through queues, banking, and arbitration}.

The preceding five sections defined the functional modules of each hardware path.
This section turns to their implementation cost: throughput constraints, resource
scale, shared-access conflicts, lifecycle, and timing closure.

\techterm{24/32-cycle throughput budget.}
The target is four-way 100GE with 400~Gbps aggregate throughput at 400~MHz.
Across four channels, 1024~bits (128~B) are processed per cycle; at a 4~KiB MTU,
each packet occupies approximately $4096/128 = 32$ aggregate data cycles, and at
3~KiB, approximately 24 cycles.
This figure describes the order of magnitude of the sustained-input service rate,
not an upper bound on the end-to-end latency of the Parser, bitmap, DMA, or
completion path.

The steady-state throughput of shared modules must not fall below the packet
arrival rate; otherwise, adding finite buffers only defers overflow.
Table~\ref{tab:parallel-vs-shared} shows the scaling approach for datapath modules
versus protocol-state modules.

\begin{table}[htb]
    \centering
    \caption{Scaling approach for datapath modules and protocol-state modules.}
    \label{tab:parallel-vs-shared}
    \small
    \setlength{\tabcolsep}{4pt}
    \renewcommand{\arraystretch}{1.15}
    \begin{tabularx}{\linewidth}{
        >{\raggedright\arraybackslash}p{0.28\linewidth}
        >{\centering\arraybackslash}p{0.18\linewidth}
        Y
    }
        \toprule
        Module & Scaling approach & Description \\
        \midrule

        Payload TX Engine, RX Entry, Packet-Parser, DMA Placement
        & $\times$4 lane parallel
        & Replicated per channel; scales payload bandwidth.
        \\

        RQ Bitmap, QP Context, Scoreboard
        & Banking / centralized pipeline
        & Shared access across multiple lanes; banks provide independent throughput,
          arbitration guarantees serialization.
        \\

        WQE Record
        & Single-BRAM reverse-lookup table
        & The SACK retransmission queue bounds the lookup volume; read-only replicas
          can be added if higher throughput is needed.
        \\

        Ordered Retire, Ordered Queue
        & Centralized
        & Global order merging; per-QP retire as a future enhancement.
        \\

        \bottomrule
    \end{tabularx}
\end{table}

The 24/32-cycle figure is the overall throughput constraint.
At the level of individual modules, resource requirements grow along different
dimensions: some grow with QP count, others with window size.
Understanding these growth factors is a prerequisite for resource planning.

\techterm{State scales along different dimensions.}
Each state component has a distinct growth dimension; resource estimation must
report them separately:

\begin{enumerate}[leftmargin=2em]
    \item \textbf{Global Slot Table}: grows linearly with the number of
          pre-configured capabilities; the base profile uses a global flat table,
          $2 \le \texttt{MAX\_SLOT} \le 2^{18}$ (even).
    \item \textbf{QP Context, Ordered Credit, Ordered Queue}: grow linearly with
          the number of QPs.
    \item \textbf{RQ Bitmap}: grows with QP count $\times$ receive window (MPR).
    \item \textbf{WQE Record, SQ Scoreboard}: grow with outstanding message count
          $\times$ PSN window.
    \item \textbf{Grouped Timer}: timing state is shared per group; state volume
          is proportional to the number of timer groups and decoupled from the
          number of outstanding PSNs.
    \item \textbf{Software recovery profile}: the sender shadow and per-QP timer
          move to the host; the NIC side incurs additional accepted-request/meta
          report bandwidth requirements.
\end{enumerate}

The two recovery profiles must report FPGA resources and CPU resources separately,
not merely compare NIC BRAM usage.

Centralized state maintenance solves the correctness problem but also introduces
a new bottleneck: four Parsers may simultaneously access the bitmap of the same QP,
and ACK/SACK processing and timeout may also concurrently update sender state.
Port contention is the primary risk in shared-state design.

\techterm{The primary risk for shared state is port contention.}
Four RX Parsers may access the bitmap of the same QP in the same cycle; ACK/SACK,
timeout, and retry completion may also concurrently update sender state.
Mitigation strategies:

\begin{enumerate}[leftmargin=2em]
    \item Work queues decouple feedback parsing from state access, preventing
          feedback processing from blocking the Parser.
    \item Banking provides independent throughput; for example, QP context is
          banked by low QPN bits to distribute concurrent accesses.
    \item Arbitration guarantees serialization on the same entry, avoiding
          write-after-write conflicts.
\end{enumerate}

Stress testing must simultaneously cover two access patterns: random multi-QP
access (which exposes bank distribution problems) and sustained high-pressure
single-QP access (which exposes hotspot arbitration bottlenecks).
If testing reveals WQE reverse-lookup or bitmap merge backlogs, banks, read-only
replicas, or pipeline stages should be added according to the measured bottleneck,
rather than pre-writing resource figures from synthesis-free analysis as protocol
constants.

State must not only be correct at runtime, but also manageable across its entire
lifecycle.
OpenRDMA defines a unified lifecycle state machine for all resources, ensuring
that no dangling references or state leaks occur across configuration, operation,
stop, and rebuild transitions.

\techterm{Explicit resource lifecycle.}
QPs, Slots, credits, replay shadows, and error epochs follow a unified state
machine: \texttt{RESET → CONFIGURE → RUN → STOP → DRAIN → RESET}.
The CPU/runtime writes QP/Slot/memory capabilities only during the CONFIGURE
stage; during RUN, it does not hot-revoke or reuse state that may still be
referenced by an accepted request.
STOP blocks new producer submissions; DRAIN covers all paths of admission, TX/RX,
retry, completion, and credit feedback; only after everything has completed or
been explicitly flushed may RESET clear counters and generations.

Any permission, DMA, shadow, credit, or queue-consistency error that cannot be
locally recovered enters fail-stop.
The hardware writes the error epoch/code and best-effort QPN/Slot diagnostic
information; the GPU producer stops submitting; the CPU decides whether to drain
or terminate in-flight work, then fully rebuilds the configuration.
Hot-fixing on existing state and continuing operation is not supported.

Throughput budget, resource scaling, port contention, and lifecycle collectively
define the feasibility boundary of the design.
But these boundaries must ultimately be closed through a hard metric: timing closure.
400\,MHz is the target clock, not a suggestion.

\techterm{Timing is closed by tool results.}
400~MHz is the target clock.
The synthesis report must separately give the frequency, slack, and resource
utilization of the following critical paths:

\begin{enumerate}[leftmargin=2em]
    \item Doorbell decode + Slot lookup (Section~\ref{sec:impl-work-submission})
    \item Bitmap merge + ACK/SACK generation (Section~\ref{sec:impl-rx-datapath})
    \item WQE Record reverse-lookup + Scoreboard arbitration
          (Section~\ref{sec:impl-reliability})
    \item DMA routing + Ordered Retire
          (Section~\ref{sec:impl-ordered-completion})
    \item Retry queue → unified TX Engine (Section~\ref{sec:impl-tx-datapath})
\end{enumerate}

The report must include real BRAM, cross-clock-domain, and debug logic.
Queue depths, bitmap widths, CC counter widths, and FPGA resource utilization are
to be finalized by synthesis and stress-test results; they are not pre-written as
protocol constants.

In summary, the implementation of OpenRDMA proceeds from Doorbell admission
through multi-path transmission, out-of-order reception, packet-level recovery,
and Ordered Flag publication, with the entire path constrained by unified resource
lifecycle and timing verification.
The boundaries described above both define the implementation responsibilities of
each module and provide verifiable acceptance criteria for subsequent synthesis,
stress testing, and end-to-end evaluation.

%% file: sections/06_evaluation.tex
\section{Evaluation}
\label{sec:evaluation}

Our evaluation examines how OpenRDMA converts packet-level parallelism into
high aggregate throughput while preserving reliable progress and ordered
completion. Our RTL--network co-simulation and modular end-to-end experiments
demonstrate four capabilities:
\textbf{380.9 Gb/s payload goodput} on four modeled 100 Gb/s paths,
\textbf{up to $3.51\times$ single-QP goodput} relative to a fixed path per QP,
\textbf{508--598 ns local drop-to-CQE recovery} across the tested single-loss
positions (620 and 682 ns for two- and three-packet losses), and
\textbf{ordered retirement under reversed DMA completion}.
We first establish the measurement conditions, then examine throughput,
path selection, request-to-packet latency, receive-side disorder, loss
recovery, and ordered completion.

\subsection{Experimental Method}
\label{sec:eval-method}

We connect Verilator simulations of the TX and RX RTL through an ns-3 network
for the throughput, path-policy, request-to-packet latency, and receive-window
experiments. Each emitted frame carries its RTL timestamp; ns-3 models
serialization, queuing, and propagation before delivering it to RX. Timestamped
advance messages coordinate the processes. This setup preserves the timing of
packet production and exposes how path selection changes arrival order.
Table~\ref{tab:eval-setup} summarizes the configuration.

\begin{table}[!htbp]
\centering
\caption{Configuration and measurement scope. Clock frequencies are simulation parameters.}
\label{tab:eval-setup}
\small
\renewcommand{\arraystretch}{1.12}
\begin{tabularx}{\linewidth}{l Y}
\toprule
Component & Configuration \\
\midrule
Tools &
BSC 2023.01; Verilator 5.048; ns-3 3.48; cocotb 2.1.0.dev0 \\
Datapath &
Four QPs, four RX lanes; 2.5 ns cycle for latency, throughput, policy,
and skew sweeps \\
Network &
Four independent 100 Gb/s point-to-point links; 200 ns delay per path
unless varied; 100-packet queues \\
Receive state &
16-packet window per QP; 16 payload slots per lane;
four Ordered Queue entries per QP \\
PMTU &
4096 B for TX and path-policy sweeps;
1024 B for receive-pressure tests; 256 B for recovery tests \\
DMA model &
Reversed completions for isolated TX latency;
in-order completions for throughput and policy sweeps;
same-source reverse diagnostic for throughput \\
Recovery tests &
Controlled single, burst, spread, and near-tail packet drops;
recovery timestamps span SACK readiness, report generation,
retry commit, and CQE \\
Wire profile &
Ethernet/IPv4/UDP/BTH/RETH; ICRC and FCS not modeled \\
\bottomrule
\end{tabularx}
\end{table}

\emph{Network payload goodput} divides unique payload bits by the interval from
first frame injection to last arrival. \emph{Mean network completion time}
measures each message from its earliest injection to the arrival of all its
frames. \emph{TX latency} starts at decoded submission acceptance and ends at
the first start of frame (SOF) or the latest end of frame (EOF) of the message.
Using the latest EOF accounts for a small tail packet finishing before an
earlier packet. \emph{Drop-to-CQE latency} starts when the test-side wire model
captures the dropped frame and ends at the RX completion event for the affected
message; it is a local RTL/end-to-end latency that excludes network
propagation, PCIe, host interrupt, and any timeout (RTO) service. We recompute
these quantities from archived frame and event records and check them against
the reported aggregates.

Rates are measured over deterministic finite batches, including startup and
drain, without a separate steady-state interval. The policy comparison uses
paired eight-message workloads; its latency metric is a mean, not a tail
estimate. The recovery experiments additionally use controlled packet drops to
isolate detection, feedback, retry, and completion costs.

The setup evaluates RTL datapath behavior with modeled DMA and network timing
for Sections~\ref{sec:eval-throughput}--\ref{sec:eval-window}. Physical
GPU/PCIe visibility, SRv6 encapsulation, congestion fairness, recovery behavior
outside the tested loss patterns, and target-FPGA timing remain outside this
measurement scope.

The recovery experiments carry four additional boundaries. First, the
end-to-end harness uses a zero-propagation wire model: frames are injected
directly into the RX lane, so recovery latency excludes network round-trip
time, switch queuing, PCIe, host interrupts, and production driver polling.
Second, the harness observes the receiver's local SACK event; on-wire
SACK/NAK encoding, ICRC, and a real RTO service are not modeled. The CPU
fetches a snapshot report by polling after a fixed interval, and the retry
engine admits and executes one request at a time, so burst and spread losses
serialize their retries. Third, the tested drop positions (indices 1--7) all
leave at least three subsequent packets, so the post-hole SACK threshold is
satisfied; first-packet and tail losses require the timeout path and are not
covered. Fourth, all recovery measurements are single deterministic runs
without seed variance or confidence intervals.

Functional regression provides the basis for interpreting the performance
results (Table~\ref{tab:eval-correctness}). We also compare three online advance
windows (100 ns, 1 $\mu$s, and 5 $\mu$s), a free-running RX clock, and offline
replay using synchronized TX RTL. All 37 consistency checks pass: per-frame
injection and arrival events agree, and the 20 per-path PCAP files match their
corresponding reference captures byte for byte. Thus these coordination choices
preserve the reference workload's observable network behavior.

\begin{table}[!htbp]
\centering
\caption{Functional checks accompanying the performance matrix. RX suites
overlap; skipped evaluation cases are exercised by dedicated ordering tests.}
\label{tab:eval-correctness}
\small
\begin{tabularx}{\linewidth}{Y l}
\toprule
Check & Result \\
\midrule
TX short regression & 67 pass \\
RX short / long regression & 37 pass, 3 skip / 38 pass, 2 skip \\
Online functional reference & 11/11 frames; 6/6 messages; zero drops \\
Five-mode timing consistency & 37 checks pass \\
Reversed DMA completion $3,2,1,0$ & Retirement $0,1,2,3$ \\
Ordered Queue capacity &
1/2/4 credits returned; fifth blocked Flag rejected \\
\bottomrule
\end{tabularx}
\end{table}

\FloatBarrier

\subsection{High Aggregate Datapath Throughput}
\label{sec:eval-throughput}

\begin{figure}[!htbp]
\centering
\includegraphics[width=\linewidth]{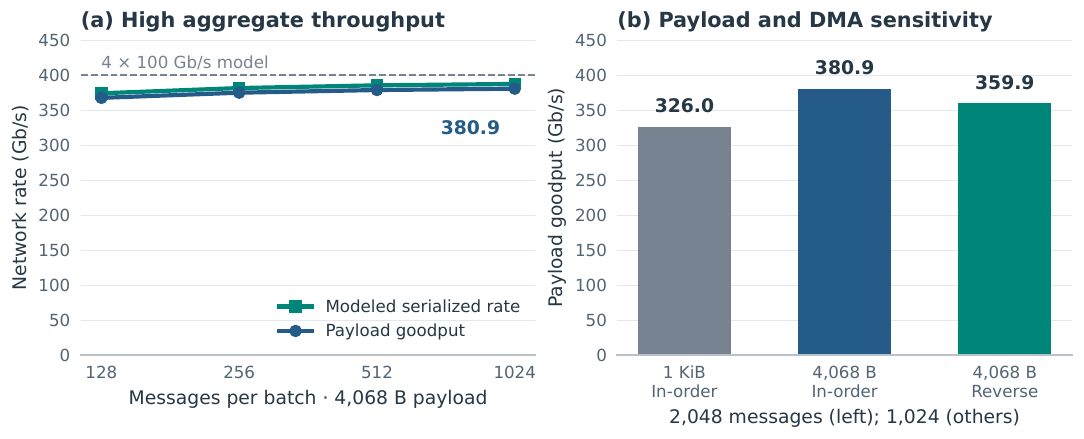}
\caption{Aggregate throughput at a 400 MHz modeled clock.
(a) All four batch sizes use 4,068 B messages, four QPs, and in-order DMA
completions. The serialized rate includes the bytes modeled by ns-3.
(b) Smaller payloads and reversed DMA completions retain high throughput;
the two 4,068 B points use identical source and workload, changing only
completion order. All bars start at zero.}
\label{fig:eval-throughput}
\end{figure}

OpenRDMA sustains high aggregate payload delivery across the four modeled links.
For 4,068 B single-packet payloads on four QPs, increasing the batch from
128 to 1,024 messages raises network goodput from 367.8 to 380.9 Gb/s
(Figure~\ref{fig:eval-throughput}(a)). The intermediate 256- and 512-message
batches achieve 375.2 and 379.0 Gb/s.

The diminishing increase in Figure~\ref{fig:eval-throughput}(a) shows that
startup and drain contribute progressively less to the measured rate.
At the largest batch, useful payload alone reaches 95.2\% of the aggregate
400 Gb/s link rate.

Including modeled framing and the ns-3 PPP header gives 387.7 Gb/s, or
96.9\% of the configured aggregate rate. This serialized-byte measure is
distinct from payload goodput and does not include every physical Ethernet
overhead. A 2,048-message batch with 1 KiB payloads achieves 326.0 Gb/s
payload goodput (Figure~\ref{fig:eval-throughput}(b)), demonstrating high
throughput beyond the large single-packet payload configuration.

The same-source DMA diagnostic quantifies sensitivity to completion order.
Changing only the completion model from in-order to reverse for the
4,068 B, 1,024-message workload yields 359.9 Gb/s, retaining 94.5\% of the
in-order result. We report this controlled comparison separately from earlier
source snapshots so that implementation changes and DMA scheduling are not
conflated.

\FloatBarrier

\subsection{Multipath Gains with Few Active QPs}
\label{sec:eval-multipath}

Packet spraying lets one QP exploit multiple paths without requiring the
application to create more connections. We compare round-robin packet
selection among ready paths with a fixed policy that maps each QP to one path
using its QPN. Each pair has the same eight offered messages, QP assignment,
4 KiB PMTU, in-order DMA model, and four equal-delay links. Actual injection
times may differ because the policy changes backpressure.

\begin{figure}[!htbp]
\centering
\includegraphics[width=\linewidth]{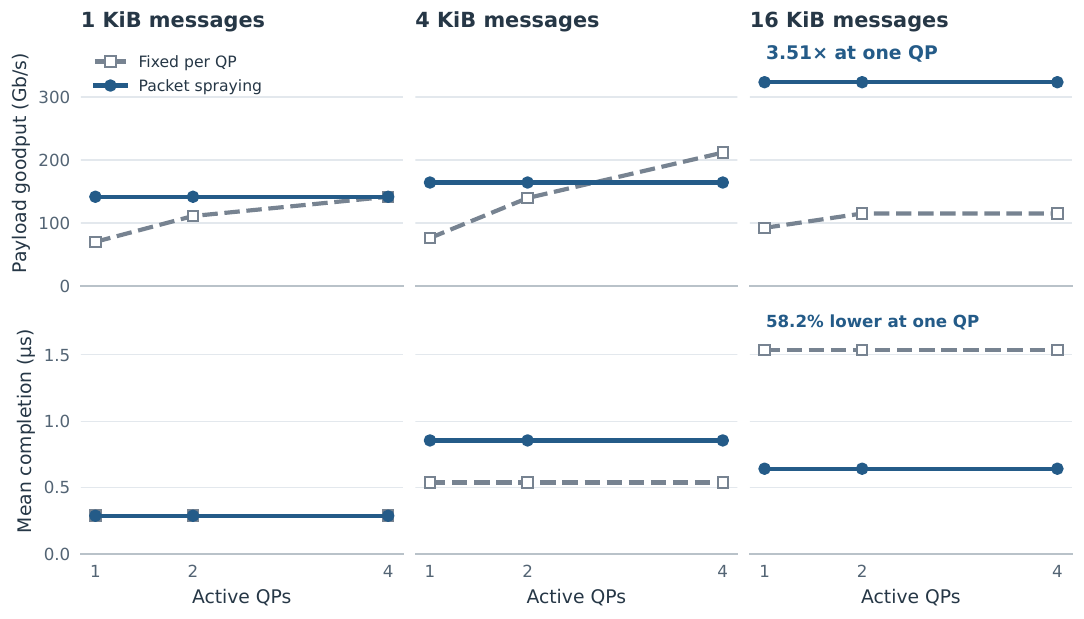}
\caption{Complete paired policy matrix: three message sizes, three QP counts,
and two policies. Upper panels report payload goodput; lower panels report
mean network completion time. Shared scales within each row preserve comparison
across message sizes. Spraying provides its largest gain for 16 KiB messages.}
\label{fig:eval-multipath}
\end{figure}

Figure~\ref{fig:eval-multipath} presents the complete policy matrix.

For one QP and 16 KiB messages, spraying increases goodput from 92.2 to
323.5 Gb/s, a $3.51\times$ improvement, while reducing mean network completion
from 1.534 to 0.642 $\mu$s (58.2\%). The 16 KiB workload remains at
323.5 Gb/s with two and four QPs, exceeding the corresponding fixed-path
results by $2.81\times$.

These results demonstrate that packet-level scheduling can expose parallelism
within the traffic of a small number of QPs. With one QP, 1 KiB and 4 KiB
messages improve by $2.02\times$ and $2.16\times$, respectively.

A 1 KiB message is a single packet: its 287.68 ns network completion time is
unchanged, while distributing successive messages raises batch goodput.
At four QPs, the 1 KiB policies converge.

For 4 KiB messages, fixed mapping reaches 211.8 Gb/s versus 164.2 Gb/s for
spraying; mean completion is 0.538 versus 0.855 $\mu$s. This case contains a
4,068 B fragment and a 28 B tail per message, so balancing packet opportunities
does not ensure balanced payload bytes or shorter completion.

The matrix therefore identifies where spraying is most effective: exploiting
parallel transfer opportunities that fixed QP placement leaves unused,
especially for the tested 16 KiB messages.

\FloatBarrier

\subsection{Request-to-Packet Latency}
\label{sec:eval-tx}

\subsubsection{GPU-Ready to First On-NIC Payload}

The packet-emission measurements below begin after the NIC has accepted a
fully decoded submission. To expose the control-path work that precedes packet
emission, we separately analyze the path from GPU data readiness to the first
payload beat returned by DMA.

The active RTL begins at \texttt{DecodedGdrpSubmission}; the GPU system-scope
release, credit reservation, 8-byte Doorbell store, PCIe transport, and
address-to-QPN and Doorbell-field decode are outside this boundary. We therefore
report the RTL portion as a code-derived cycle budget and retain the external
terms symbolically rather than presenting them as measured latency.

The RTL path is serialized by data dependencies. The Slot lookup returns the
local key, remote address, and length bound; the MR lookup validates the key and
returns the page-table offset; QP reservation supplies PMTU and PSN state; and
the PGT lookup translates the first physical page before the burst splitter can
emit a DMA request.

Counting the intervening rule and FIFO boundaries under no stalls and
one-cycle table lookups gives Table~\ref{tab:eval-submission-cycles}.

\begin{table}[!htbp]
\centering
\caption{Code-derived submission-path cycle budget. The count starts when a
fully decoded submission is visible to M0 and ends at the first exported DMA
request handshake. It is a structural lower bound, not a post-route timing
measurement.}
\label{tab:eval-submission-cycles}
\small
\begin{tabularx}{\linewidth}{Y r}
\toprule
Serialized segment & Cycles \\
\midrule
M0 Slot, MR, QP, and PGT processing plus burst split & 12 \\
M0 adapter and context/chunk synchronization & 3 \\
M1 normalization, scheduling, and reorder installation & 3 \\
Exported \texttt{DmaReadRequest} handshake & 1 \\
\midrule
Total to first external DMA request & 19 \\
\bottomrule
\end{tabularx}
\end{table}

Thus, from decoded-submission visibility to the first external DMA request,
\begin{equation}
C_{\mathrm{req}} = 12 + 3 + 3 + 1 = 19 \text{ cycles}.
\end{equation}

The 12-cycle M0 component consists of three cycles each for Slot, MR, and PGT
issue/lookup/join, two cycles for QP reservation and its join, and one cycle
for burst splitting. Target BRAMs or rule conflicts may add cycles, so this
count describes the current functional organization under the stated
assumptions.

M1 does not release a chunk until all completion beats for that chunk have
arrived. Let $R$ be the external DMA request-to-first-completion delay in NIC
cycles and $K$ the number of 128-byte completion beats in the first chunk.
With one completion beat per cycle and no downstream backpressure, the first
payload beat and the complete chunk reach \texttt{payloadOut} at

\begin{align}
C_{\mathrm{first}}
&= C_{\mathrm{req}} + R + K + 3, \\
C_{\mathrm{chunk}}
&= C_{\mathrm{req}} + R + 2K + 2.
\end{align}

For a 4-KiB-aligned 1-KiB first chunk, $K=8$, yielding $R+30$ cycles to the
first payload beat and $R+37$ cycles to the complete chunk. At the 400-MHz
model clock used by the performance sweeps, the fixed portions correspond to
75 and 92.5 ns, respectively, before adding $R$; 400 MHz remains a simulation
parameter rather than an achieved FPGA frequency.

The complete data-ready-to-first-payload path includes terms outside the
active RTL. Expressed in time rather than by mixing GPU and NIC clock domains,

\begin{equation}
\begin{split}
T_{\mathrm{ready\rightarrow first}} ={}&
T_{\mathrm{release}}
+ T_{\mathrm{credit}}
+ T_{\mathrm{Doorbell/PCIe}} \\
&+
\frac{
D_{\mathrm{decode}} + 19 + R + K + 3
}{
f_{\mathrm{NIC}}
}.
\end{split}
\end{equation}

Here $D_{\mathrm{decode}}$ covers address/QPN extraction and Doorbell-field
decode. This decomposition makes the evidence boundary explicit: the RTL
establishes the 12/19-cycle NIC-side structure, but the GPU release, atomic
credit reservation, MMIO write, and PCIe transport still require a matched
physical measurement.

\paragraph{Comparison with IBGDA.}

IBGDA and GDAKI move queue operation onto the GPU but retain GPU-side WQE
construction, Doorbell-record management, and a dependent NIC WQE fetch before
payload DMA~\cite{ibgda}. Baydamirli et al. report p50 issue and
put-plus-completion times for an 8-byte inline RDMA Write on an H200/ConnectX-7
200-Gb/s InfiniBand platform.\footnote{
J. Baydamirli, I. Ismayilov, K. Oktay, and D. Unat,
``GPU-Initiated Communication: Dissecting Down to the Bone,''
2026, \url{https://arxiv.org/abs/2610.01380}.}

Table~\ref{tab:eval-ibgda-cycles} also converts these times using an assumed
2.0-GHz SM clock. The cycle values are our conversion, not cycle counts
reported by the authors.

\begin{table}[!htbp]
\centering
\caption{Published p50 GPU-side latency and its cycle-equivalent conversion at
an assumed 2.0-GHz SM clock. These GPU cycles are not directly comparable with
OpenRDMA NIC RTL cycles.}
\label{tab:eval-ibgda-cycles}
\small
\begin{tabular}{lrrrr}
\toprule
Path &
Issue ($\mu$s) &
Put+comp. ($\mu$s) &
Issue (kcycles) &
Put+comp. (kcycles) \\
\midrule
mini-gda, inline & 0.70 & 4.03 & 1.40 & 8.06 \\
GDAKI, inline & 1.60 & 6.02 & 3.20 & 12.04 \\
NVSHMEM, internal & 4.26 & 9.18 & 8.52 & 18.36 \\
NVSHMEM, public & 5.31 & 11.10 & 10.62 & 22.20 \\
\bottomrule
\end{tabular}
\end{table}

The meaningful comparison is architectural rather than a subtraction of cycle
counts. For one operation, IBGDA publishes a WQE and Doorbell record and rings
the Doorbell; the NIC must fetch that WQE before fetching the payload.
OpenRDMA publishes only an 8-byte Doorbell token.

Slot, MR, QP, and PGT state already reside on the NIC, so the dependent WQE
fetch is replaced by the 12-cycle M0 lookup and split path. Isolating only
this mechanism gives

\begin{equation}
T_{\mathrm{IBGDA}} - T_{\mathrm{OpenRDMA}}
\approx
T_{\mathrm{WQE\ build}}
+
T_{\mathrm{dbrec}}
+
T_{\mathrm{WQE\ fetch}}
-
\frac{12}{f_{\mathrm{NIC}}}.
\end{equation}

The common GPU release, MMIO/PCIe transport, and payload DMA terms are omitted
from this expression. Consequently, the supported result is narrow:
OpenRDMA removes GPU-side WQE construction and one dependent WQE fetch and
replaces them with a 12-cycle on-NIC lookup path.

The 19-cycle request-issue boundary and the $R+30$/$R+37$ payload boundaries
quantify the resulting RTL pipeline. Establishing an end-to-end latency
advantage over IBGDA still requires both designs to be measured on the same
GPU, NIC, PCIe link, payload, fence scope, and completion endpoint.

\subsubsection{Accepted Request to Packet Emission}

\begin{figure}[!htbp]
\centering
\includegraphics[width=\linewidth]{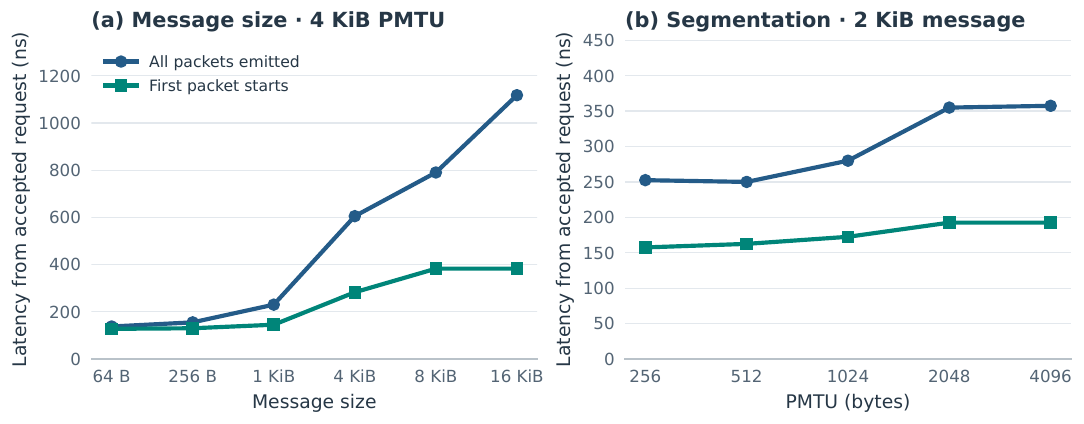}
\caption{Isolated TX latency with reversed modeled DMA completions at 400 MHz.
(a) Message-size sweep at 4 KiB PMTU.
(b) PMTU sweep for one 2 KiB message.
Complete emission uses the latest EOF across all packets.
Horizontal positions represent the tested categories;
connecting lines guide the eye.}
\label{fig:eval-tx-latency}
\end{figure}

The transmitter exposes parallelism while keeping first-packet startup short.
With a 4 KiB PMTU, a 64 B request starts after 127.5 ns and finishes after
137.5 ns; a 1 KiB message starts after 145 ns and finishes after 230 ns.
Figure~\ref{fig:eval-tx-latency} summarizes the latency sweep.

First-packet latency is 382.5 ns for both 8 KiB and 16 KiB, while complete
emission takes 790 and 1,117.5 ns. The pipeline can therefore begin
transmission before the entire message has drained.

Segmentation offers another way to expose concurrent work. For a fixed 2 KiB
message, a 512 B PMTU produces five packets and completes emission in 250 ns,
compared with 357.5 ns for a 4 KiB PMTU. At 256 B, nine packets take 252.5 ns.

This sweep illustrates the interaction between parallel packet output and
packet-processing overhead; network-level PMTU selection additionally depends
on framing and receive-side pressure.

\FloatBarrier

\subsection{Receive Progress Under Path Skew}
\label{sec:eval-window}

The receiver completes all 15 short-burst tests across five path-delay profiles
and one, two, or four active QPs. Each 1 KiB message is segmented using a
1 KiB PMTU.

\begin{figure}[!htbp]
\centering
\includegraphics[width=\linewidth]{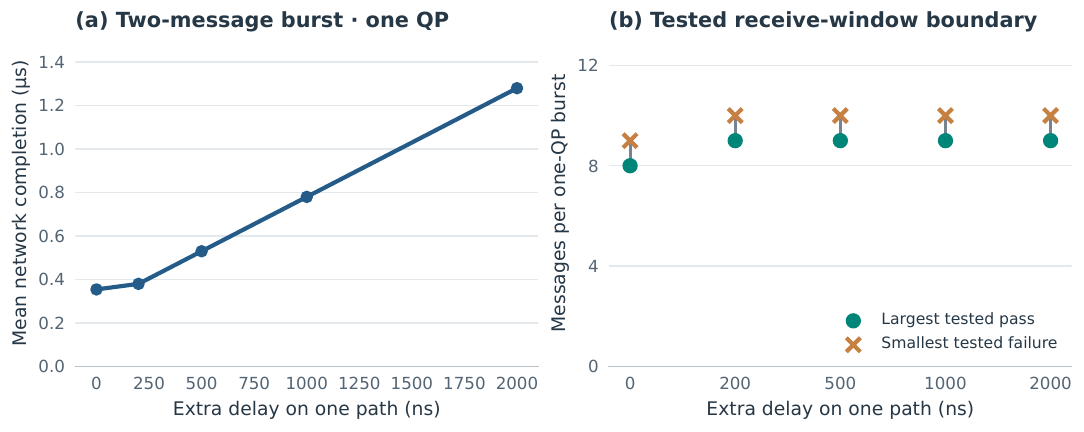}
\caption{Receive progress with three 200 ns paths and one path with added delay.
(a) Mean network completion for a two-message, one-QP burst.
(b) Largest tested passing and smallest tested failing bursts for the same
message size and PMTU. Markers show tested workloads, not measured buffer
occupancy or a universal capacity limit. Failures report RX out-of-window
errors.}
\label{fig:eval-window}
\end{figure}

With one QP and two messages, mean network completion increases from
0.354 $\mu$s on equal-delay paths to 1.280 $\mu$s when one path receives an
additional 2 $\mu$s delay. Figure~\ref{fig:eval-window} shows both the
short-burst latency and the observed receive-window boundary.

The four-QP cases use four messages and serve as functional coverage rather
than an equal-load scaling comparison.

The longer-burst sweep makes the finite receive window explicit. Eight messages
pass on equal-delay paths and nine reach the window boundary; for each nonzero
skew tested, nine pass and ten reach it.

All 16 archived failing cases are retained as boundary observations, while
ns-3 delivers their frames. These thresholds reflect packet order and window
advancement, not a cumulative message-count limit. They identify the operating
condition that receive-window admission or a recovery loop must enforce for
longer segmented bursts.

The receive-window experiment isolates disorder rather than loss. Once a
packet is actually missing, forward progress additionally depends on detecting
the hole, fetching a snapshot report, and retransmitting the corresponding
packet. We evaluate that recovery path next.

\FloatBarrier

\subsection{Loss Recovery and Retransmission Cost}
\label{sec:eval-recovery}

We evaluate recovery under controlled packet drops and divide the recovery
timeline into four stages: drop to SACK readiness, SACK readiness to
report generation, report to retry commit, and retry commit to completion
of the affected message. A drop-to-CQE sample covers the local RTL and
end-to-end wire-model path only; it is not a network round-trip recovery time.

\begin{figure}[!htbp]
\centering
\includegraphics[width=\linewidth]{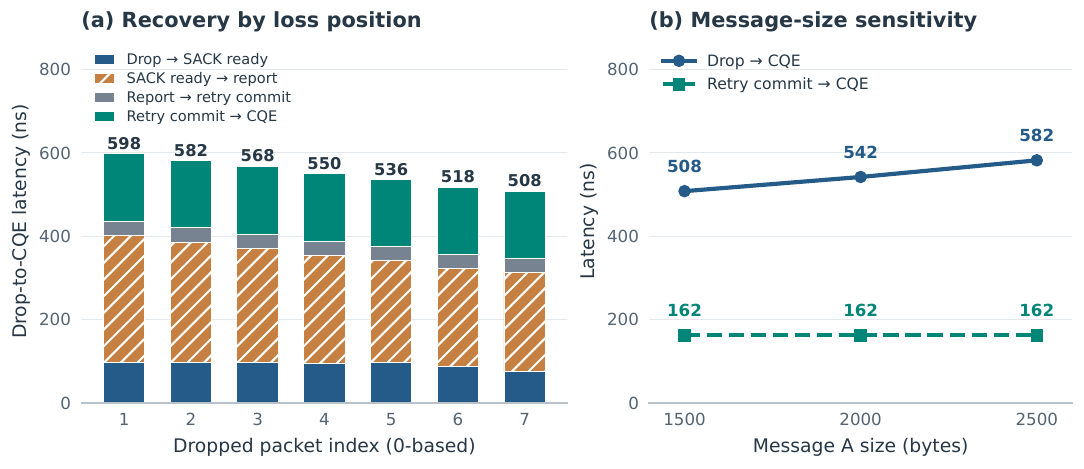}
\caption{Recovery latency under controlled single-packet loss in the modular
end-to-end harness. (a) Drop-to-CQE latency decomposed into SACK readiness,
report generation (labeled ``report'' in the figure), retry commit, and
post-commit completion stages for drop positions 1--7 (2,500 B message A,
256 B PMTU, 900 B message B, 400 MHz). (b) Message-size sensitivity of total
recovery latency and the retry-commit-to-CQE interval. The harness uses zero
network propagation, local SACK events, a fixed polling interval for the
report, one active retry, and a single deterministic run; times exclude PCIe,
host interrupt, and RTO.}
\label{fig:eval-recovery}
\end{figure}

Figure~\ref{fig:eval-recovery} first examines how the recovery latency changes
with loss position and message size.

Across dropped-packet indices 1 through 7, total drop-to-CQE latency decreases
from 598 to 508 ns. The measured values are 598, 582, 568, 550, 536, 518,
and 508 ns, respectively. Each tested position leaves at least three subsequent
packets, so the runtime's post-hole SACK threshold of three is satisfied;
first-packet and tail losses require the timeout path and are not covered.

The decomposition shows that recovery cost is not confined to retransmission
itself: hole detection, feedback, and report generation before retry commit account
for a substantial fraction of the total delay. Across positions 1--7, SACK
readiness takes 76--96 ns, the report generation takes 236--306 ns, CPU
decision plus retry commit takes about 34 ns, and retry commit to CQE takes
160--164 ns.
The report segment dominates because the prototype polls a snapshot after a
fixed interval; it is a policy cost, not a hardware lower bound.
First-frame-to-CQE remains 616--634 ns across all positions, so message
completion is stable; the drop-to-CQE trend mainly reflects how much of
message A remains when the drop occurs, not a faster recovery path.

Figure~\ref{fig:eval-recovery}(b) varies the size of message A from
1,500 to 2,500 B. Total drop-to-CQE latency increases from 508 to 542 and
582 ns, whereas the retry-commit-to-CQE interval remains 162 ns in all three
cases.

Thus, for these workloads, message size changes the pre-retry portion of the
recovery path while the post-commit retry interval remains stable. Multiple
losses extend the same local latency: two adjacent dropped packets raise
drop-to-CQE to 620 ns, and three spread drops raise it to 682 ns. The retry
engine executes one request at a time, so these cases serialize retry
admission, fetch, and ejection; the values characterize the current prototype,
not a concurrent-retry design.

\begin{figure}[!htbp]
\centering
\includegraphics[width=\linewidth]{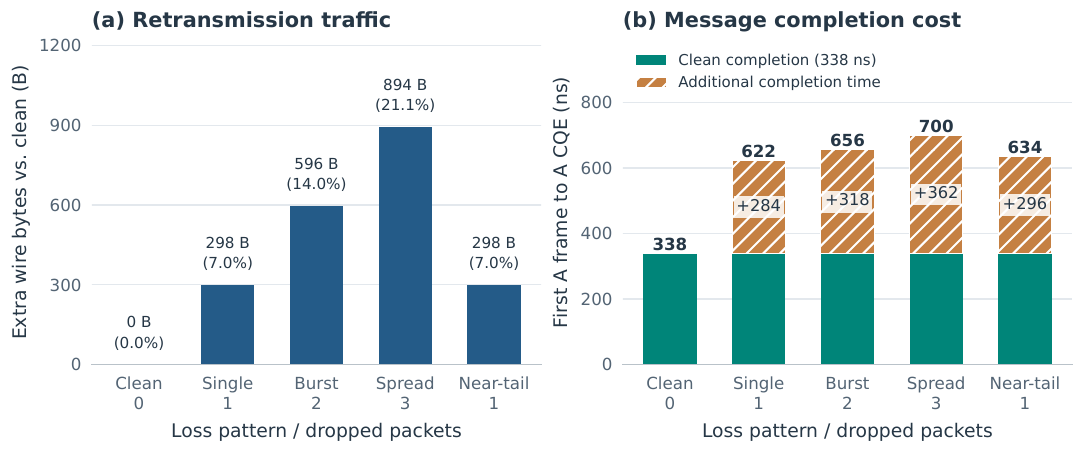}
\caption{Retransmission cost under controlled loss patterns (2,500 B message A,
900 B message B). (a) Additional wire bytes relative to the 4,242 B clean
trace; these are single-message event costs, not steady-state network
overhead. (b) First-A-frame-to-CQE completion, separated into the 338 ns clean
completion and the additional recovery delay. Burst and spread delays include
serialized retries.}
\label{fig:eval-retransmission-overhead}
\end{figure}

We next quantify the additional network traffic and message-completion cost
introduced by retransmission in
Figure~\ref{fig:eval-retransmission-overhead}.

A single dropped packet adds 298 B, two dropped packets in the burst pattern
add 596 B, and three dropped packets in the spread pattern add 894 B.
The near-tail single-loss case again adds 298 B.

These results increase in 298-B steps with the number of dropped packets in the
tested workloads, consistent with retransmitting the missing packet rather than
replaying the complete message.

The completion cost depends on more than retransmission volume. The clean
message completes in 338 ns. The single, burst, spread, and near-tail loss
patterns complete in 622, 656, 700, and 634 ns, corresponding to additional
delays of 284, 318, 362, and 296 ns.

The three-packet spread pattern therefore produces the largest completion
penalty in this set, while two different single-packet loss positions incur
the same wire overhead but different completion delays. Loss position and its
interaction with message progress therefore matter in addition to the number
of retransmitted bytes.

Retransmission also must not disturb the main data stream. In an interleaving
test, two packets of message B are injected before the retry request. The
retry still commits between B's first and last commit, preserving B's packet
order and final data, at the cost of a 142 ns increase in drop-to-CQE
(710 versus 568 ns at the same drop position).

Together, these experiments show that the evaluated recovery path restores
progress with local drop-to-CQE latency of 508--598 ns for single losses and
620--682 ns for two or three losses, while limiting additional traffic to the
retransmitted packets. They do not establish performance under arbitrary loss
rates, timeout-driven recovery, first-packet or tail losses, concurrent
retries, or congestion. Those conditions require a broader stochastic loss and
congestion sweep.

\FloatBarrier

\subsection{Ordered Completion and Blocking Cost}
\label{sec:eval-ordering}

Successful packet recovery restores receive progress, but it does not by itself
authorize an Ordered Flag. Publication additionally depends on retirement
progress reaching the Flag's barrier.

\begin{figure}[!htbp]
\centering
\includegraphics[width=\linewidth]{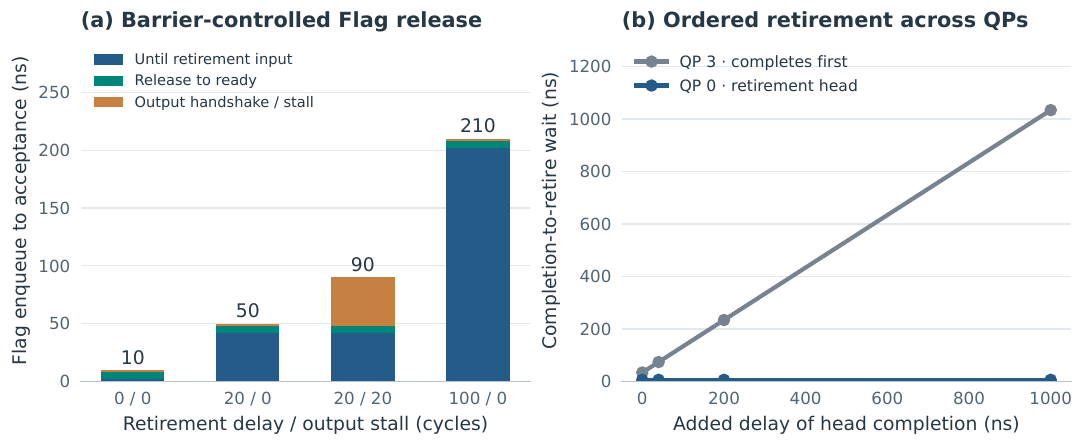}
\caption{Controlled ordering tests, using a 2 ns module-test cycle.
(a) Flag release separates barrier waiting, release-to-ready, and output
handshake or stall time.
(b) Four QPs complete in order $3,2,1,0$ and always retire in order
$0,1,2,3$. Times include the testbench drive and drain cadence.}
\label{fig:eval-ordering}
\end{figure}

We therefore validate completion ordering independently of packet arrival and
loss recovery. Figure~\ref{fig:eval-ordering} separates Flag publication delay
from the cost of a shared global retirement sequence.

We enqueue a Flag in the Ordered Queue and control when retirement evidence
reaches its barrier. This evidence is an injected module input, which isolates
the release rule.

With input delays of 0, 20, and 100 cycles, observed enqueue-to-acceptance
times are 10, 50, and 210 ns; the Flag becomes ready 6 ns after retirement
input acceptance. Adding 20 cycles of output stall to the 20-cycle delay case
extends acceptance to 90 ns
(Figure~\ref{fig:eval-ordering}(a)).

Credit return follows Flag request acceptance by 2 ns in the recorded
sequence, before the modeled Flag completion response. Dedicated capacity
tests return one, two, and four credits for the corresponding released Flags;
a fifth Flag presented to a blocked four-entry queue triggers the expected
queue-full error.

These checks exercise both dependency-controlled publication and bounded
resource accounting.

Finally, we reverse completion across four DMA lanes. All tested head delays
preserve retirement order $0,1,2,3$. The head request's observed
completion-to-retire interval remains 6 ns, while an additional 1,000 ns delay
at the head increases the earlier-completed QP 3's wait from 34 to 1,034 ns
(Figure~\ref{fig:eval-ordering}(b)).

This quantifies the cross-QP waiting cost of the shared retirement sequence
while confirming that completion disorder does not violate its ordering rule.

\FloatBarrier

\subsection{Implementation Evidence}
\label{sec:eval-state}

The implementation also passes RTL generation and generic synthesis checks.
Yosys synthesis of the four-channel TX scheduler reports 261,337 generic cells,
including 38,512 generic DFF and 56,899 generic MUX cells, with four BRAM2
instances. The RX top generates 21,533 lines of production Verilog.

These results establish elaboration and synthesis evidence for the evaluated
modules; target-device resource utilization and achievable frequency require
FPGA mapping and place-and-route.

Together, the experiments establish high aggregate datapath throughput,
substantial few-QP multipath gains, controlled selective-loss recovery, and
explicit receive and ordered-completion behavior under packet and DMA
disorder.

The remaining system-level evidence includes physical GPU/PCIe visibility,
broader congestion and stochastic-loss experiments, SRv6 encapsulation,
and target-device FPGA place-and-route.

\FloatBarrier

%% file: sections/08_conclusion.tex
\section{Conclusion}
\label{sec:conclusion}

OpenRDMA starts from a simple observation: a GPU communication fast path does not need to expose the full CPU-oriented WQE/CQE lifecycle. The GPU primarily needs to express a pre-authorized communication intent and learn when the resulting data is safe to consume. OpenRDMA therefore adopts a doorbell-in, flag-out interface: an 8-byte Doorbell triggers NIC-side request reconstruction, while a GPU-visible Flag is published only after the required payload placement and ordering conditions are satisfied.

The architecture combines this lightweight interface with packet-level multipath execution and explicit reliability state. Packets may be sprayed across multiple paths and arrive out of order; receiver-side bitmap state preserves useful progress, selective feedback drives retransmission, and ordered completion separates packet arrival from the point at which data can safely become visible to the GPU. Fast-path correctness state remains bounded in hardware, while configuration and low-frequency policy remain with the CPU/runtime.

Our RTL--network co-simulation reaches 380.9 Gb/s payload goodput across four modeled 100 Gb/s paths. Packet spraying improves single-QP goodput by up to $3.51\times$ over fixed per-QP path selection. For the tested single-loss positions, local drop-to-CQE recovery completes within 508--598 ns; two- and three-packet losses raise it to 620 and 682 ns, while retransmitting only missing packets. The experiments also expose the limits of aggressive parallelism: path skew creates receive-window pressure, and the current shared retirement sequence can introduce cross-QP head-of-line blocking.

The remaining work is to validate these mechanisms with physical GPU/PCIe measurements, broader loss and congestion workloads, SRv6 encapsulation, and target-FPGA place-and-route. More broadly, OpenRDMA suggests that GPU-native networking should not simply move a traditional CPU queue protocol onto the GPU. A cleaner interface is to let the GPU express \emph{intent}, let the NIC execute and preserve transport semantics, and expose only the final \emph{readiness} of the data.

%% file: bibliography/references.bib
@manual{rdma-aware-programming,
  title        = {{RDMA Aware Networks Programming User Manual}},
  author       = {{NVIDIA Corporation}},
  organization = {NVIDIA Corporation},
  edition      = {Version 1.7},
  year         = {2023},
  note         = {Last updated May 23, 2023},
  url          = {https://docs.nvidia.com/networking/display/rdmaawareprogrammingv17}
}

@manual{gpudirect-rdma,
  title        = {{GPUDirect RDMA}},
  author       = {{NVIDIA Corporation}},
  organization = {NVIDIA Corporation},
  edition      = {CUDA Toolkit 13.3},
  year         = {2026},
  note         = {CUDA Toolkit Documentation; accessed July 23, 2026},
  url          = {https://docs.nvidia.com/cuda/gpudirect-rdma/}
}

@manual{ibgda,
  title        = {{NVSHMEM Documentation: Using GPUDirect Async
                   Kernel-Initiated Communication}},
  author       = {{NVIDIA Corporation}},
  organization = {NVIDIA Corporation},
  edition      = {NVSHMEM 3.7.0},
  year         = {2026},
  note         = {Documentation for the IBGDA and GPUNetIO GDAKI transports;
                  accessed July 23, 2026},
  url          = {https://docs.nvidia.com/nvshmem/api/using.html}
}

@manual{nvshmem-signaling,
  title        = {{NVSHMEM Documentation: Signaling Operations}},
  author       = {{NVIDIA Corporation}},
  organization = {NVIDIA Corporation},
  edition      = {NVSHMEM 3.7.0},
  year         = {2026},
  note         = {Documentation for put-with-signal and signal-wait operations;
                  accessed July 28, 2026},
  url          = {https://docs.nvidia.com/nvshmem/api/gen/api/signal.html}
}

@inproceedings{potluri2017gpu,
  author    = {Potluri, Sreeram and
               Goswami, Anshuman and
               Rossetti, Davide and
               Newburn, Chris J. and
               Venkata, Manjunath Gorentla and
               Imam, Neena},
  title     = {{GPU-Centric Communication on NVIDIA GPU Clusters
                with InfiniBand: A Case Study with OpenSHMEM}},
  booktitle = {2017 IEEE 24th International Conference on
               High Performance Computing (HiPC)},
  pages     = {253--262},
  year      = {2017},
  month     = dec,
  publisher = {IEEE},
  doi       = {10.1109/HiPC.2017.00037}
}

@inproceedings{mprdma,
  author    = {Lu, Yuanwei and
               Chen, Guo and
               Li, Bojie and
               Tan, Kun and
               Xiong, Yongqiang and
               Cheng, Peng and
               Zhang, Jiansong and
               Chen, Enhong and
               Moscibroda, Thomas},
  title     = {{Multi-Path Transport for RDMA in Datacenters}},
  booktitle = {15th USENIX Symposium on Networked Systems
               Design and Implementation (NSDI 18)},
  pages     = {357--371},
  year      = {2018},
  month     = apr,
  publisher = {USENIX Association},
  address   = {Renton, WA},
  isbn      = {978-1-939133-01-4},
  url       = {https://www.usenix.org/conference/nsdi18/presentation/lu}
}

@inproceedings{gangidi2024rdma,
  author    = {Gangidi, Adithya and
               Miao, Rui and
               Zheng, Shengbao and
               Bondu, Sai Jayesh and
               Goes, Guilherme and
               Morsy, Hany and
               Puri, Rohit and
               Riftadi, Mohammad and
               Shetty, Ashmitha Jeevaraj and
               Yang, Jingyi and
               Zhang, Shuqiang and
               Fernandez, Mikel Jimenez and
               Gandham, Shashidhar and
               Zeng, Hongyi},
  title     = {{RDMA over Ethernet for Distributed AI Training
                at Meta Scale}},
  booktitle = {Proceedings of the ACM SIGCOMM 2024 Conference},
  pages     = {57--70},
  year      = {2024},
  month     = aug,
  publisher = {Association for Computing Machinery},
  doi       = {10.1145/3651890.3672233}
}

@manual{mrc-spec,
  author       = {Sohan, Rip and
                  Spada, Eric and
                  Davis, Eric and
                  Handley, Mark and
                  Burstein, Idan and
                  Hurson, Tony and
                  Jose, Jithin and
                  Kashyap, Vivek and
                  Pan, Rong and
                  Sur, Sayantan},
  title        = {{Multipath Reliable Connection (MRC) Specification}},
  organization = {Open Compute Project Foundation},
  edition      = {Revision 1.0},
  year         = {2026},
  month        = mar,
  note         = {Joint contribution from AMD, Broadcom, Intel,
                  Microsoft, NVIDIA, and OpenAI;
                  publication date March 21, 2026},
  url          = {https://www.opencompute.org/documents/ocp-mrc-1-0-pdf}
}

@inproceedings{left,
  author    = {Huang, Peihao and
               Zhang, Xin and
               Chen, Zhigang and
               Liu, Can and
               Chen, Guo},
  title     = {{LEFT: LightwEight and FasT Packet Reordering for RDMA}},
  booktitle = {Proceedings of the 8th Asia-Pacific Workshop
               on Networking (APNet 2024)},
  pages     = {67--73},
  year      = {2024},
  month     = aug,
  publisher = {Association for Computing Machinery},
  doi       = {10.1145/3663408.3663418}
}

@inproceedings{irn,
  author    = {Mittal, Radhika and
               Shpiner, Alexander and
               Panda, Aurojit and
               Zahavi, Eitan and
               Krishnamurthy, Arvind and
               Ratnasamy, Sylvia and
               Shenker, Scott},
  title     = {{Revisiting Network Support for RDMA}},
  booktitle = {Proceedings of the 2018 Conference of the ACM
               Special Interest Group on Data Communication
               (SIGCOMM '18)},
  pages     = {313--326},
  year      = {2018},
  month     = aug,
  publisher = {Association for Computing Machinery},
  doi       = {10.1145/3230543.3230557}
}

@inproceedings{dctcp,
  author    = {Alizadeh, Mohammad and
               Greenberg, Albert and
               Maltz, David A. and
               Padhye, Jitendra and
               Patel, Parveen and
               Prabhakar, Balaji and
               Sengupta, Shubhashis and
               Sridharan, Murari},
  title     = {{Data Center TCP (DCTCP)}},
  booktitle = {Proceedings of the ACM SIGCOMM 2010 Conference},
  pages     = {63--74},
  year      = {2010},
  month     = aug,
  publisher = {Association for Computing Machinery},
  doi       = {10.1145/1851182.1851192}
}

@inproceedings{dcqcn,
  author    = {Zhu, Yibo and
               Eran, Haggai and
               Firestone, Daniel and
               Guo, Chuanxiong and
               Lipshteyn, Marina and
               Liron, Yehonatan and
               Padhye, Jitendra and
               Raindel, Shachar and
               Yahia, Mohamad Haj and
               Zhang, Ming},
  title     = {{Congestion Control for Large-Scale RDMA Deployments}},
  booktitle = {Proceedings of the 2015 ACM Conference on
               Special Interest Group on Data Communication
               (SIGCOMM '15)},
  pages     = {523--536},
  year      = {2015},
  month     = aug,
  publisher = {Association for Computing Machinery},
  doi       = {10.1145/2785956.2787484}
}

@manual{nccl-docs,
  title        = {{NVIDIA Collective Communication Library (NCCL)
                  Documentation}},
  author       = {{NVIDIA Corporation}},
  organization = {NVIDIA Corporation},
  edition      = {NCCL 2.30.3},
  year         = {2026},
  note         = {Accessed July 28, 2026},
  url          = {https://docs.nvidia.com/deeplearning/nccl/user-guide/index.html}
}

@manual{nccl-device-api,
  title        = {{NCCL Device API -- GIN}},
  author       = {{NVIDIA Corporation}},
  organization = {NVIDIA Corporation},
  edition      = {NCCL 2.30.3},
  year         = {2026},
  note         = {Device-initiated communication documentation; accessed July 28, 2026},
  url          = {https://docs.nvidia.com/deeplearning/nccl/archives/nccl_2303/user-guide/docs/api/device_gin.html}
}

@manual{doca-gpunetio,
  title        = {{DOCA GPUNetIO}},
  author       = {{NVIDIA Corporation}},
  organization = {NVIDIA Corporation},
  year         = {2026},
  note         = {Last updated March 2, 2026; accessed July 28, 2026},
  url          = {https://docs.nvidia.com/doca/sdk/doca-gpunetio/}
}

@manual{cuda-c-programming-guide,
  title        = {{CUDA C++ Programming Guide}},
  author       = {{NVIDIA Corporation}},
  organization = {NVIDIA Corporation},
  edition      = {CUDA Toolkit 13.0},
  year         = {2026},
  note         = {Memory model and memory fence functions; accessed July 28, 2026},
  url          = {https://docs.nvidia.com/cuda/archive/13.0.0/cuda-c-programming-guide/index.html}
}

@manual{nvshmem-env,
  title        = {{NVSHMEM Environment Variables}},
  author       = {{NVIDIA Corporation}},
  organization = {NVIDIA Corporation},
  edition      = {NVSHMEM Documentation},
  year         = {2026},
  note         = {IBGDA configuration variables; accessed July 28, 2026},
  url          = {https://docs.nvidia.com/nvshmem/api/gen/env.html}
}

@manual{rfc8754,
  title        = {{IPv6 Segment Routing Header (SRH)}},
  author       = {Filsfils, Clarence and
                  Dukes, Darren and
                  Previdi, Stefano and
                  Leddy, John and
                  Matsushima, Satoru and
                  Voyer, Daniel},
  organization = {RFC Editor},
  year         = {2020},
  month        = mar,
  note         = {RFC 8754},
  doi          = {10.17487/RFC8754},
  url          = {https://www.rfc-editor.org/info/rfc8754}
}

@manual{rfc8986,
  title        = {{Segment Routing over IPv6 (SRv6) Network Programming}},
  author       = {Filsfils, Clarence and
                  Camarillo, Pablo and
                  Leddy, John and
                  Voyer, Daniel and
                  Matsushima, Satoru and
                  Li, Zhenbin},
  organization = {RFC Editor},
  year         = {2021},
  month        = feb,
  note         = {RFC 8986},
  doi          = {10.17487/RFC8986},
  url          = {https://www.rfc-editor.org/info/rfc8986}
}

@manual{rfc6437,
  title        = {{IPv6 Flow Label Specification}},
  author       = {Amante, Shane and
                  Carpenter, Brian and
                  Jiang, Sheng and
                  Rajahalme, Jarno},
  organization = {RFC Editor},
  year         = {2011},
  month        = nov,
  note         = {RFC 6437},
  doi          = {10.17487/RFC6437},
  url          = {https://www.rfc-editor.org/info/rfc6437}
}

@manual{rfc2992,
  title        = {{Analysis of an Equal-Cost Multi-Path Algorithm}},
  author       = {Hopps, Christian E.},
  organization = {RFC Editor},
  year         = {2000},
  month        = nov,
  note         = {RFC 2992},
  doi          = {10.17487/RFC2992},
  url          = {https://www.rfc-editor.org/info/rfc2992}
}

@manual{uec-spec,
  title        = {{Ultra Ethernet Specification}},
  author       = {{Ultra Ethernet Consortium}},
  organization = {Ultra Ethernet Consortium},
  edition      = {Version 1.0.3},
  year         = {2026},
  month        = jul,
  note         = {Specification history; current published version dated July 16, 2026; accessed July 28, 2026},
  url          = {https://ultraethernet.org/specification-history/}
}

@manual{intel-ftile-ethernet,
  title        = {{F-Tile Ethernet Hard IP User Guide}},
  author       = {{Altera Corporation}},
  organization = {Altera},
  edition      = {Version 25.3.1},
  year         = {2026},
  month        = jun,
  note         = {Document ID 683023; accessed July 28, 2026},
  url          = {https://docs.altera.com/r/docs/683023/current}
}

@misc{mrc-srv6-paper,
  author        = {Araujo, Joao and
                   Chow, Alex and
                   Handley, Mark and
                   Lewis, Ryder and
                   Paasch, Christoph and
                   Padhye, Jitendra and
                   Papamichael, Michael and
                   Steinbrecher, Greg and
                   Tootoonchian, Amin and
                   Yuan, Lihua and others},
  title         = {{Resilient AI Supercomputer Networking using MRC and
                   SRv6}},
  year          = {2026},
  month         = may,
  note          = {arXiv: 2605.04333},
  archivePrefix = {arXiv},
  eprint        = {2605.04333},
  primaryClass  = {cs.NI},
  doi           = {10.48550/arXiv.2605.04333},
  url           = {https://arxiv.org/abs/2605.04333}
}

@manual{openai-mrc,
  title        = {{Supercomputer Networking to Accelerate Large Scale AI
                  Training}},
  author       = {{OpenAI}},
  organization = {OpenAI},
  year         = {2026},
  month        = may,
  note         = {Accessed July 28, 2026},
  url          = {https://openai.com/index/mrc-supercomputer-networking/}
}

@inproceedings{timely,
  author    = {Mittal, Radhika and
               Lam, Terry and
               Dukkipati, Nandita and
               Blem, Emily and
               Wassel, Hassan and
               Ghobadi, Monia and
               Vahdat, Amin and
               Wang, Yaogong and
               Wetherall, David and
               Zats, David},
  title     = {{TIMELY: RTT-based Congestion Control for the Datacenter}},
  booktitle = {Proceedings of the 2015 ACM Conference on Special Interest
               Group on Data Communication (SIGCOMM '15)},
  pages     = {537--550},
  year      = {2015},
  month     = aug,
  publisher = {Association for Computing Machinery},
  doi       = {10.1145/2785956.2787510}
}

@inproceedings{hpcc,
  author    = {Li, Yuliang and
               Miao, Rui and
               Liu, Hongqiang and
               Zhuang, Yan and
               Feng, Fei and
               Tang, Lingbo and
               Cao, Zheng and
               Zhang, Ming and
               Kelly, Frank and
               Alizadeh, Mohammad and
               Yu, Minlan},
  title     = {{HPCC: High Precision Congestion Control}},
  booktitle = {Proceedings of the ACM Special Interest Group on Data
               Communication (SIGCOMM '19)},
  pages     = {44--58},
  year      = {2019},
  month     = aug,
  publisher = {Association for Computing Machinery},
  doi       = {10.1145/3341302.3342085}
}

@inproceedings{swift,
  author    = {Kumar, Gautam and
               Dukkipati, Nandita and
               Jang, Keon and
               Wassel, Hassan M. G. and
               Wu, Xian and
               Montazeri, Behnam and
               Wang, Yaogong and
               Springborn, Kevin and
               Alfeld, Christopher and
               Ryan, Michael and
               Wetherall, David and
               Vahdat, Amin},
  title     = {{Swift: Delay is Simple and Effective for Congestion Control
                in the Datacenter}},
  booktitle = {Proceedings of the Annual Conference of the ACM Special
               Interest Group on Data Communication on the Applications,
               Technologies, Architectures, and Protocols for Computer
               Communication (SIGCOMM '20)},
  pages     = {514--528},
  year      = {2020},
  month     = aug,
  publisher = {Association for Computing Machinery},
  doi       = {10.1145/3387514.3406591}
}

@inproceedings{presto,
  author    = {He, Keqiang and
               Rozner, Eric and
               Agarwal, Kanak and
               Felter, Wes and
               Carter, John and
               Akella, Aditya},
  title     = {{Presto: Edge-based Load Balancing for Fast Datacenter
                Networks}},
  booktitle = {Proceedings of the 2015 ACM Conference on Special Interest
               Group on Data Communication (SIGCOMM '15)},
  pages     = {465--478},
  year      = {2015},
  month     = aug,
  publisher = {Association for Computing Machinery},
  doi       = {10.1145/2785956.2787507}
}

@misc{megatron-lm,
  author        = {Narayanan, Deepak and
                   Shoeybi, Mohammad and
                   Casper, Jared and
                   LeGresley, Patrick and
                   Patwary, Mostofa and
                   Korthikanti, Vijay Anand and
                   Vainbrand, Dmitri and
                   Kashinkunti, Prethvi and
                   Bernauer, Julie and
                   Catanzaro, Bryan and
                   Phanishayee, Amar and
                   Zaharia, Matei},
  title         = {{Efficient Large-Scale Language Model Training on GPU
                   Clusters Using Megatron-LM}},
  year          = {2021},
  note          = {arXiv: 2104.04473},
  archivePrefix = {arXiv},
  eprint        = {2104.04473},
  primaryClass  = {cs.CL},
  doi           = {10.48550/arXiv.2104.04473},
  url           = {https://arxiv.org/abs/2104.04473}
}

@inproceedings{zero,
  author    = {Rajbhandari, Samyam and
               Rasley, Jeff and
               Ruwase, Olatunji and
               He, Yuxiong},
  title     = {{ZeRO: Memory Optimizations Toward Training Trillion
                Parameter Models}},
  booktitle = {SC20: International Conference for High Performance
               Computing, Networking, Storage and Analysis},
  pages     = {1--16},
  year      = {2020},
  publisher = {IEEE},
  doi       = {10.1109/SC41405.2020.00024},
  url       = {https://arxiv.org/abs/1910.02054}
}

@misc{zero-plus-plus,
  author        = {Wang, Guanhua and
                   Qin, Heyang and
                   Jacobs, Sam Ade and
                   Holmes, Connor and
                   Rajbhandari, Samyam and
                   Ruwase, Olatunji and
                   Yan, Feng and
                   Yang, Lei and
                   He, Yuxiong},
  title         = {{ZeRO++: Extremely Efficient Collective Communication for
                   Giant Model Training}},
  year          = {2023},
  note          = {arXiv: 2306.10209},
  archivePrefix = {arXiv},
  eprint        = {2306.10209},
  primaryClass  = {cs.DC},
  doi           = {10.48550/arXiv.2306.10209},
  url           = {https://arxiv.org/abs/2306.10209}
}

@misc{deepspeed-ulysses,
  author        = {Jacobs, Sam Ade and
                   Tanaka, Masahiro and
                   Zhang, Chengming and
                   Zhang, Minjia and
                   Song, Leon and
                   Rajbhandari, Samyam and
                   He, Yuxiong},
  title         = {{DeepSpeed Ulysses: System Optimizations for Enabling
                   Training of Extreme Long Sequence Transformer Models}},
  year          = {2023},
  note          = {arXiv: 2309.14509},
  archivePrefix = {arXiv},
  eprint        = {2309.14509},
  primaryClass  = {cs.LG},
  doi           = {10.48550/arXiv.2309.14509},
  url           = {https://arxiv.org/abs/2309.14509}
}

@inproceedings{gpipe,
  author    = {Huang, Yanping and
               Cheng, Youlong and
               Bapna, Ankur and
               Firat, Orhan and
               Chen, Mia Xu and
               Chen, Dehao and
               Lee, HyoukJoong and
               Ngiam, Jiquan and
               Le, Quoc V. and
               Wu, Yonghui and
               Chen, Zhifeng},
  title     = {{GPipe: Efficient Training of Giant Neural Networks using
                Pipeline Parallelism}},
  booktitle = {Advances in Neural Information Processing Systems 32
               (NeurIPS 2019)},
  year      = {2019},
  url       = {https://arxiv.org/abs/1811.06965}
}

@misc{deepseek-v3,
  author        = {{DeepSeek-AI}},
  title         = {{DeepSeek-V3 Technical Report}},
  year          = {2024},
  note          = {arXiv: 2412.19437},
  archivePrefix = {arXiv},
  eprint        = {2412.19437},
  primaryClass  = {cs.CL},
  doi           = {10.48550/arXiv.2412.19437},
  url           = {https://arxiv.org/abs/2412.19437}
}

@manual{pytorch-distributed,
  title        = {{PyTorch Distributed Overview}},
  author       = {{PyTorch Contributors}},
  organization = {PyTorch},
  year         = {2026},
  note         = {Stable documentation; accessed July 28, 2026},
  url          = {https://docs.pytorch.org/docs/stable/distributed.html}
}

@manual{pytorch-fsdp,
  title        = {{PyTorch FullyShardedDataParallel}},
  author       = {{PyTorch Contributors}},
  organization = {PyTorch},
  year         = {2026},
  note         = {Stable documentation; accessed July 28, 2026},
  url          = {https://docs.pytorch.org/docs/stable/fsdp.html}
}

@manual{megatron-core,
  title        = {{Megatron-Core Developer Guide}},
  author       = {{NVIDIA Corporation}},
  organization = {NVIDIA Corporation},
  year         = {2026},
  note         = {API guide for tensor, pipeline, context, and expert parallelism; accessed July 28, 2026},
  url          = {https://docs.nvidia.com/megatron-core/developer-guide/latest/api-guide/}
}

@misc{demystifying-nccl,
  author        = {Hu, Zhiyi and
                   Shen, Siyuan and
                   Bonato, Tommaso and
                   Jeaugey, Sylvain and
                   Alexander, Cedell and
                   Spada, Eric and
                   Dinan, James and
                   Hammond, Jeff and
                   Hoefler, Torsten},
  title         = {{Demystifying NCCL: An In-depth Analysis of GPU
                   Communication Protocols and Algorithms}},
  year          = {2025},
  note          = {arXiv: 2507.04786},
  archivePrefix = {arXiv},
  eprint        = {2507.04786},
  primaryClass  = {cs.DC},
  doi           = {10.48550/arXiv.2507.04786},
  url           = {https://arxiv.org/abs/2507.04786}
}
